\documentclass{aa}
\usepackage{graphicx,natbib,color}
\usepackage[british]{babel}

\bibpunct{(}{)}{;}{a}{}{,}
\newcommand{\ignore}[1]{}
\newcommand{\eq}[1]{\begin{equation} #1 \end{equation}}

\newcommand{\dmath}[1]{\begin{displaymath} #1 \end{displaymath}}

\newcommand{\kpc}{\mathrm{kpc}}
\newcommand{\ccm}{\mathrm{cm}^{-3}}
\newcommand{\rj}{R_\mathrm{j}}
\newcommand{\vj}{v_\mathrm{j}}

\newcommand{\add}[1]{{#1}}

\begin{document}
\title{Very Light Jets II: \\ Bipolar large scale simulations in King atmospheres}
\titlerunning{Very Light Jets II}
\author{Martin~Krause \inst{1,2}}
\institute{Landessternwarte K\"onigstuhl, D-69117 Heidelberg, Germany 
\and 
Astrophysics Group, Cavendish Laboratory, Madingley Road, Cambridge CB3 0HE, United Kingdom}
\offprints{M.Krause,
\email{M.Krause@lsw.uni-heidelberg.de}}
\date{Received \today / Accepted <date>}
\abstract{
Hydrodynamic jets, underdense with respect to their environment by a factor of up to
$10^4$, were computed in axisymmetry as well as in 3D. They finally reached a size of 
up to 220 jet radii, corresponding to a 100~kpc sized radio galaxy.
The simulations are "bipolar", involving both jets.
These are injected into a King type density profile with small stochastic density variations.
The back-reaction of the cocoons on the beams in the center produces armlength asymmetries
of a few percent, with the longer jets on the side with the higher average density.
Two distinguishable
bow shock phases were observed: an inner elliptical part,
and a later cylindrical, cigar-like phase, which is known from previous simulations.
The sideways motion of the inner elliptical bow shock part is shown to follow the 
law of motion for spherical blast waves also in the late phase, where the aspect ratio is high,
with good accuracy.
X-ray emission maps are calculated and the two bow shock phases are shown to appear as rings and 
elongated or elliptical regions, depending on the viewing angle.
Such structures are observed in the X-ray data of several radio galaxies (e.g. in Abell~2052 
and Hercules~A), the best example being 
Cygnus~A.
In this case,
an elliptical bow shock is infered from the observations, 
a jet power of $10^{47}$~erg/s is derived, and
the Lorentz factor can be limited to $\Gamma>10$.
Based on the simulation results and the comparison to the observations, 
the emission line gas producing the alignment effect in radio galaxies at high
redshift is suggested to be cooled gas entrained over the cocoon boundary.
} 

\authorrunning{M. Krause}
\maketitle

\section{Introduction}
The research on simulations of extragalactic jets has been extended recently into the 
regime of very light jets 
\citep[compare e.g.][]{mypap02a,Saxea02,CO02a,CO02b,Zanea03,BSS03},
down to density contrasts (jet over ambient density) of $\eta=10^{-5}$
\citep{mypap03a}. The interest has been stimulated not only by the difficulty to 
fix that parameter for real sources, but also by a first success in explaining 
some parameters of observed extragalactic jets. Low jet densities are needed
to get large radio cocoon and bow shock widths. Based on the results 
from a grid of simulations and comparison to the radio and X-ray data of the
radio galaxy Cygnus~A \citep{CarBar96,Sea01}, a density contrast of roughly
$\eta<10^{-3}$ has been claimed for this source \citep{Rosea99,mypap03a}.
All these simulations have been carried out in axisymmetry and a scale 
of a few dozen jet radii.
The aim of this paper is to verify and extend the earlier simulation results on the 100~kpc 
(=200 jet radii, $\rj$) scale and in three dimensions. 

\citet{mypap03a} found that very light jets start their life in a blastwave-like phase.
During that phase, terminating when the bow shock reaches
$R_1 \approx \rj /2 \eta^{1/4}$, the bow shock is spherical because of pressure 
predominance, obeying the blastwave equation of motion:
\begin{equation}\label{globeqmot}
\int_0^r{\cal M}(r^\prime)r^\prime\; \mathrm{d}r^\prime=
2\int_0^t \mathrm{d}t^\prime \int_0^{t^\prime} E(t^{\prime\prime}) 
\mathrm{d}t^{\prime\prime}\enspace.
\end{equation}
Here, ${\cal M}(r)$ is an arbitrary spherically symmetric
ambient gas mass distribution and $E(t)$ is the energy injection law.
For stationary energy injection $E=Lt$ and constant ambient density $\rho_0$,
one gets:
\begin{equation}\label{constden}
r=\left(\frac{5Lt^3}{4\pi\rho_0}\right)^{1/5}\enspace.
\end{equation}
This equation relates the total luminosity, $L$, of the jet to observables
like the bow shock radius and its velocity (via time derivation of (\ref{constden})).
With information on the bow shock propagation at later phases, it can be hoped 
that the jet power could be reconstructed from the bow shock shape,
which is probably already observed in the case of Cygnus~A by Chandra
(see sect.~\ref{obs}).
It is therefore important to check the evolution of the bow shock at later phases.
 This was difficult for most 
simulations so far because of the assumed reflection symmetry in the equatorial plane 
of the system. Interaction of the backflow with this boundary disturbs the sideways
evolution of the bow shock. Therefore, the symmetry assumption has been dropped
in the present work, and
the simulations are bipolar, evolving back-to-back jets in opposite 
directions, a technique that has emerged only recently
\citep{RHB01,mypap02b,mypap02d,BA03}. 
A fully 3D simulation is presented in sect.~\ref{3D}, and a
large scale axisymmetric simulation is presented in sect.~\ref{2p5d}.
Comparisons to observations are presented in sect.~\ref{obs}. Part of this is an update of previously 
published results, using a more involved analysis.

\section{Numerics}
\label{num}
The magneto-hydrodynamic (MHD) code
{\em NIRVANA} was employed \citep{ZY97}. It solves the hydrodynamic
equations in three dimensions (3D) for density $\rho$, velocity $\vec{v}$,
and internal energy $e$:
\begin{eqnarray}
\frac{\partial \rho}{\partial t} + \nabla \cdot \left( \rho \vec{v}\right)&
 = & 0 \label{conti}\\
\frac{\partial \rho \vec{v}}{\partial t} + \nabla \cdot\left( \rho \vec{v}
\vec{v} \right) & = &
- \nabla p - \rho \nabla \Phi \label{mom}\\
\frac{\partial e}{\partial t} + \nabla \cdot \left(e \vec{v} \right) & = &
- p \; \nabla \cdot \vec{v} -\rho^2 \Lambda\label{ie}\enspace ,
\end{eqnarray}
where $\Phi$ denotes an external gravitational potential and $\Lambda$ 
is a cooling function. Here, bremsstrahlung 
($\Lambda=7.5 \times 10^{20} \sqrt{T}(1.+4.4\times10^{-10}T)$~cm$^3$erg/s)
has been used for the 3D computation.


The code was vectorised and parallelised by OpenMP like methods,
and successfully tested on the NEC~SX-5 \citep{mypap02b}
at the high performance computing center in Stuttgart (Germany), where the computations
have been carried out. 

\subsection{Boundary conditions for the 3D cylindrical grid}
\label{3Dcylbc}
For the 3D simulation, a cylindrical grid was employed.
The disadvantage of the cylindrical coordinates $(Z,R,\phi)$ compared to the Cartesian 
ones
is the appearance of internal boundaries. The grid somehow has to be connected across
these boundaries. In the $\phi$ direction, 
periodic boundary conditions were applied.
For the boundary on the axis, no analogue could be found in the literature.
The boundary condition here is similar to the periodic case: One side of the grid should
know about the other side. Therefore, three cells were used below the axis,
to which consequently the index $i_R=3$ was assigned.
With this choice and the staggered mesh, equations (\ref{conti}-\ref{ie}) are
well defined everywhere on the grid.
For the scalar quantities, the following boundary conditions were applied:
\begin{eqnarray*}
f(i_Z,i_R,i_\phi)&:=&f(i_Z,6-i_R,i_\phi+i_\pi)\hspace{.2cm},i_R=0,1,2\\
f(i_Z,3,i_\phi)&:=&<f(i_Z,3,i_\phi)>
\left|_{i_Z=\mathrm{const}}\right.,
\end{eqnarray*}
where $i_\pi$ denotes the number of grid cells corresponding to the angle $\pi$.
The second equation indicates that directly on the axis , the scalars were averaged
over $i_\phi$ for constant $i_Z$,
since all the different $i_\phi$ refer to the same physical point.
Due to the staggered mesh, the $Z$ and $\phi$ components of vectors are shifted by
half a grid cell away from the axis. Therefore for these components, there arise no special
problems, and the boundary condition is:
\begin{displaymath}
v_{Z,\,\phi}(i_Z,i_R,i_\phi):=v_{Z,\,\phi}(i_Z,5-i_R,i_\phi+i_\pi)\hspace{.5cm},i_R=0,1,2
\end{displaymath}
The radial vector components are not shifted away from the axis. So
in principal, they all denote the same physical point. However, the flow should be allowed
to cross the axis from one side to the other. This is only possible if the radial velocity
takes a reasonable value there. Therefore, two possibilities
arise for the boundary conditions:
\begin{eqnarray*}
v_{R}(i_Z,i_R,i_\phi)&:=&-v_{R}(i_Z,6-i_R,i_\phi+i_\pi)\hspace{.5cm},i_R=0,1,2\\
v_{R}(i_Z,3,i_\phi)&:=&\left\{\begin{array}{r@{\quad:\quad}r}
0&\mathrm{I} \\ (v_{R}(i_Z,4,i_\phi)-v_{R}(i_Z,2,i_\phi))/2&
\mathrm{II}\end{array}\right.
\end{eqnarray*}
In order to check the influence of these two different boundary 
conditions on the simulation,
the 3D jet was simulated twice, the first time with the case I, and
the second time with the case II boundary condition. For both runs, the integrated
quantities (directionally split energy and momentum, and mass) and also the timesteps were
equal. Also, from comparison of the contour plots, no difference could be found.
Hence, the flow seems to have enough possibility to flow past the axis, and the detailed
behaviour at that line does not influence the result by much. 
The simulation result also shows
that this approach in general works in regions of undisturbed jet flow.
Where the jet is dominated by instabilities and wants to bend,
the axis looks like an obstacle. Therefore, the method seems to be problematic if details
about the jet beam are of special interest. Nevertheless, the results confirm that the
representation of the beam is acceptable, and for the other regions the output was fine.
\begin{figure}
\centering 
\includegraphics[width=0.5\textwidth]{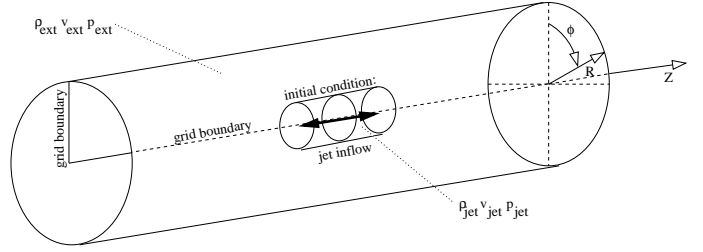}
\caption{\small Sketch of the cylindrical grid.
  \label{cylgrid}}
\end{figure}
\begin{figure}[b]
\begin{center}
\rotatebox{0}{\includegraphics[width=.5\textwidth]{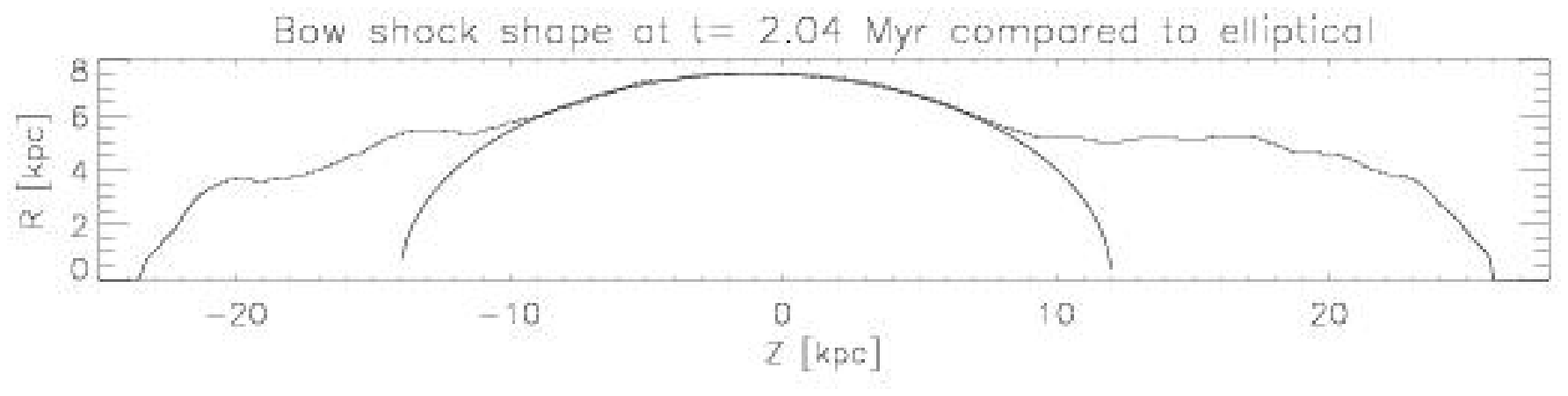}}\\
\rotatebox{0}{\includegraphics[width=.5\textwidth]{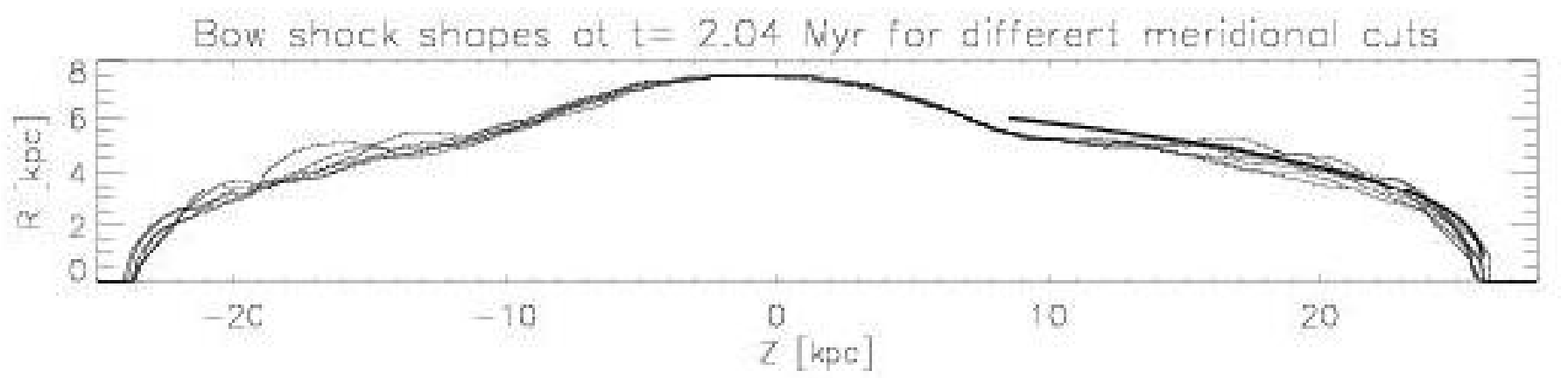}}\\
 \caption{\small Bow shock shapes for the 3D run at $t=2.04$~Myr. On the top panel 
        the shape in one meridional plane is compared to the elliptical 
        $R=7.5\sqrt{1-(z+1)^2/169}$.
        The bottom panel compares the bow shock shape for different meridional planes 
        to the parabola $R=2.3(26-Z)^{1/3}$ (thick line). 
        The bow shock is axisymmetric in the middle, where 
        it can be well represented by an ellipse. 
        For $|Z|>10$, the bow shock is bumpy and not axisymmetric. 
        That part is called cigar-like.
        \label{bs11}}
\end{center}
\end{figure}
\begin{figure}[bth]
\begin{center}
\rotatebox{-90}{\includegraphics[width=.3\textwidth]{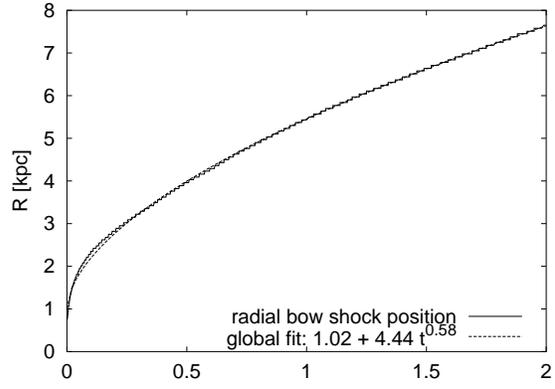}}
 \caption{\small Radius where the bow shock hits the line ($\phi=0, Z=0$) 
	versus time,
        with a globally fitting function. \label{bowfits3d}}
\end{center}
\end{figure}
The big advantage of the cylindrical grid is the reduction of the number of cells to compute.
The reduction of the physical computational 
volume is 22\%. Since the $\phi$-direction can be covered with fewer cells than a third 
dimension in Cartesian coordinates, this approach saves a significant amount of memory
and CPU time. The computation would not have been possible on a 3D Cartesian grid of 
similar central resolution.

\section{3D simulation}
\label{3D}

\subsection{Simulation setup}
\begin{figure*}[t]
\begin{center}
\includegraphics[width=0.49\textwidth]{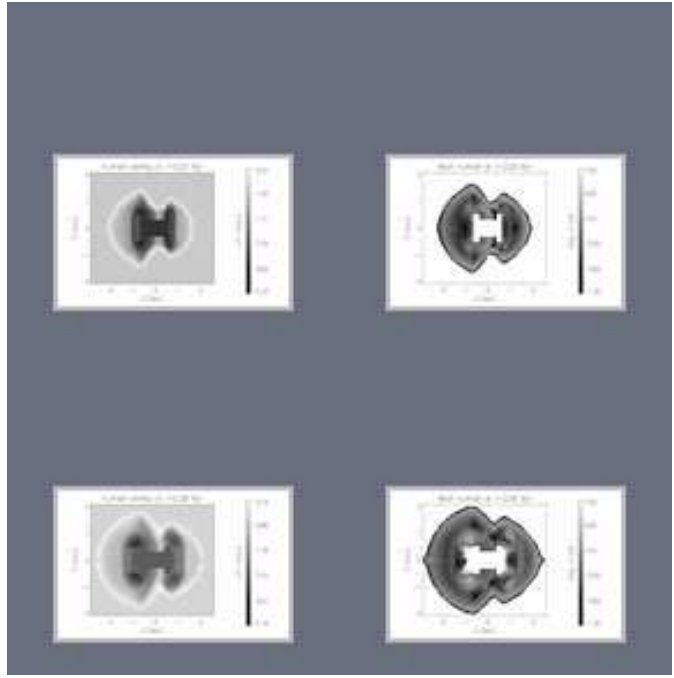}
 \caption{\small 
        The very early evolution of the 3D simulation. 
        The meridional slices at $\phi=0,\pi$, joined at the axis, of number density
        (left) and Mach number (right) are shown at 0.03~Myr (top) and 0.06~Myr (bottom).
        Values below 0.1 were omitted in the plot of the Mach number.
        The bottom row shows that sometimes small 
        numerical artefacts at the axis are present,
        especially in the Mach number, which includes the radial velocity.
        \label{cyga_early}}
\end{center}
\end{figure*}
\begin{figure*}[t]
\begin{center}
\rotatebox{0}{\includegraphics[width=.24\textwidth]{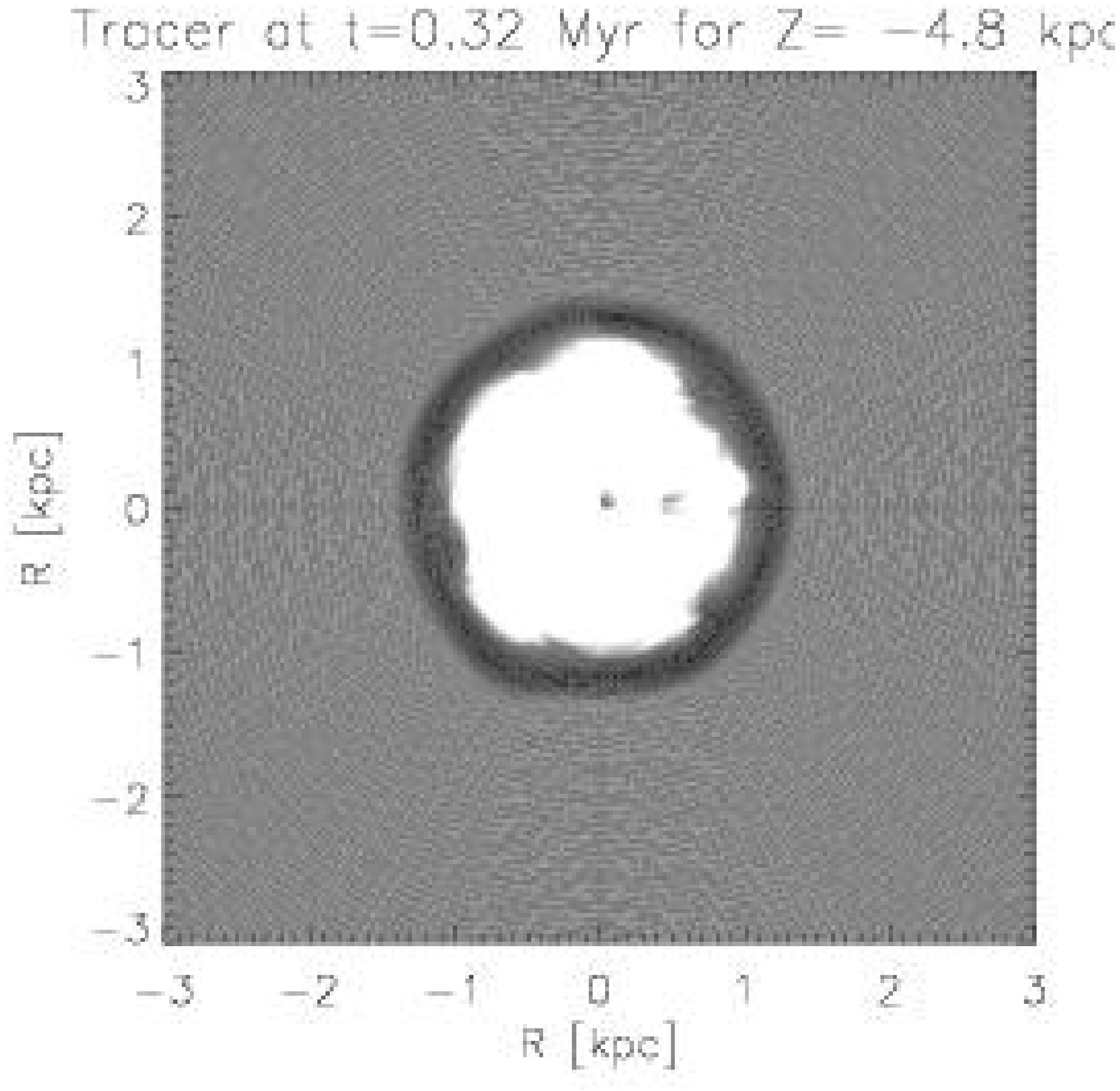}}
\rotatebox{0}{\includegraphics[width=.24\textwidth]{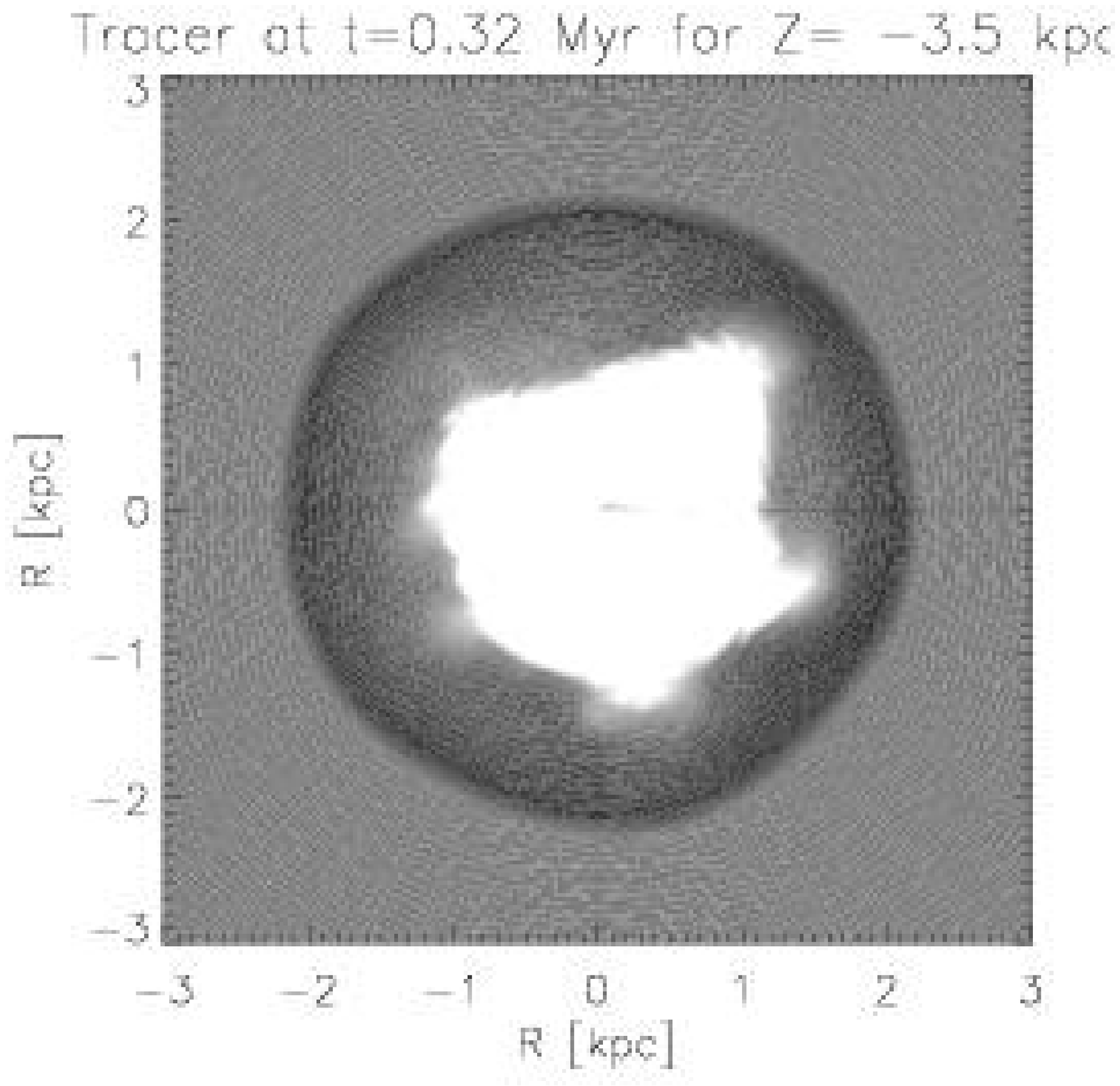}}
\rotatebox{0}{\includegraphics[width=.24\textwidth]{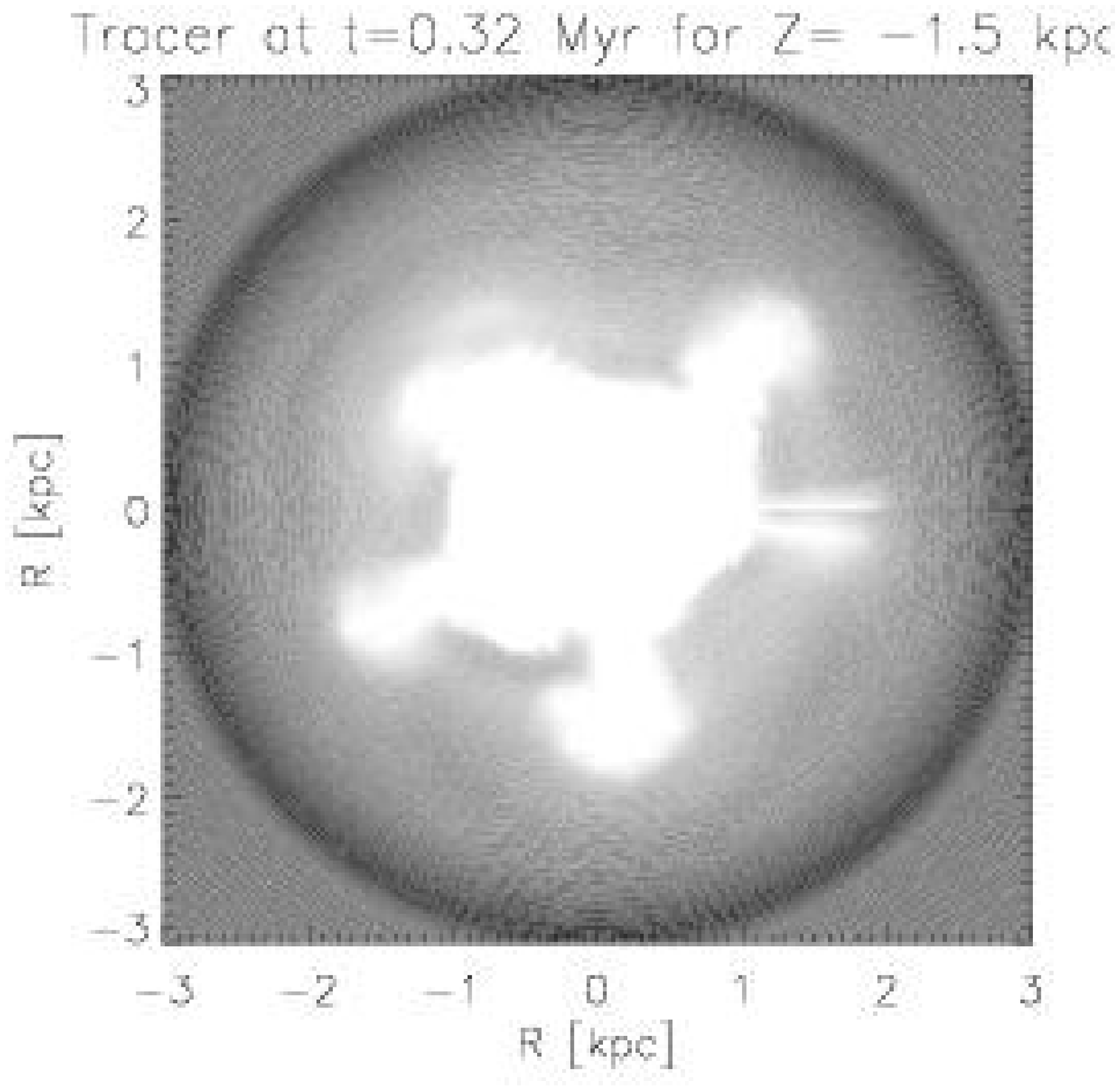}}
\rotatebox{0}{\includegraphics[width=.24\textwidth]{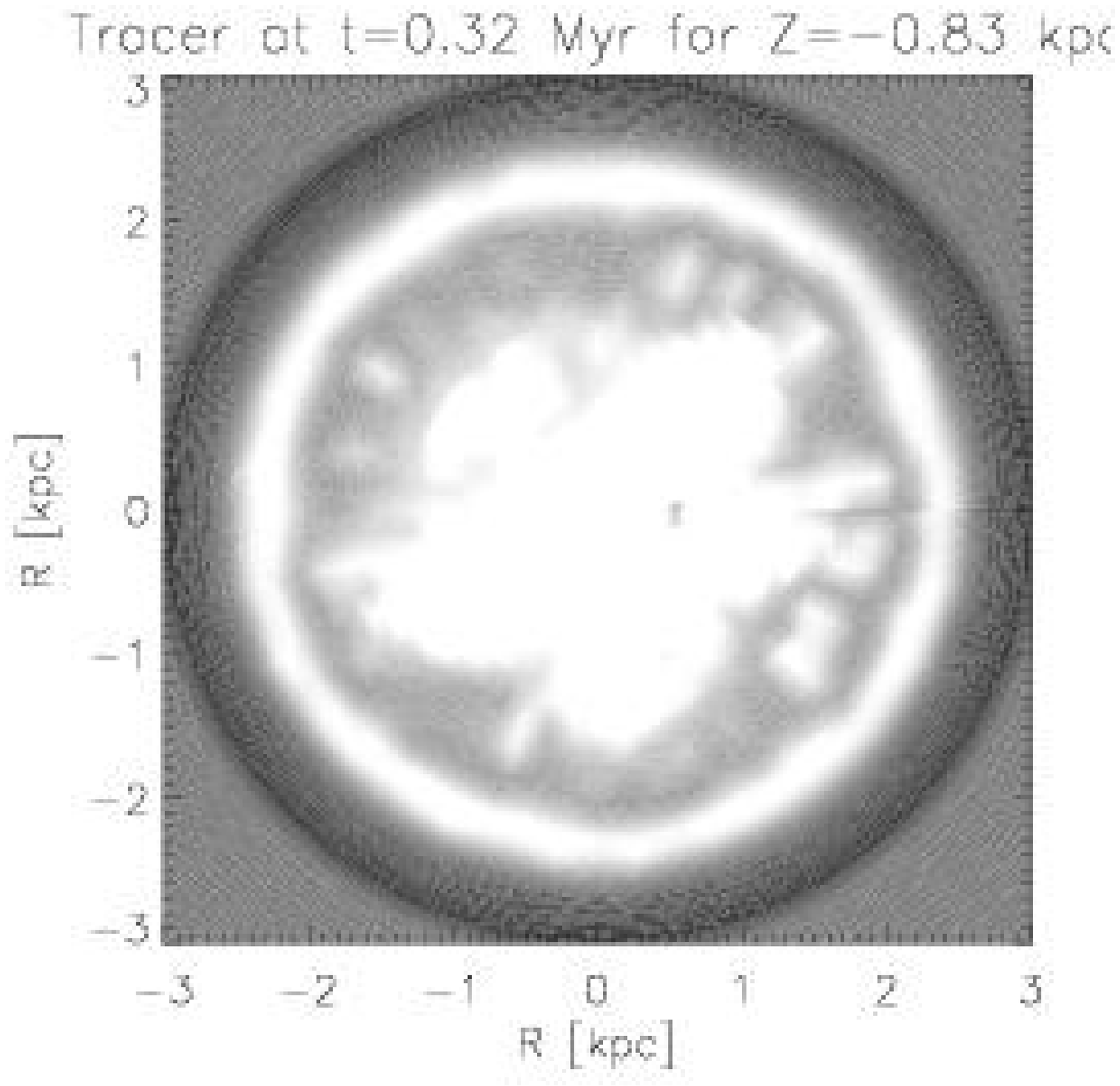}}\\
\rotatebox{0}{\includegraphics[width=.24\textwidth]{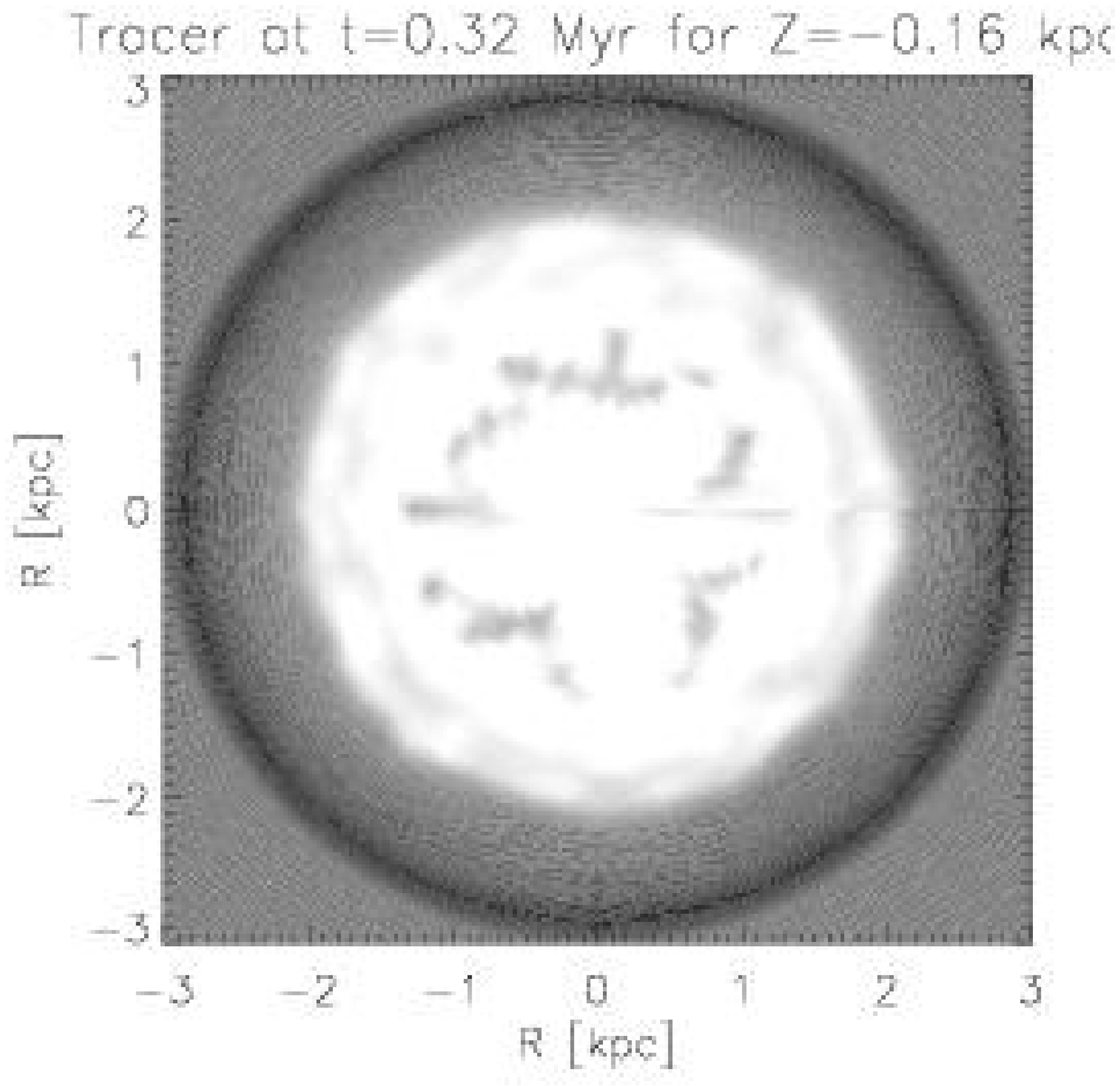}}
\rotatebox{0}{\includegraphics[width=.24\textwidth]{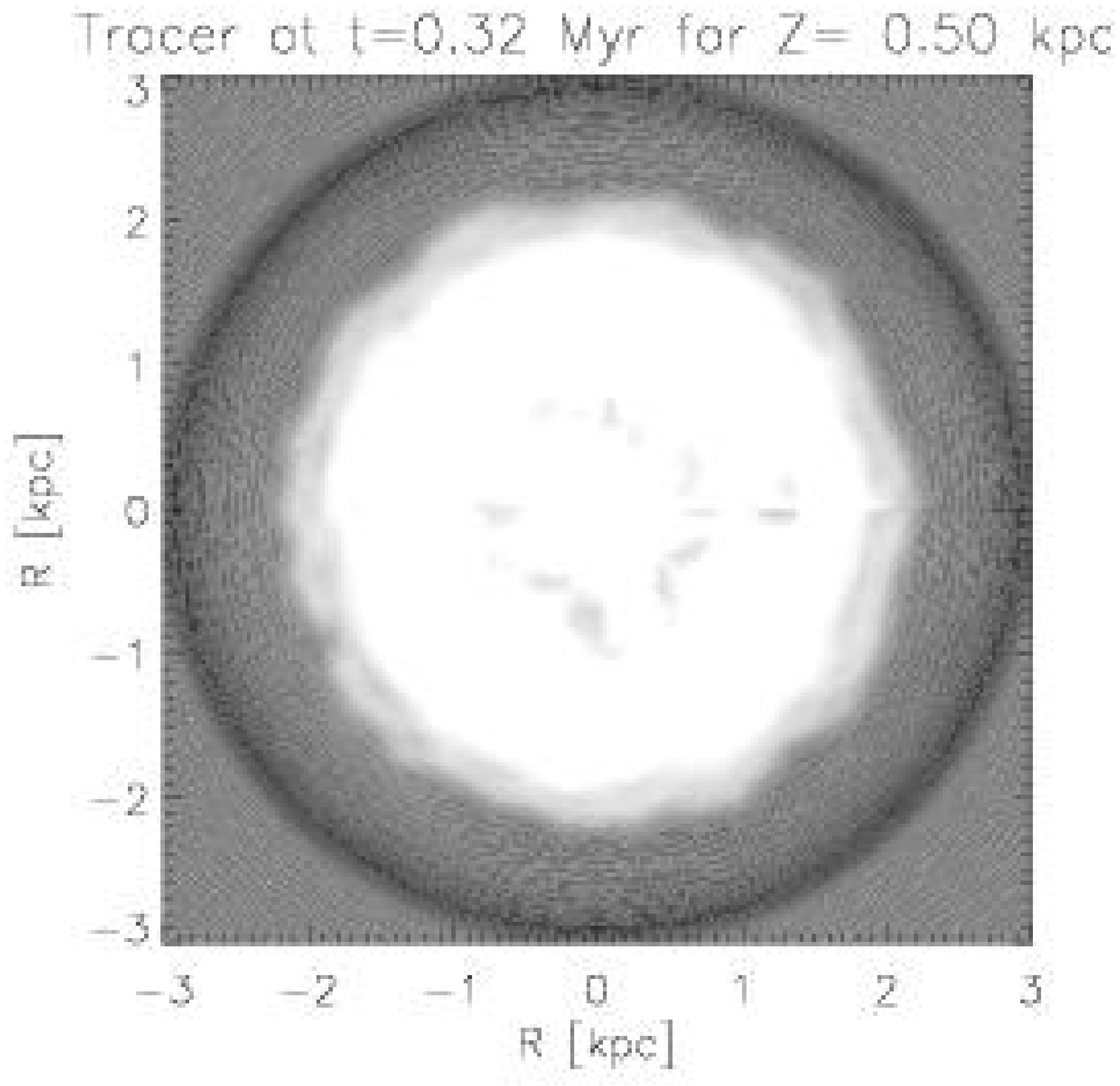}}
\rotatebox{0}{\includegraphics[width=.24\textwidth]{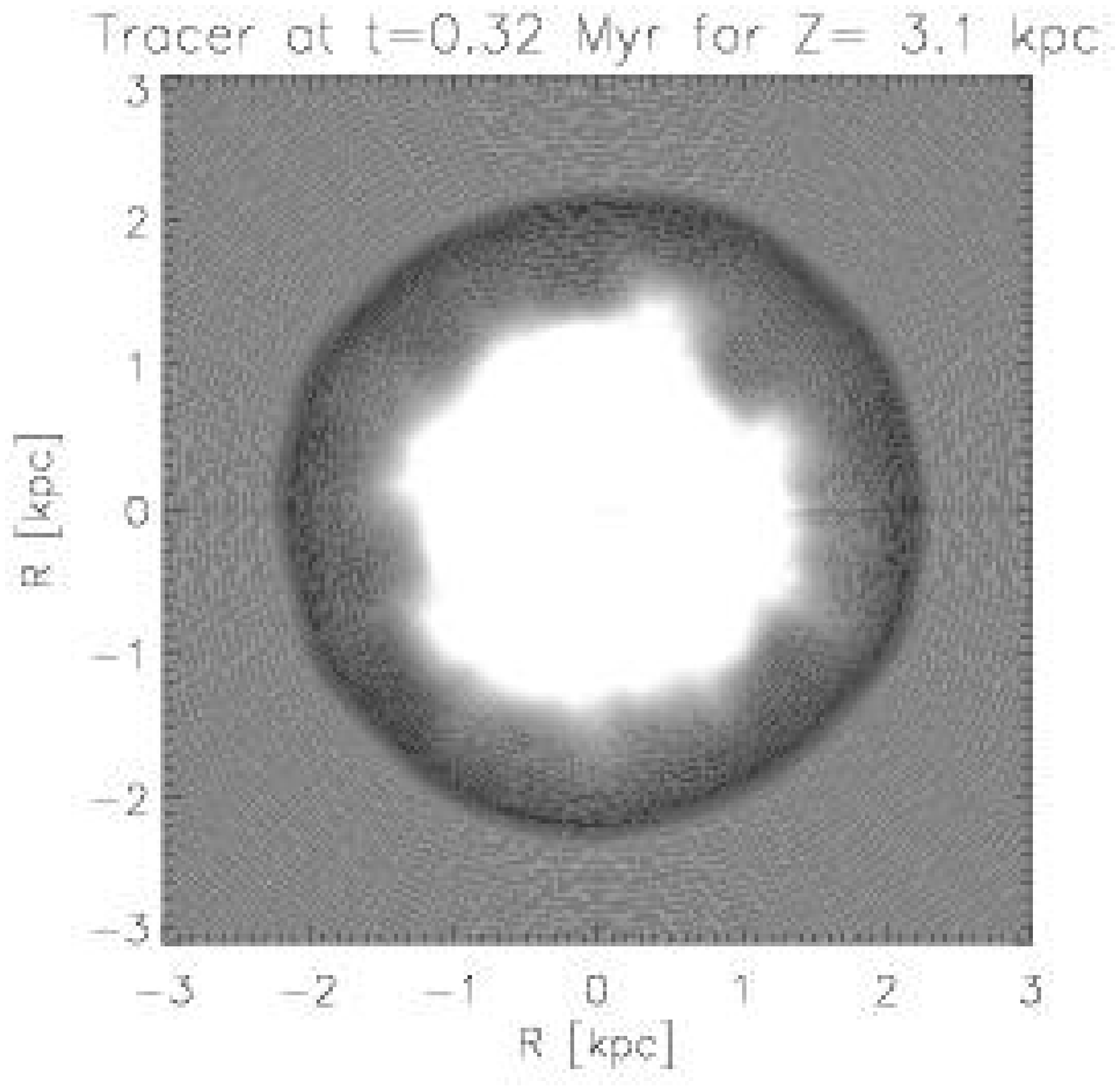}}
\rotatebox{0}{\includegraphics[width=.24\textwidth]{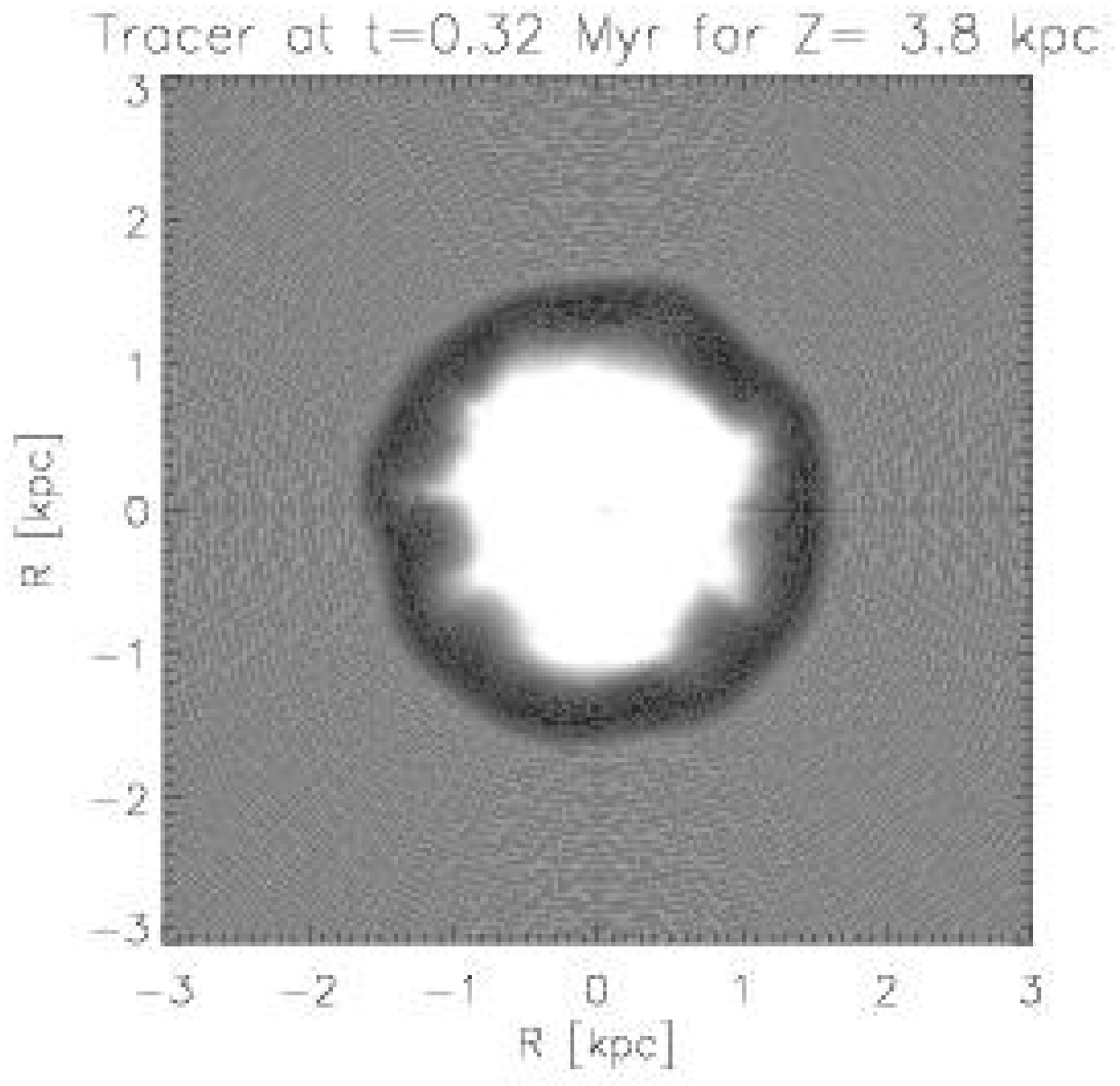}}\\
 \caption{\small Slices of the tracer at $t=0.32$~Myr for different axial positions.
        The shocked ambient gas is shown dark. Compressed regions are darker, rarefied and diluted 
        regions appear lighter. Cells that contain less then roughly $10\%$ ambient gas, i.e the 
        beam plasma, are shown in white. \label{jet1b}}
\end{center}
\end{figure*}
\begin{figure*}[th]
\begin{center}
\rotatebox{0}{\includegraphics[width=.48\textwidth]{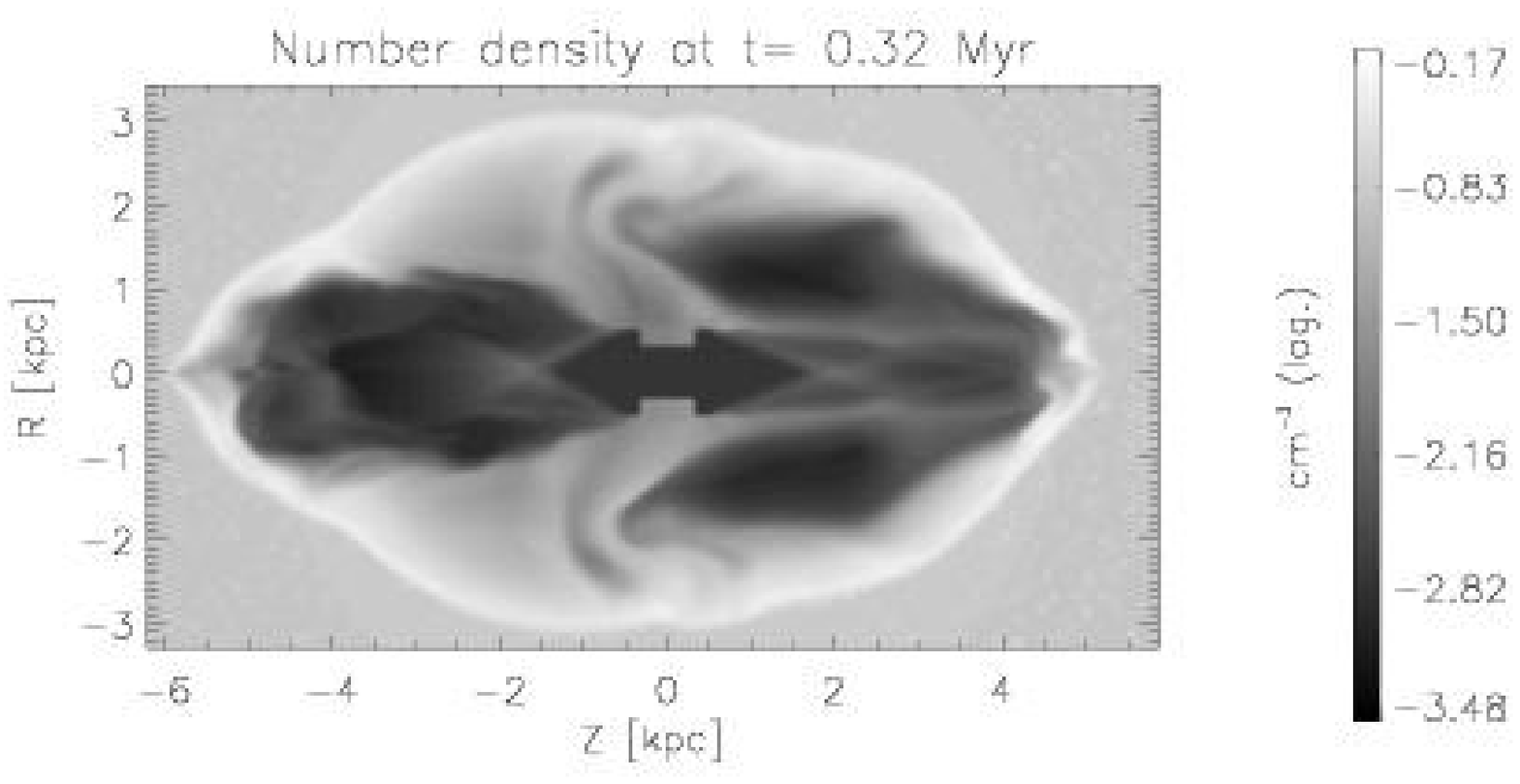}}
\rotatebox{0}{\includegraphics[width=.48\textwidth]{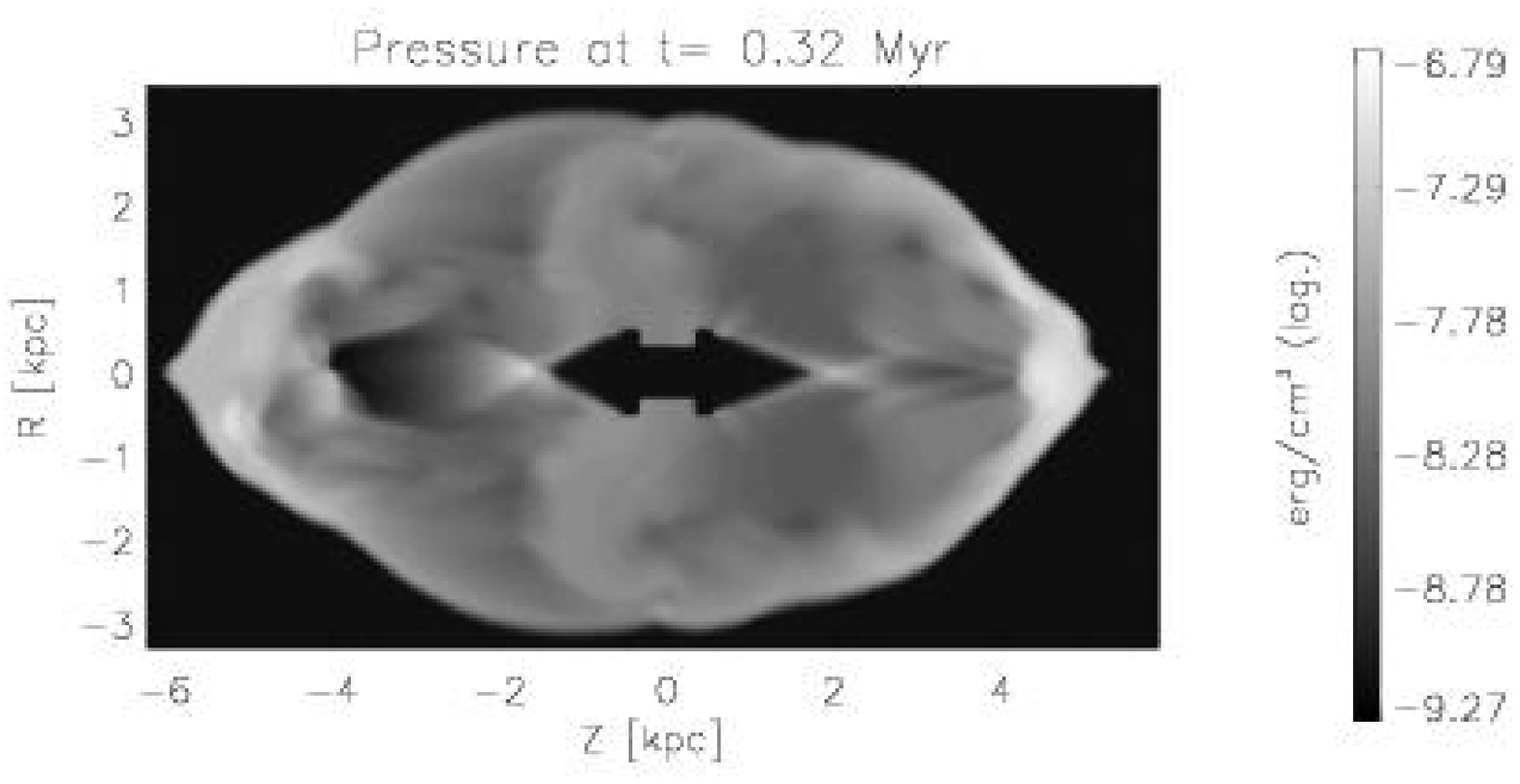}}\\
\rotatebox{0}{\includegraphics[width=.48\textwidth]{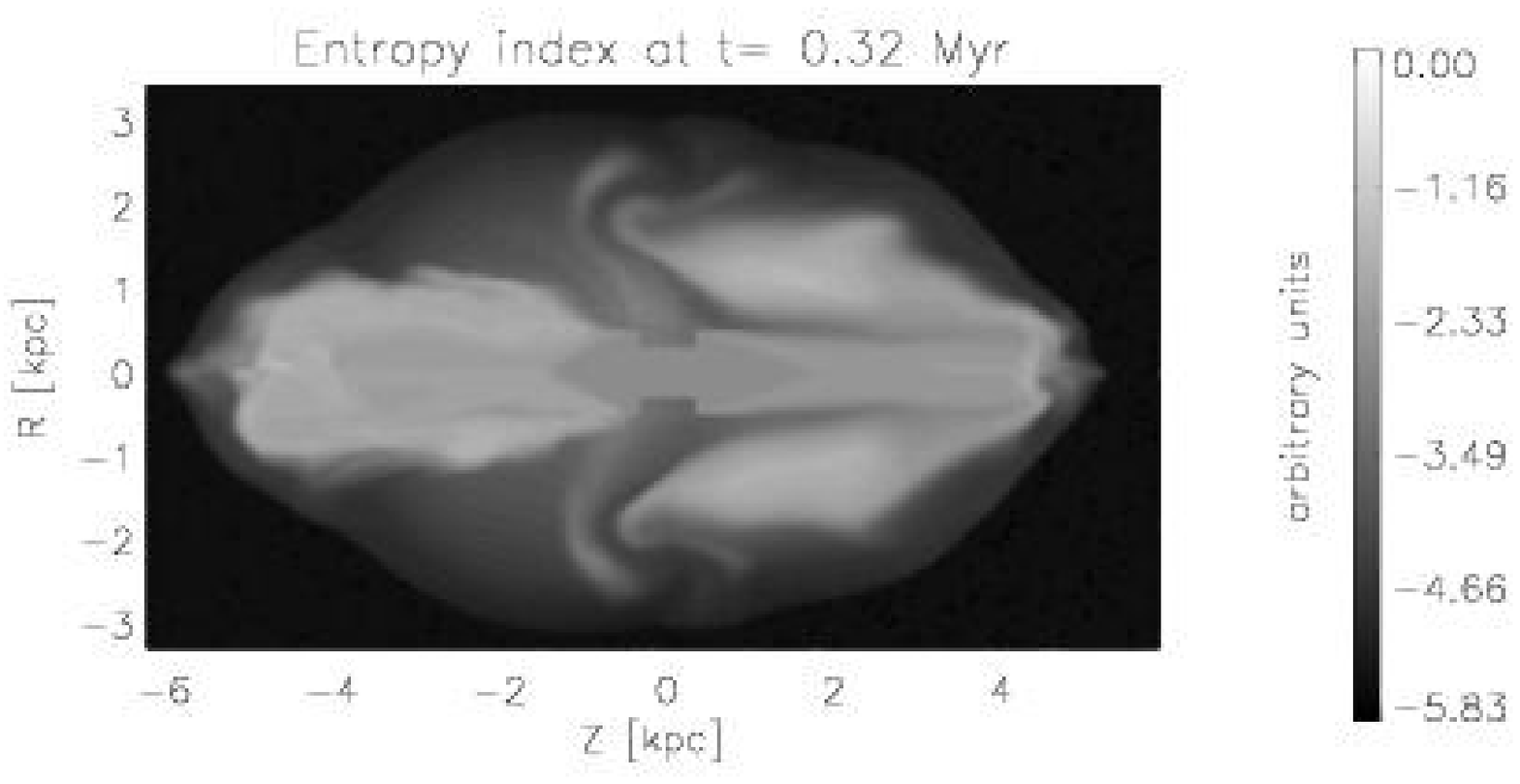}}
\rotatebox{0}{\includegraphics[width=.48\textwidth]{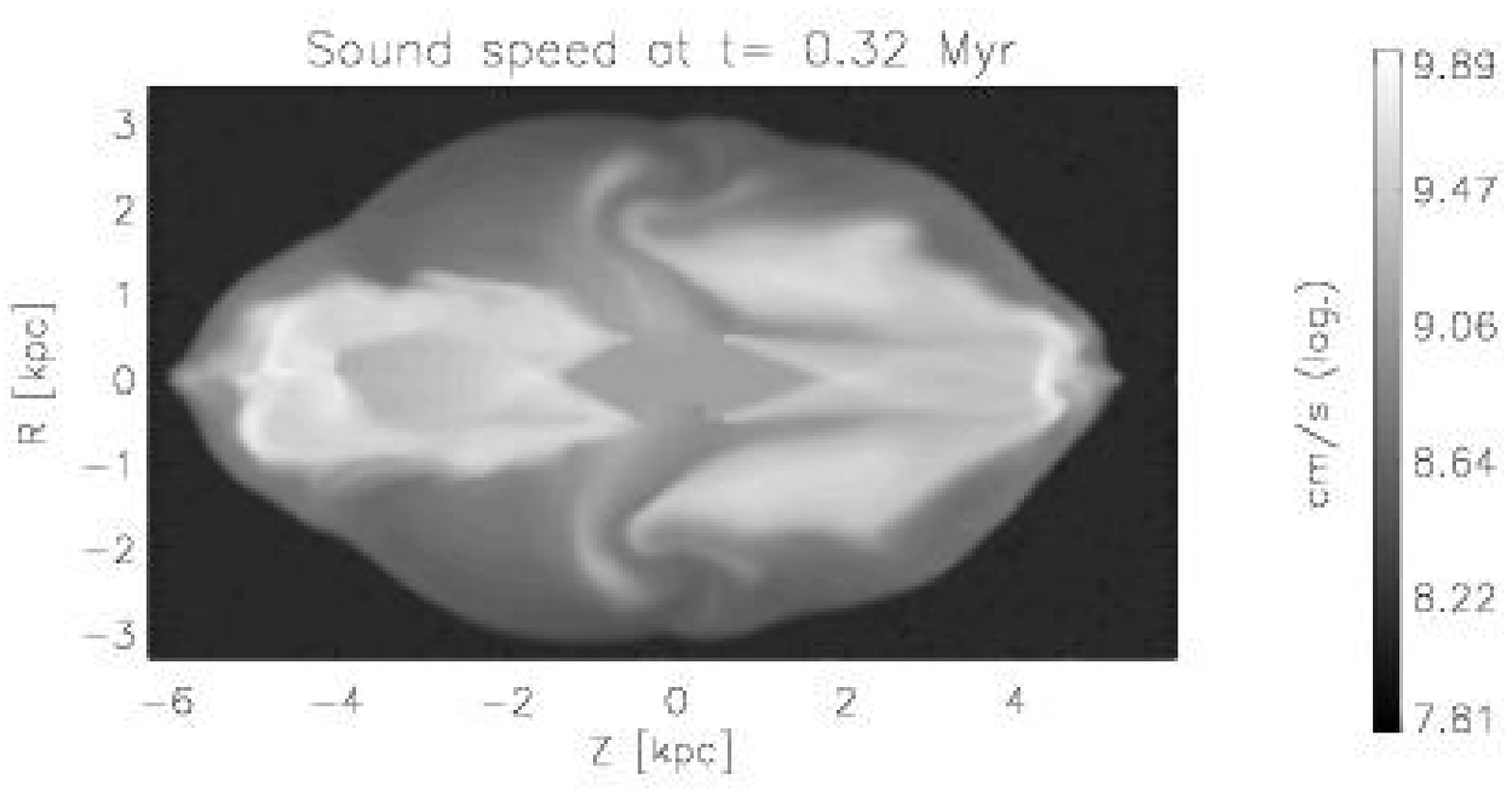}}\\
\rotatebox{0}{\includegraphics[width=.48\textwidth]{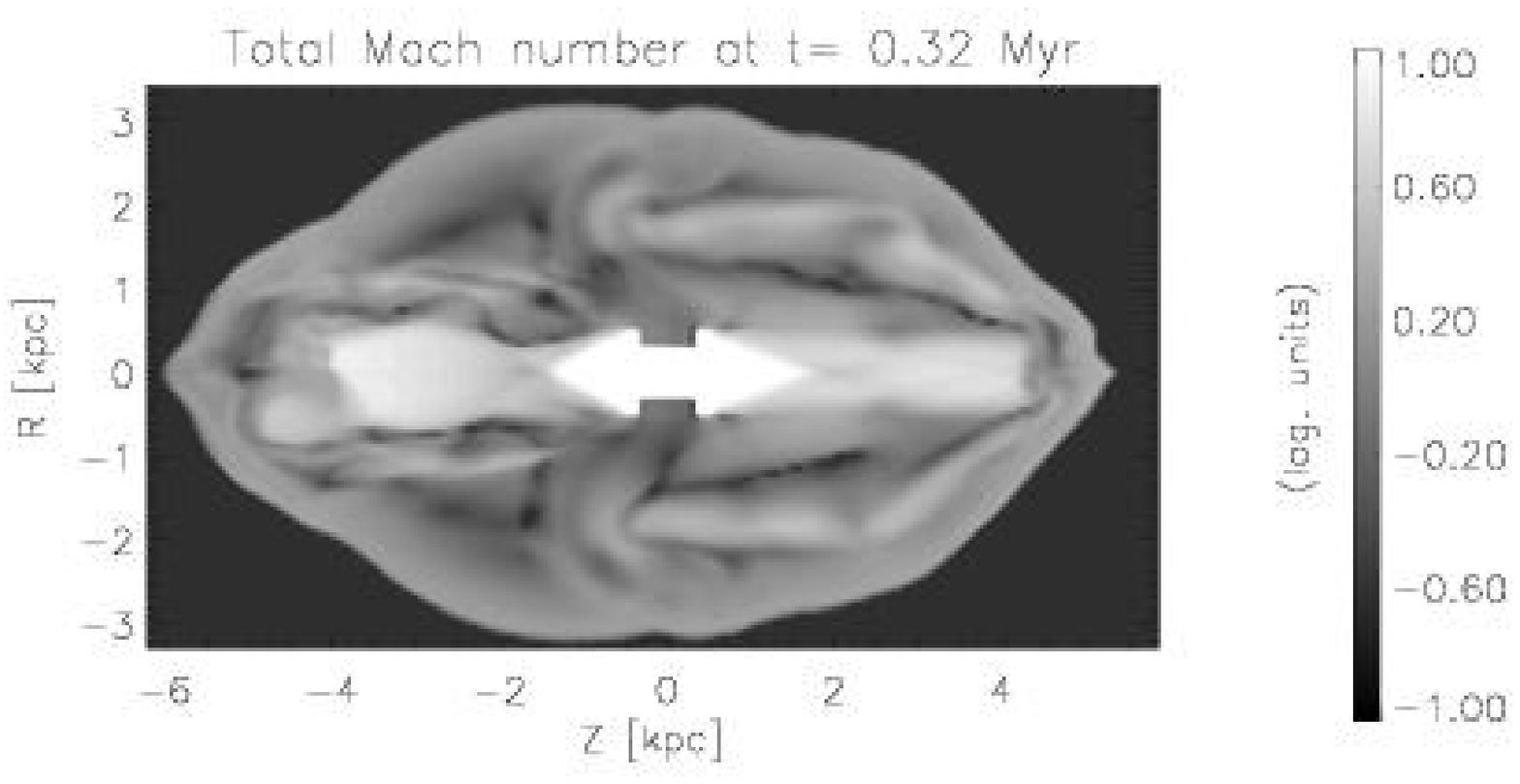}}
\rotatebox{0}{\includegraphics[width=.48\textwidth]{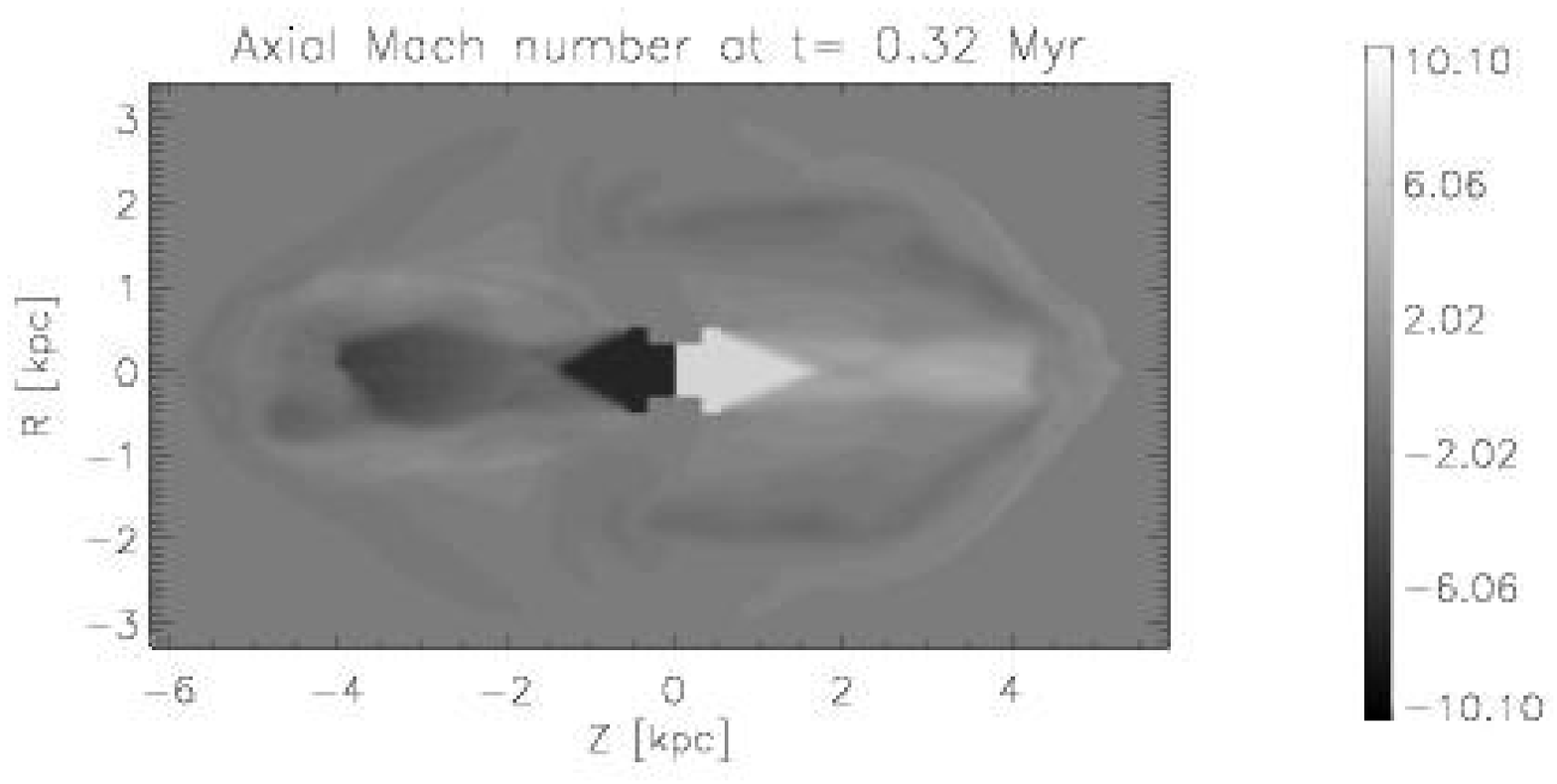}}\\
\rotatebox{0}{\includegraphics[width=.48\textwidth]{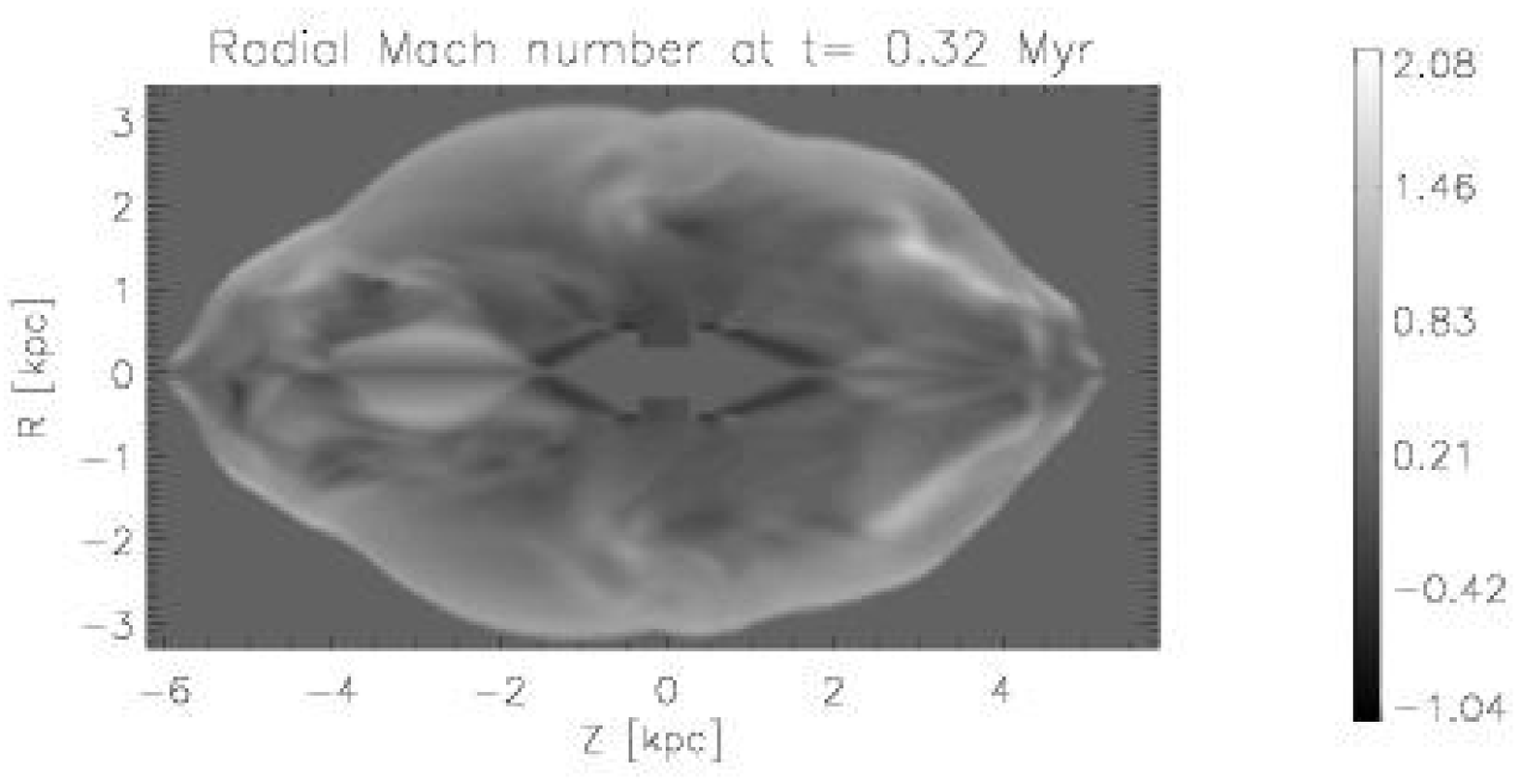}}
\rotatebox{0}{\includegraphics[width=.48\textwidth]{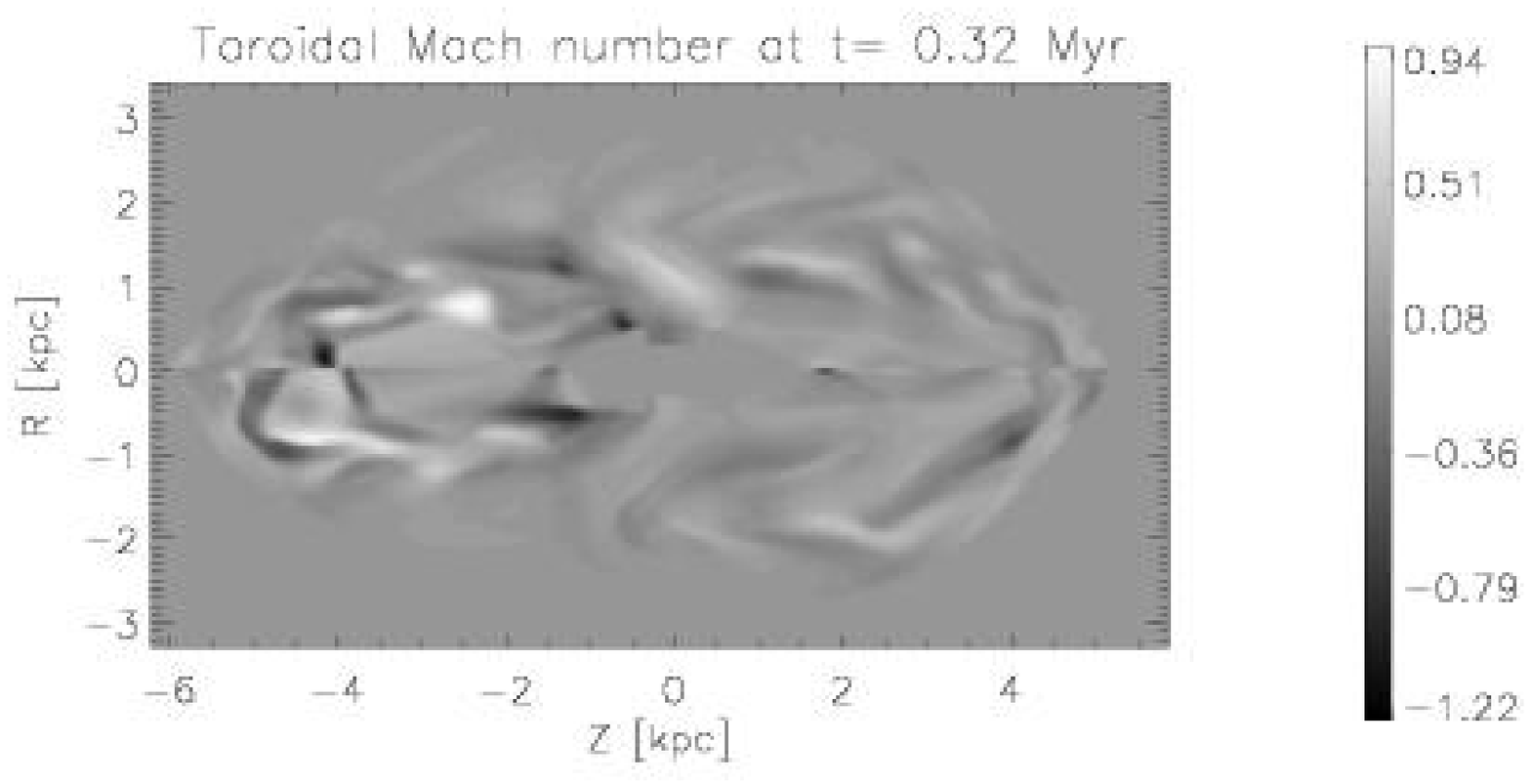}}
 \caption{\small 
        Meridional slices for $\phi=0/\pi$ at 0.32 Myr. The quantity plotted is indicated
        on top of the individual figures, units can be found next to the bar.
        The radial Mach number is defined to be positive away from the Z-axis.
        A positive toroidal Mach number is intended to mean motion into the plane
        for $r>0$, and out of the plane for $r<0$. The beam shows rotation
        in some places and translation in some others. Jet beam material can be
        decerned by its high entropy index.  
        \label{jet1}}
\end{center}
\end{figure*}
\begin{figure*}[thb]
\begin{center}
\rotatebox{0}{\includegraphics[width=\textwidth]{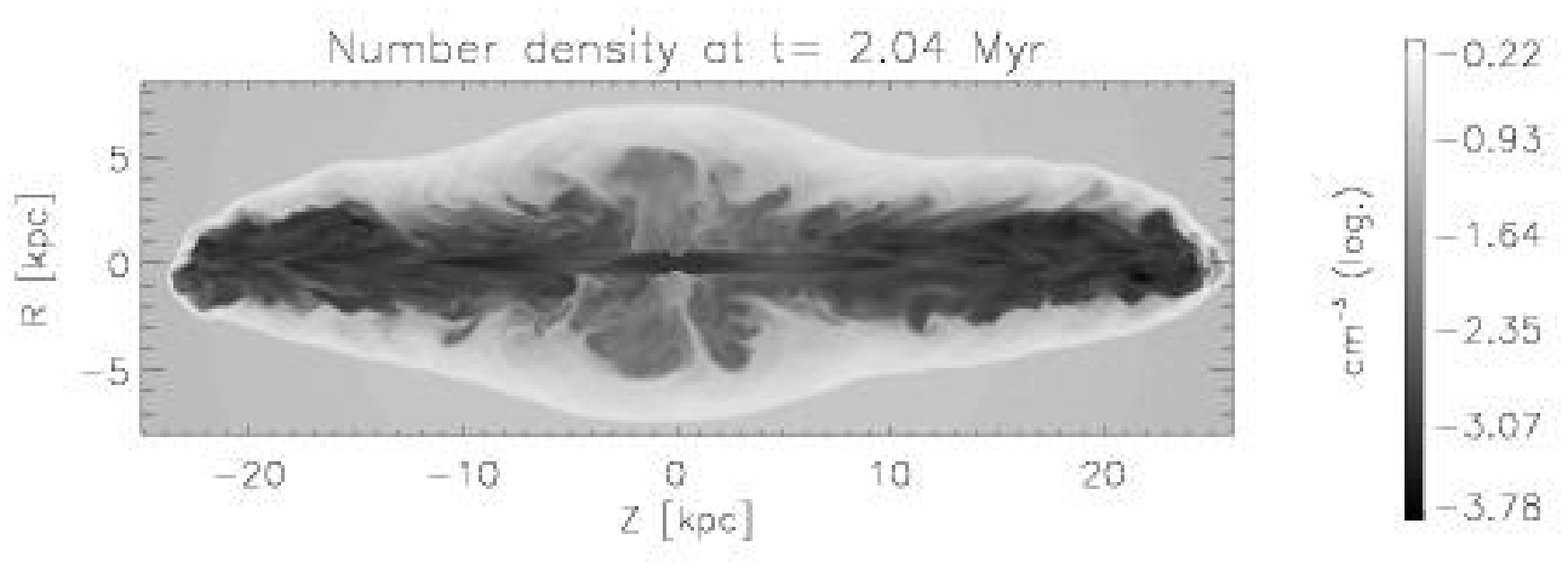}}\\
\rotatebox{0}{\includegraphics[width=\textwidth]{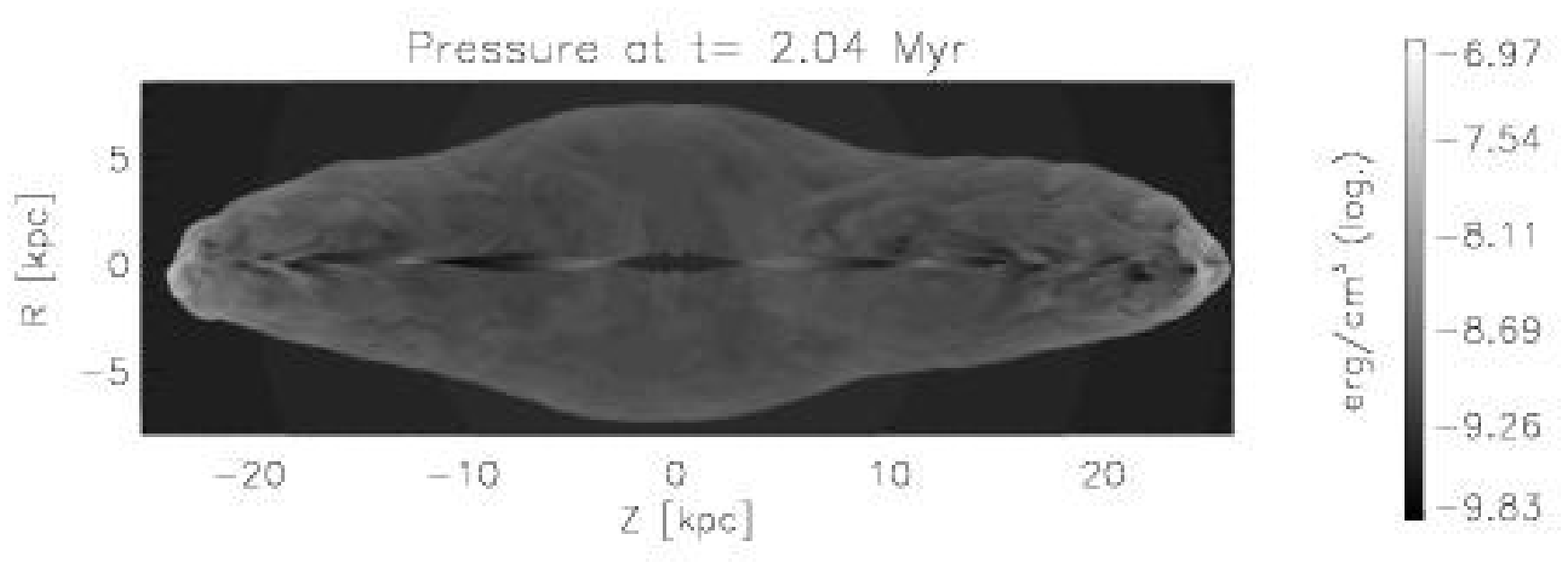}}\\
\rotatebox{0}{\includegraphics[width=\textwidth]{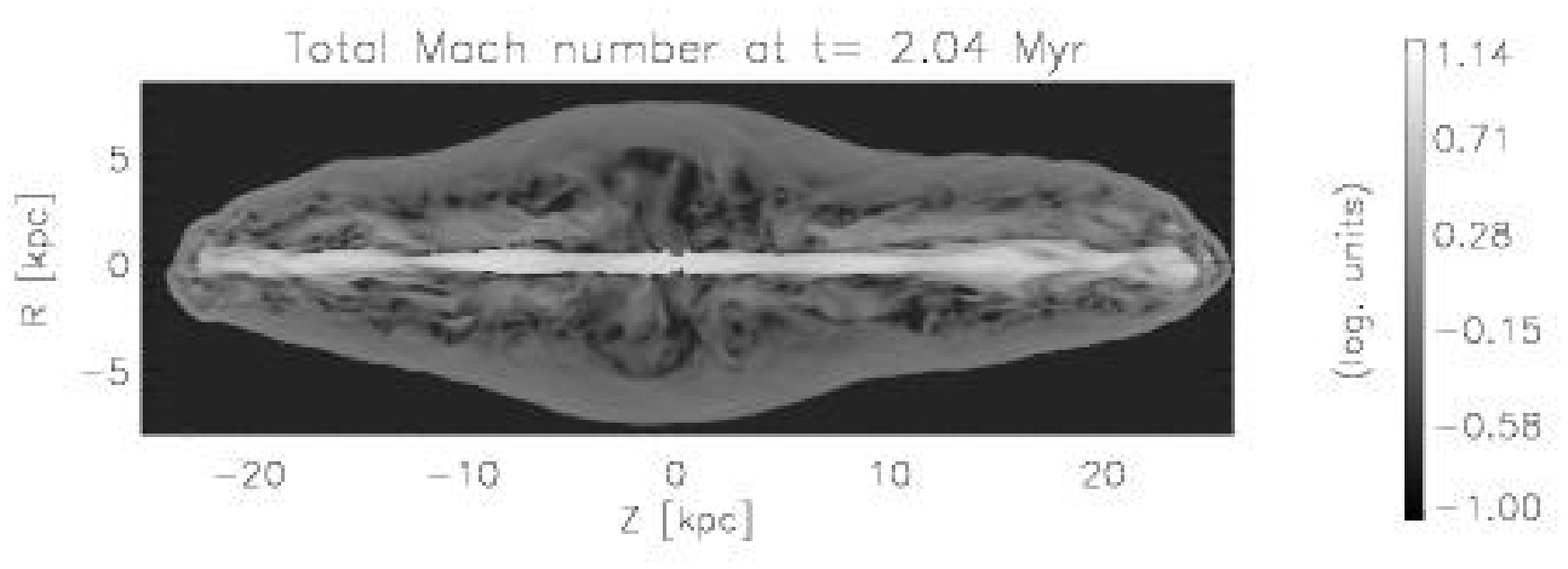}}\\
 \caption{\small 
        The same as Fig.~\ref{jet1}, but for $t=2.04$. 
        (Only some of the variables are shown.)
        \label{ml11}}
\end{center}
\end{figure*}
A cylindrical grid was used for the jet simulation (compare section \ref{3Dcylbc}).
The size of the computational domain was:
$Z \in [-69\;\mathrm{kpc},69\;\mathrm{kpc}]$, $R \in [0,57\;\mathrm{kpc}]$ and
$\phi \in [0, 2\pi]$. 2042, 805, and 57 grid points were used in the Z,R and
$\phi$ directions, respectively. 
With a jet radius of $r_\mathrm{j}=0.55$ kpc, this gives a resolution of 8 
points per beam radius (ppb).
The resolution in $\phi$ direction scales linearly down from
16 ppb on the jet boundary to 0.2 ppb on the edge of the grid.
The grid was initialised with an isothermal King cluster atmosphere:
\begin{equation}\label{kprof}
\rho_\mathrm{e}(r) = \rho_\mathrm{e,0}\left(1+\frac{r^2}{a^2}
  \right)^{-3\beta/2}\enspace ,
\end{equation}
where $r=\sqrt{R^2+Z^2}$ denotes the spherical radius,
$\rho_\mathrm{e,0} = 1.2 \times 10^{-25} \; {\mathrm{g}/\mathrm{cm}^3} $ 
is the characteristic density,
$\beta=0.75$ and $a=35\;\mathrm{kpc}$ is the core radius.
In order to break the bipolar and axial symmetry, this density
profile was modified by random perturbations of the following kind:
\begin{enumerate}
\item With 10\% probability the density was increased by a factor between 1
  and 1.4.
\item With 5\% probability the density was increased by a factor between 1
  and 2, only if the cell was unmodified in the first process and the Z
  coordinate was positive.
\end{enumerate}
The jet is injected in the middle of the grid in the region
$Z \in [-0.55,0.55\;{\mathrm{kpc}}]$, $R \in [0,0.55\;{\mathrm{kpc}}]$, and
$\phi \in [0,2\pi]$. This region has the constant values:
$\rho_{{\mathrm{jet}}}=6.68\times10^{-28}\;{\mathrm{g}/\mathrm{cm}^3}$,
$v_Z=\pm0.4c$, where c denotes the speed of light.
The plus sign applies for
the positive Z region and the minus sign for the negative one.
The kinetic jet luminosity is $L_\mathrm{kin}=1.04 \times 10^{46}$ erg/s 
for both jets together.
The pressure was set in order to match the external pressure at that position.
This gives a slightly varying density contrast across the grid of
$\eta =\rho_{{\mathrm{jet}}}/\rho_{{\mathrm{ext}}} \approx 7 \times10^{-3}$
and an internal Mach number $M=10$.
The jacket of the injection cylinder is a further boundary. This boundary is open, so the 
material can leave the grid here into the central kiloparsec of the simulated radio galaxy,
which is not attempted to model here.
The temperature in the external medium is set to
$T=3\times10^7$ K. 
The cooling time in the shocked cluster gas is approximately 100 Myr. The jet is expected
to propagate through the whole volume in 10 Myr. So, cooling by bremsstrahlung
marginally influences
the state of the gas. This was taken into account (comp. (\ref{ie}) and sect.~\ref{num}).
Thus the calculation is not scalable anymore, formally.
But given the smallness of the effect, scaling should be possible, in practice. 
Because the atmosphere is isothermal, the pressure varies in the same way as the density.
In order to keep the system in hydrostatic equilibrium, gravity by an assumed dark matter
distribution had to be applied.
The gravitational potential, necessary to prevent the King atmosphere from exploding is:
\begin{equation}
\Phi_\mathrm{DM}=\frac{3 \beta k T}{2 \mu m_\mathrm{H}} \log\left(1+(r/a)^2\right) \enspace ,
\end{equation}
where $\mu$ is the number of particles per proton mass.

\subsection{Early evolution \& verification of boundary conditions}
A testrun was performed on a ten times smaller grid, in order to verify the boundary 
conditions.
This early evolution is shown in Fig.~\ref{cyga_early}.
The images show the formation of a bow shock and backflows, for both jets.

In that phase, the jet almost ignores the stochastic nature of the density.
The bow shock has a round and regular shape, and 
the density varies smoothly along its surface. 
However, the evolution on the two sides is different.
The jet on the left-hand side, where the average background 
density is smaller, evolves faster: It produces a bigger bow shock,
and a faster backflow at equal times. At $t=0.06$ Myr, the two bow shocks are nearly 
joined together. As the later evolution shows, a single round bubble forms, 
soon after.
The upper and lower halfs of the pictures fit quite good together, although
there is sometimes a suspect spike at the apex of the bow shock on both sides.
This numerical artefact reminds us about the imperfect treatment of the axis.
The effect is rather small, and even absent for most of the simulation time.
This supports the choice of boundary conditions on the axis
(compare section \ref{3Dcylbc}).

\subsection{Medium term evolution}
A snapshot of several quantities at $t=0.32$ Myr is shown in Fig.~\ref{jet1}.
The jet plasma takes the lowest density, and the highest temperature values.
At that time, the evolution continued in the same way as in the early phase:
the right-hand side, propagating into the on average denser medium, develops a broader cocoon,
and is slower. The once separated bow shocks have united. 
The shape of the bow shock in that phase is oval, a sign of the blastwave phase.
The bow shock shows two extensions in jet direction. Due to their later appearance,
these parts of the bow shock will be called {\em cigar-like}.
The aspect ratio of the bow shock is nearly 2, thanks to the extensions, which contribute
approximately 0.5 to the aspect ratio at that time.


The plots of the Mach number show a slightly supersonic backflow. The radial Mach number is 
positive for motion away from the axis, the toroidal Mach number changes sign at $R=0$
for motion out of (into) the plane. The flow in R~and~Z directions
is well ordered, whereas in the azimuthal direction, arbitrary fluctuations are seen.
The jet beam shows both rotation (same colour above and below $R=0$) and translation
away from the axis (colour change at $R=0$). The latter supports the chosen approach
as being able to represent beam motions away from the axis of symmetry.

The pressure plot shows a high pressure at shocks in the beam, especially on the axis.
While this is in principal correct, the exact amount of the pressure could be influenced by the choice
of the cylindrical coordinate system. The first shock in the beam is stronger on the left-hand side.
This follows from the higher pressure there, but also from the higher inclination angle
to the axis. Since the beams have identical conditions, why are the shocks not identical?
The shocks in the beam are driven by whatever hits it. From the density plot,
this seems to be the cocoon on the left-hand side, and the entrained cluster gas
on the right-hand side. The backflows collide approximately at the center, forming a region
of enhanced pressure. This region is asymmetric at the time shown here. Due to the stronger
backflow from the right-hand side, the region has moved to the left. The higher pressure there
drives a stronger shock into the beam. This explains why the first shocks are different
on both sides.

A passive tracer variable was advected with the flow, set initially to unity in the shocked ambient gas,
to zero in the right beam, and to $-1$ in the left one. This tracer enables differentiation
between the beam and the ambient gas. Slices at constant axial position of the tracer at $t=0.32$~Myr
are shown in Fig.~\ref{jet1b}. The right side has an approximately 
axisymmetric cocoon, while the left one is clearly not (compare slices at $Z=-3.5$~kpc; $Z=-1.5$~kpc).
The slice at $Z=-0.83$~kpc demonstrates how the cocoons manage to merge together:
The left one is smaller and slips inside the right one. This also brings shocked ambient gas
inside the cocoon, which can still be spotted in the $Z=0.5$~kpc slice.  

The stronger shock on the axis on the left-hand side causes the beam to widen,
due to high pressure. On the right-hand side, less energy has been converted into heat,
the beam stays narrow and delivers more power to the terminal shock, where the pressure
is consequently higher. 

Fingers 
of shocked cluster gas, reaching down to the beam surface,
are present just as they are in the unipolar 2.5D simulations \citep[compare e.g.][]{mypap01a}.
They are generated in the following way: Kelvin-Helmholtz instabilities are excited at the boundary
between cocoon and shocked external medium. The backflow advects those instabilities,
while they are amplified. Hence, they are biggest in the symmetry plane.
The material in the symmetry plane could in principal flow outward, creating bumps in the bow shock,
or inward. The simulation clearly shows no sign of outward motion. Instead, the gas is transported
far down into the radio cocoon in geometrically thin fingers. Soon, their extension falls
below the resolution limit of the simulation, and the gas mixes with cocoon plasma.
However, in reality, the two phases may remain separate. The magnetic field of the radio plasma
further supports the separation of the two phases.
In principle, in that way new fuel could be channeled to the central source.
However, the dynamics of the central kpc is beyond the scope of this work.
The central high pressure region pushes the entrained shocked cluster gas
to the right, where it slips between beam and cocoon, thereby widening the right cocoon.

It has already been pointed out above that at the time the simulation is shown in Fig.~\ref{jet1},
0.32~Myr, the left jet converts more kinetic energy into heat than the jet on the right-hand side.
At that time the tip of the jet has already advanced approximately 20~\% further towards the left
than towards the right. The difference of the average density is only 2.5~\%, which cannot
explain the fast propagation of the left jet, considering the estimate by the one dimensional 
force balance: 
$v_\mathrm{head} \approx \sqrt{\eta \epsilon} v_\mathrm{beam}$, where
$\epsilon$ is the ratio of beam to head area.
One important result is therefore that
the nonlinear dynamics amplifies the effect of the different density on both sides by more than
a factor of ten. 
The situation at that time is unstable, and the later timesteps show that the right
jet catches up, and outruns the left one not later than at $t=0.95$ Myr.
The situation stays that way until the end of the simulation. On average, the right jet
is approximately 10\% faster than the left one, at late times.
This is in conflict with naive intuition, but is readily explained, considering
that the stronger backflow from the right jet shifts the central pressure enhancement
to the left, where stronger oblique shocks are driven into the beam slowing the left jet down,
as pointed out above.

\subsection{Long term evolution}
The final snapshot at $t=2.04$ Myr is shown in Fig.~\ref{ml11}.
The plots show the usual picture of a hydrodynamic jet simulation, largely consistent
with FR II radio galaxies: The cocoon is now nicely placed around the jet beam. The
aspect ratio of the bow shock is~3.6. The Kelvin-Helmholtz instabilities show up prominently.
They still grow towards the center, and develop into long fingers at the innermost positions.
The pressure shows a regular spacing of shock compression and rarefaction zones in the beam.
High pressure regions are small and show up only at the end of the beams, where the Mach disk
is located. The oblique shocks in the beam are now weaker than at $t=0.32$ Myr. This follows
from the smaller angle with the jet axis. The central region, with a diameter of roughly 10 kpc,
is now dynamically calm. No large Mach numbers are observed there, and the pressure is
approximately constant. The density takes intermediate values. This is now a relaxed
region where jet plasma and shocked cluster gas are mixed (mixing
may not happen in nature, see above).

\subsubsection{The shape of the bow shock}
\label{bowshap}
Figure~\ref{bs11} shows the bow shock shape for the final snapshot in detail.
It has an axisymmetric part in the middle, where it can be well represented by an ellipse.
The center of this ellipse is not the origin of the grid, but is shifted to the left by one kpc.
Such a shift of the center is also found in the larger 2.5D simulation.
The elliptical shape ends at $|Z|\approx10$~kpc where two cigar-like extensions join the bow shock.
These extensions are 3D in nature with several bumps. They can be represented, on average, by a parabola
of rank three.

\subsubsection{The law of motion of the bow shock}
\label{bowprop3d}
In all the plots in the long term evolution, the two bow shock phases are easily discernable.
The inner bow shock structure is a remnant from the blastwave phase. In its early phase,
the radius of the bow shock should have obeyed equation (\ref{constden}), i.e. $r \propto t^{0.6}$.
(This part of the bow shock is far inside the core radius wherefore the density profile
is roughly constant.)
In the following, this law is checked for early and later evolution using the bow shock position
in the $Z=0$ plane.
For every hundredth timestep, the positive radial bow shock position for $\phi=0$ and $Z=0$
was saved, up to the final timestep 220200, corresponding to  $t=2.04$ Myr.
These values are plotted against time in Fig.~\ref{bowfits3d}.
The positions were fitted with a function of the type: $a+bt^c$. 
The constant $a$ was allowed for 
in order to take into account an artificial offset due to
the initial conditions imposed by a jet with a final radius and non-developed 
structure at $t=0$. The fit was carried out 
globally, and separately for the early and the late evolution. The result is given in
Table~\ref{bowfits3dtab}.
\begin{table}[h]
\caption{\label{bowfits3dtab}Fit parameters for the bow shock position.
The star denotes a fit with fixed $a=0$.}\nopagebreak
\begin{center} \noindent
\begin{tabular}{lccc}  \hline \hline
\multicolumn{4}{c}{\vspace{-3mm}}\\
time range [Myr]  & a       & b & c     \\ \hline
  global         & 1.02 & 4.44 & 0.58    \\ 
  {$[0:0.5]$}        & 0.41 & 4.52 & 0.38   \\
  {$[1:2  ]$}        & 1.25 & 4.21 & 0.60   \\ 
  {$[1:2  ]^*$}        & 0 & 5.45 & 0.49   \\ \hline \hline
\end{tabular}
\end{center}
\end{table}
The global fit gives an exponent $c$ of 0.58, quite close to the exponent
analytically derived for the blastwave phase (0.6). 
However, the behaviour is quite different at late times compared to the early evolution.
At late times, the exponent is in the range $0.5$ to $0.6$ (compare below), 
whereas for the early evolution,
$c$ is close to 0.4, which would be the value expected for a supernova blastwave 
(initial energy input
with no additional supply, compare~(\ref{globeqmot})):
\begin{equation}\label{rsupnov}
r=\left(\frac{15 E_0 t^2}{4 \pi \rho_0}\right)^{1/5}
\end{equation}
 This is due to the initial conditions
of the simulation: in order to get a propagating jet, 
one has to start with a short propagating jet.
The energy of that initial jet  is quite high. 
It can be estimated by calculating the energy
stored in the initial injection box:
$E_0\approx \pi R_\mathrm{j}^2 h \rho_\mathrm{j} v_\mathrm{j}^2$, 
where h is the height of the 
the cylindrical region. This amounts to approximately $10^{57}$ erg. 
The energy is also given by the
parameter $b$, according to equation (\ref{rsupnov}), and since the density is known:
\eq{E_0=2.35 \times 10^{80} \rho_0 b^5 \approx 10^{58}\, \mathrm{erg}.} 
The difference of a factor of ten
could be attributed to increased heat production in the very early evolution,
where the unphysical initial condition, which is always present in such simulations,
relaxes into a regular jet structure.
\ignore{
The normalised differences of the fits for the early and the late evolution are also given
in Fig.~\ref{bowfits3d}.
For the early phase fit, the deviation from the measured values is in the percent regime for the
time between 0.01 Myr and 0.5 Myr. Prior to that the fit worsens. Looking at the contour plots for that
time (Fig.~\ref{cyga_early}), one recognises that the jet first develops a bow shock feature,
before the beam propagates at all. Thus, at that time a hot thermal bubble develops with an 
infinite reservoir of energy. From Fig.~\ref{bowfits3d}, this phase should be finished by 
$t\approx 0.01$~Myr,
which is in agreement with Fig.~\ref{cyga_early}, where the jet is seen really to get started
around that time. The anomalous initial behaviour is probably
also responsible for the
differing $a$ parameter of the fits: Since the jet radius is 0.55 kpc, one should have expected
the offset to be not too far from that value.
The fit for the late evolution, adapted to $t=(1-2)$ Myr, is a good approximation down to 
$t=0.5$ Myr. This indicates that at this point the constant energy injection begins to dominate
over the startup energy.}

At late times, the initial conditions may be regarded as relaxed.
In that case, the fit parameter $a$ could be set to zero. This would
change the best fitting
exponent from $0.6$ to $0.5$. 
An exponent lower than $0.6$ would be expected due to the 
increase of the aspect ratio. However, the effect is quite small.
 
\subsubsection{Implications on the jet power}\label{pow3d}
From the $b$ parameter of the late evolution, one can calculate the jet power,
according to equation (\ref{constden}): 
\eq{L_\mathrm{kin}=2.23 \times10^{67} \rho_0 b^5 \approx 3.75 \times 10^{45}\, \mathrm{erg}.}
This is approximately 40\% of the true jet power 
(it would be 50\% taking b from the global fit, and 130\% for the late fit with $a=0$).
One can also use the bow shock position and velocity at e.g. $t=2.04$~Myr 
in order to compute its power. From (\ref{constden}) one gets:
\begin{equation}
L=100\pi \rho_0 (r_\mathrm{bow})^2 (v_\mathrm{bow}/3)^3\enspace.
\end{equation}
This yields a jet power of $L\approx10^{46}$~erg/s, accurate to a factor of a few because of
the slow bow shock velocity and the low spatial resolution. The true jet power is 
\mbox{$1.04 \times 10^{46}$~erg/s}.  
It follows that the information on the
jet power is conserved in the law of motion
of the central part of the bow shock at least with an accuracy of a factor of a few.

\ignore{
Summarizing, the situation is as follows: 
In the very early phase, where we should have the $r\propto t^{0.6}$ law, 
the numerical effects
dominate, and we cannot see the expected behaviour. 
But soon the expansion turns into the $r\proptot^{0.6}$
power law. 
Interestingly, the geometry change, i.e. the rising aspect ratio that 
increasingly violates the spherical symmetry approximation, seems to have 
at most a slight effect on 
the exponent $c$.
The bubble size is a direct indicator of the energy content, if the density is known:
When a real jet starts its kpc evolution, it produces a spherical bow shock. 
At that time it prints an identity card,
containing date of birth, 
luminosity, and useful limits on almost any other interesting quantity.
This information is transported outward with no loss of accuracy, as long as the $r\propto t^{0.6}$
law holds. The simulation shows that the law holds for the radial extention, which is
the only dimension that is not affected by projection effects.}

\begin{figure*}[bth]
\begin{center}
\rotatebox{-90}{\includegraphics[height=.38\textwidth]{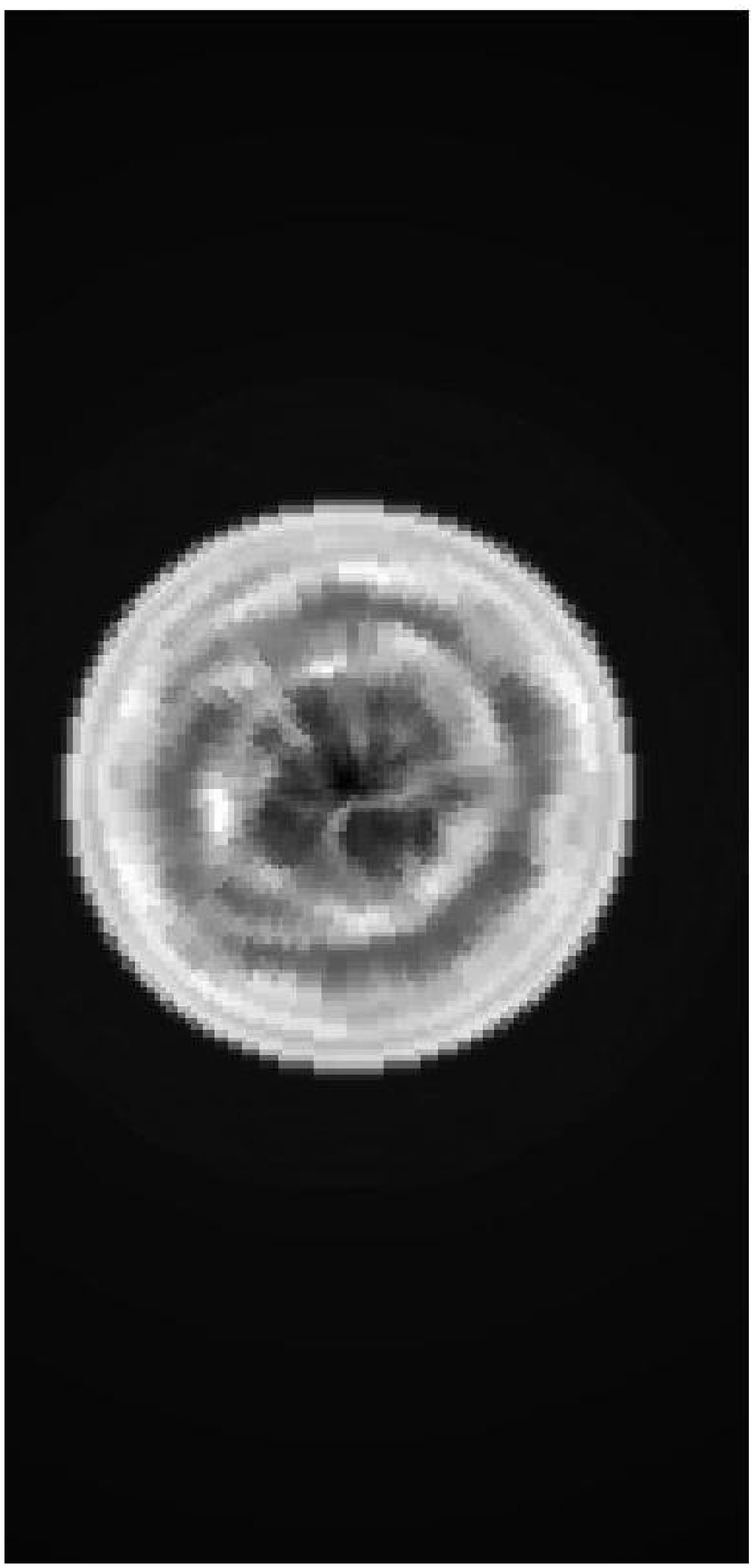}}
\mbox{\rotatebox{-90}{\includegraphics[height=.30\textwidth]{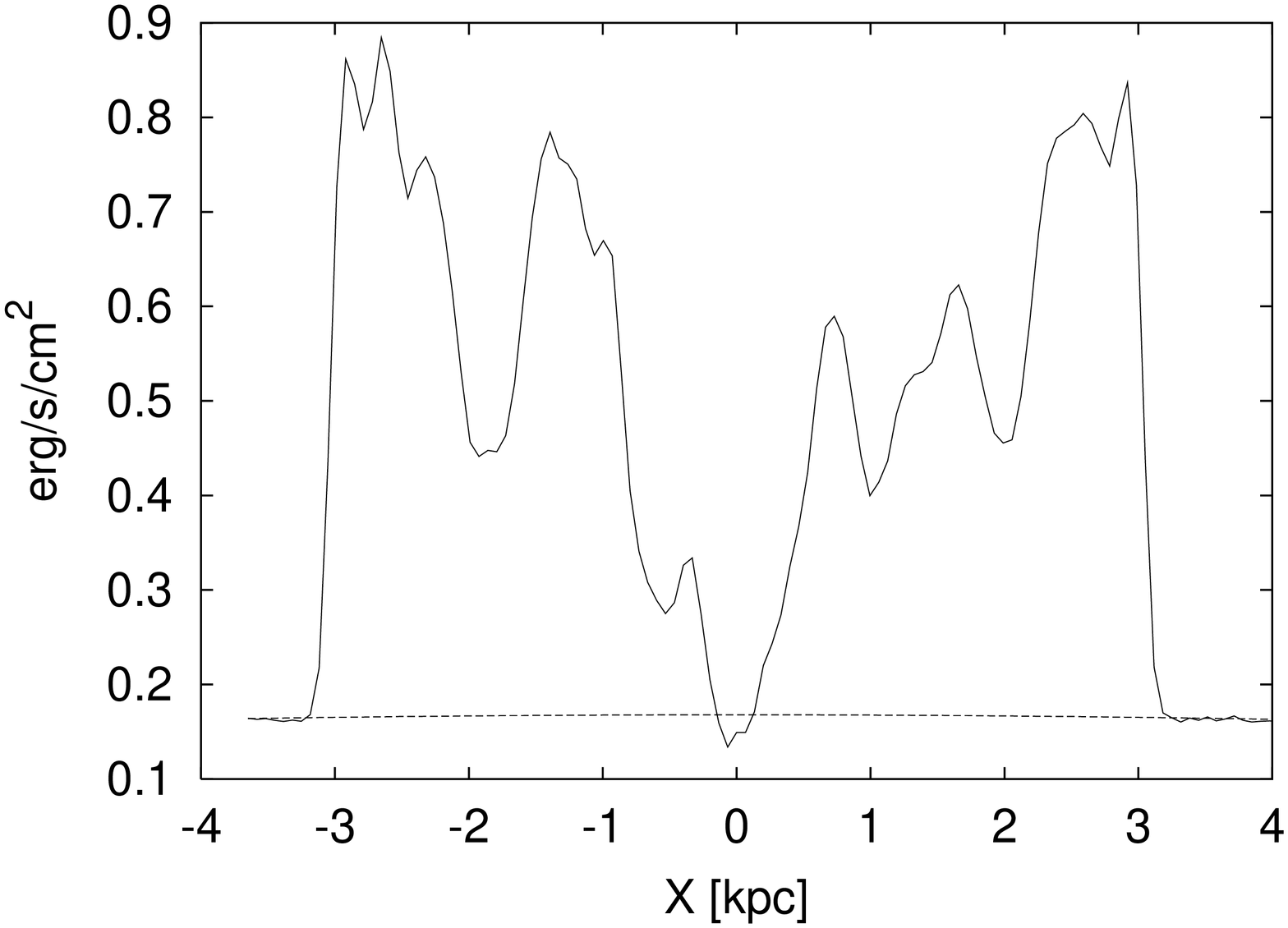}}
\rotatebox{-90}{\includegraphics[height=.30\textwidth]{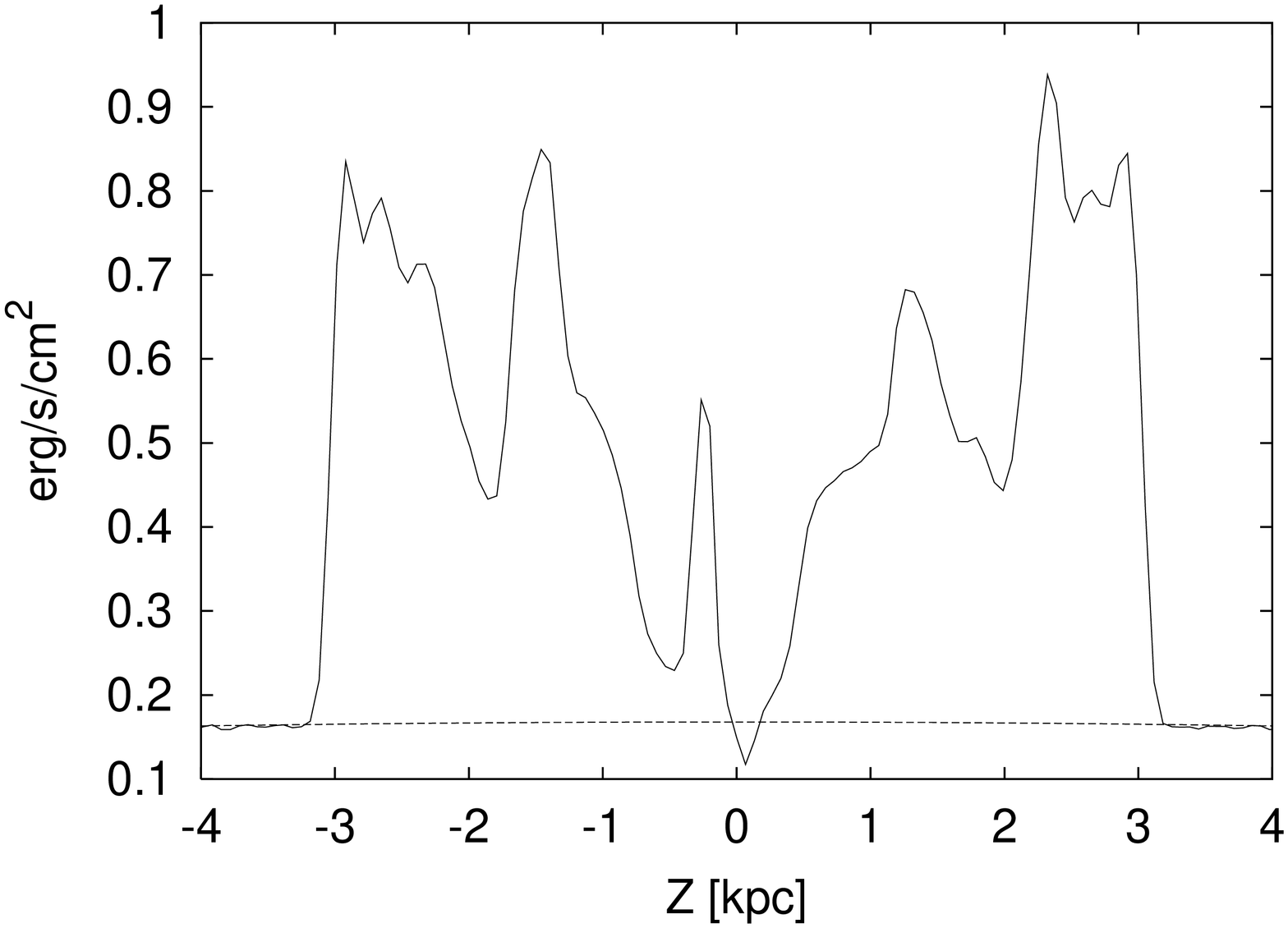}}}\\
\rotatebox{-90}{\includegraphics[height=.38\textwidth]{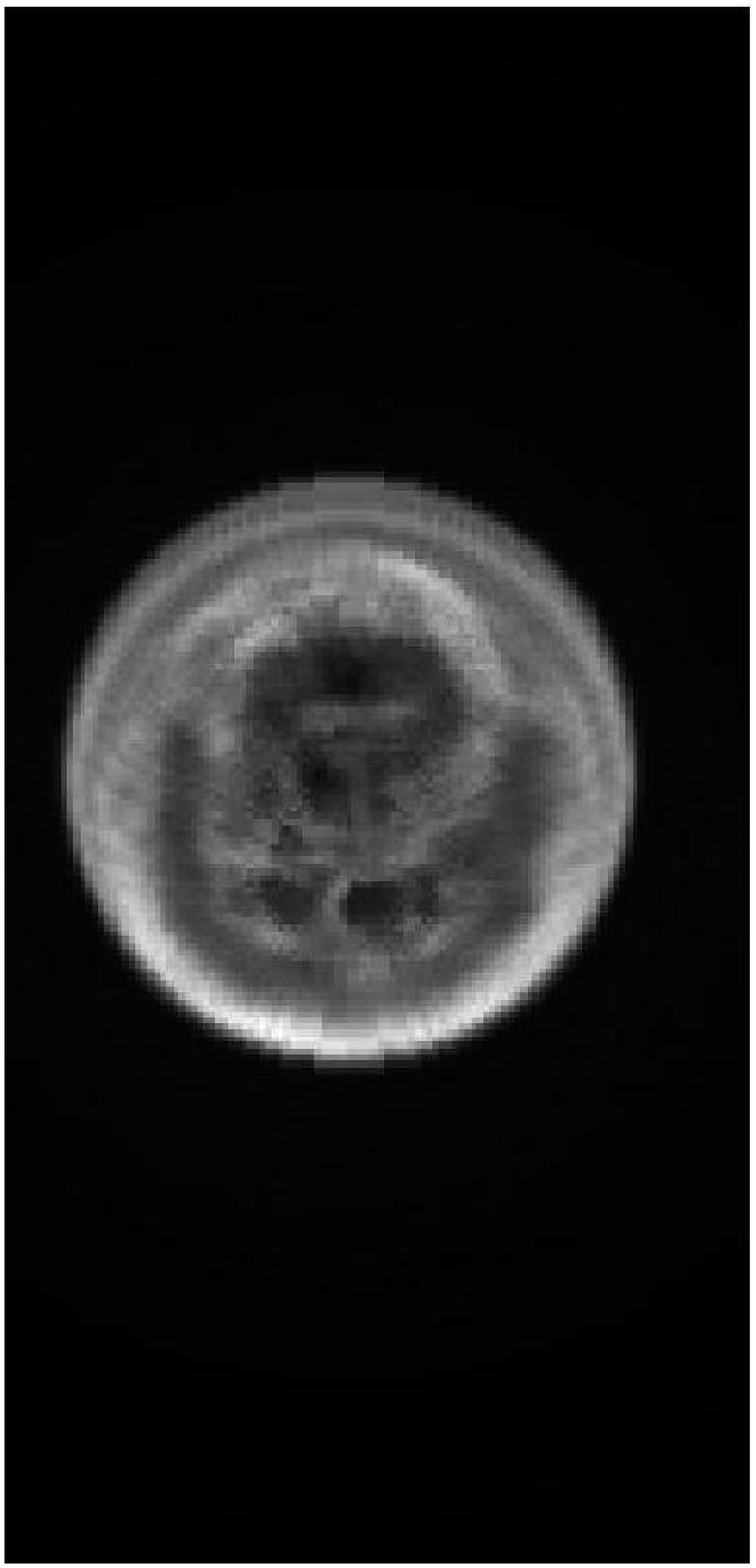}}
\rotatebox{-90}{\includegraphics[height=.30\textwidth]{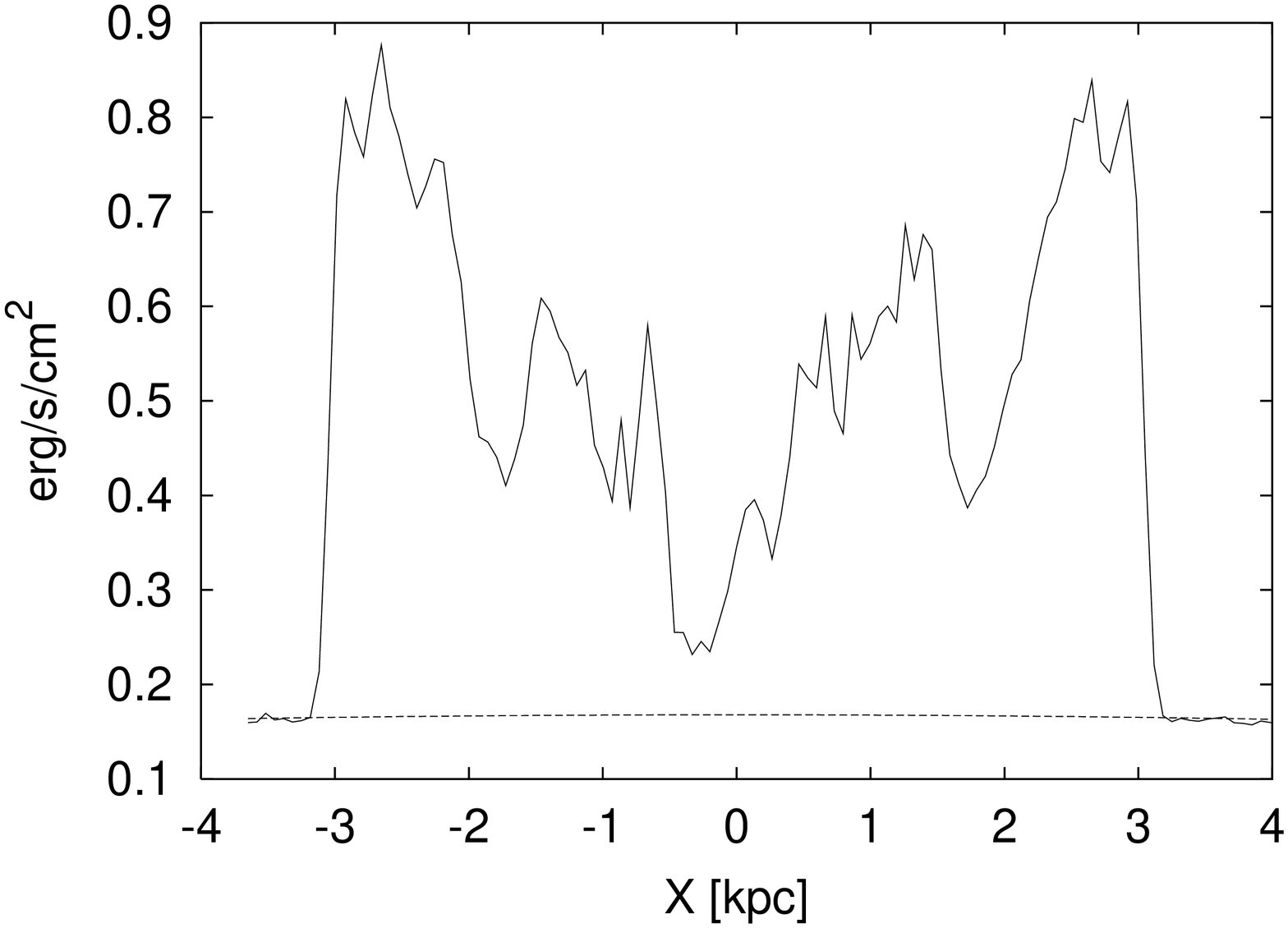}}
\rotatebox{-90}{\includegraphics[height=.30\textwidth]{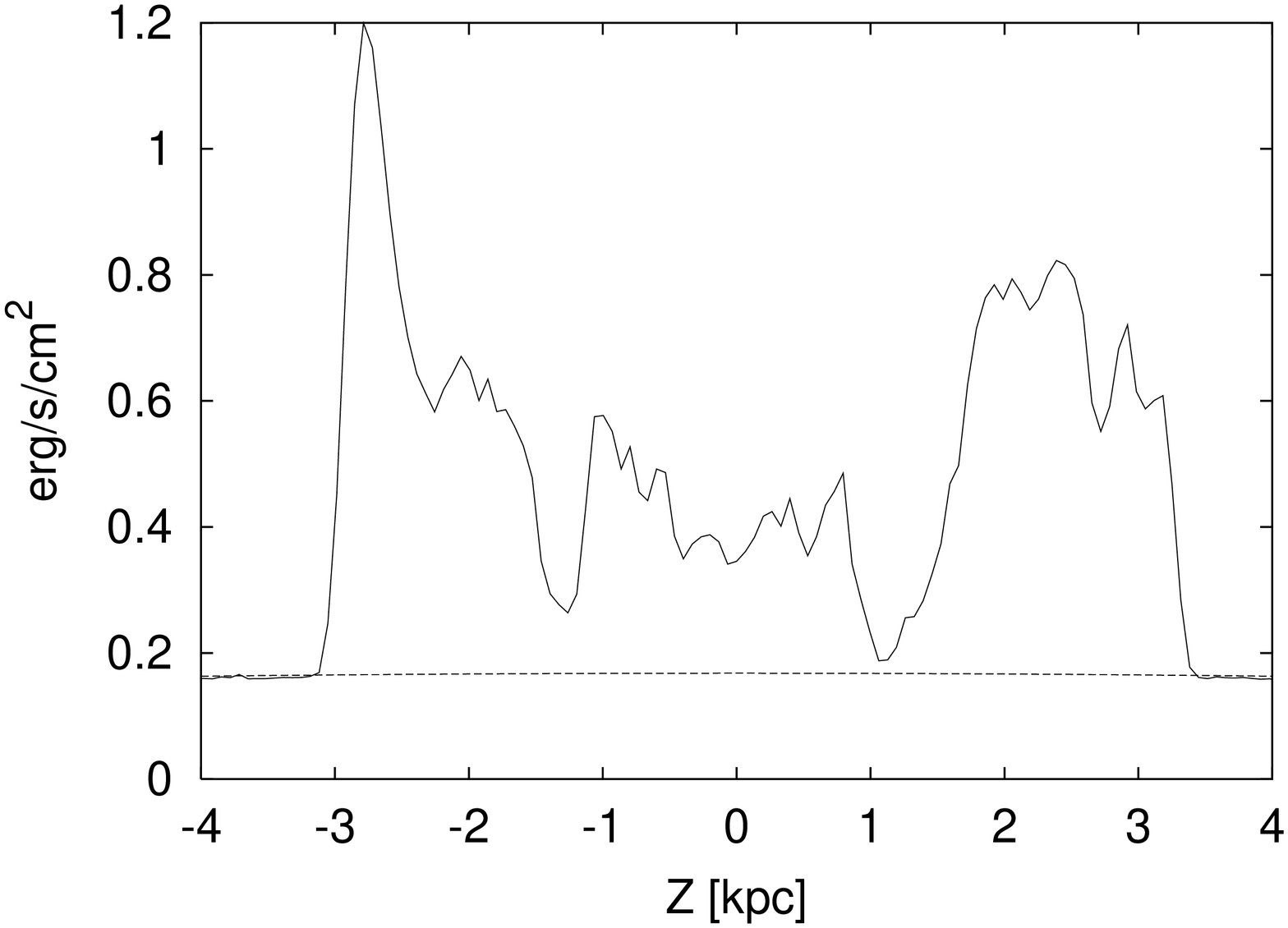}}\\
\rotatebox{-90}{\includegraphics[height=.38\textwidth]{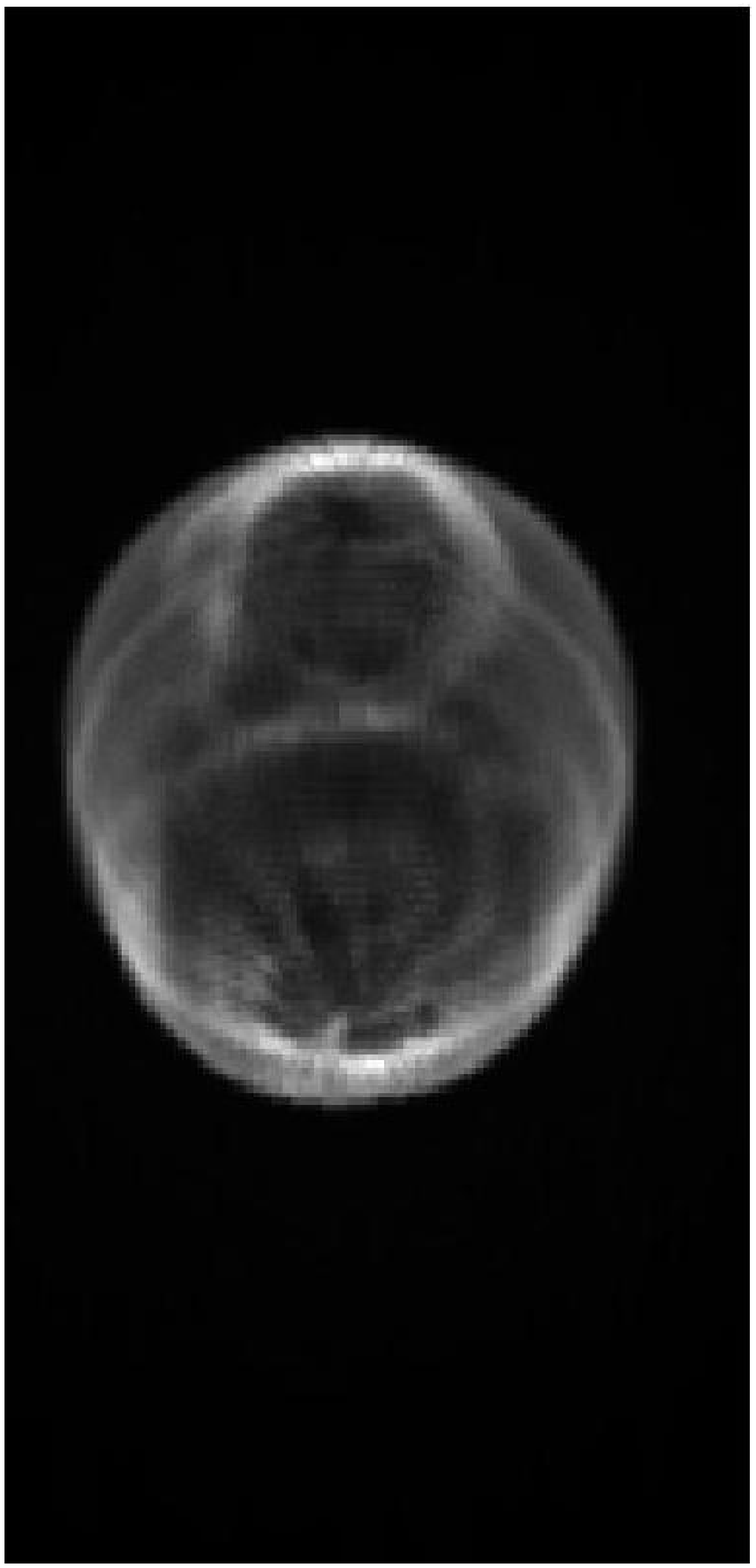}}
\rotatebox{-90}{\includegraphics[height=.30\textwidth]{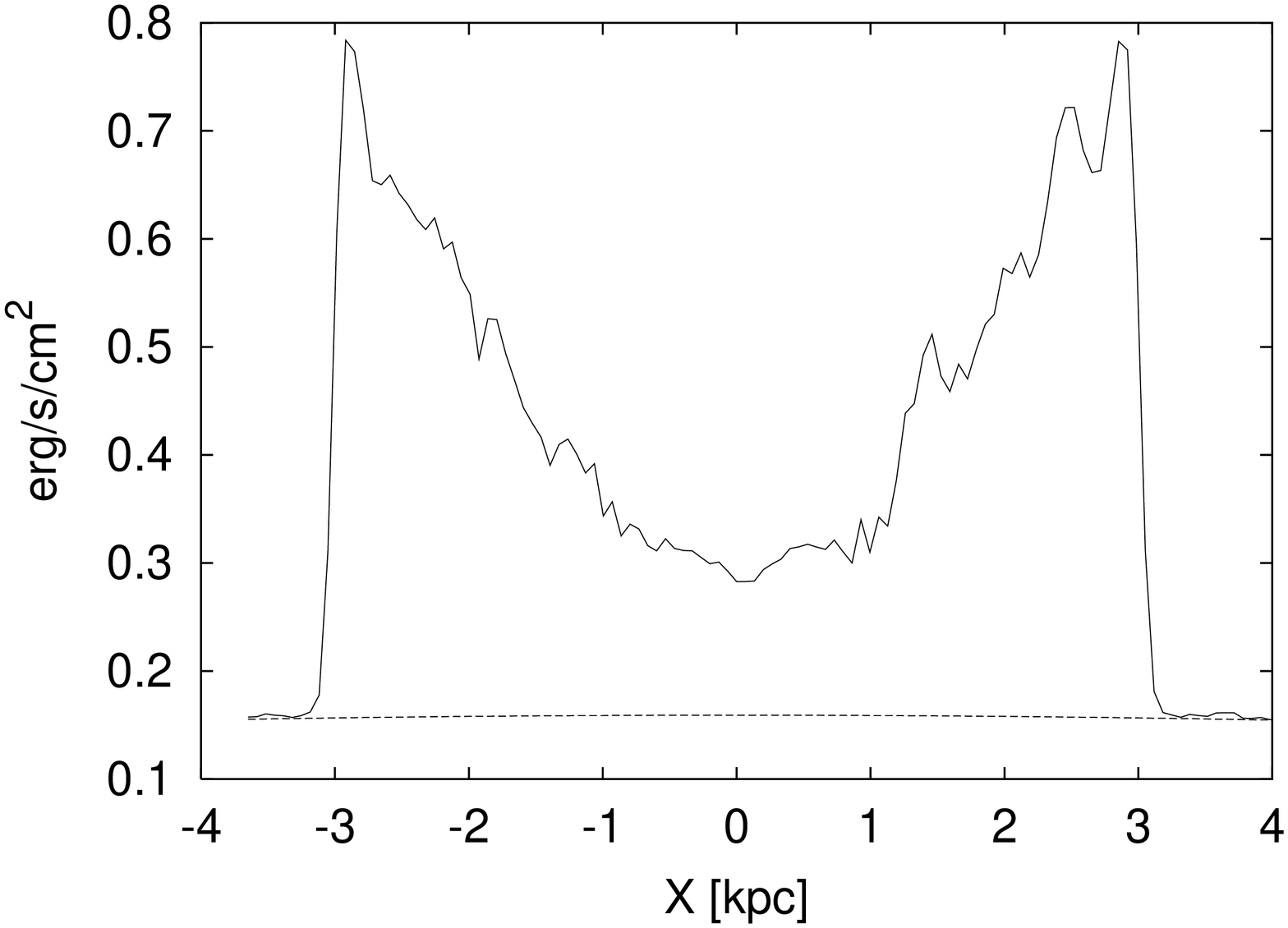}}
\rotatebox{-90}{\includegraphics[height=.30\textwidth]{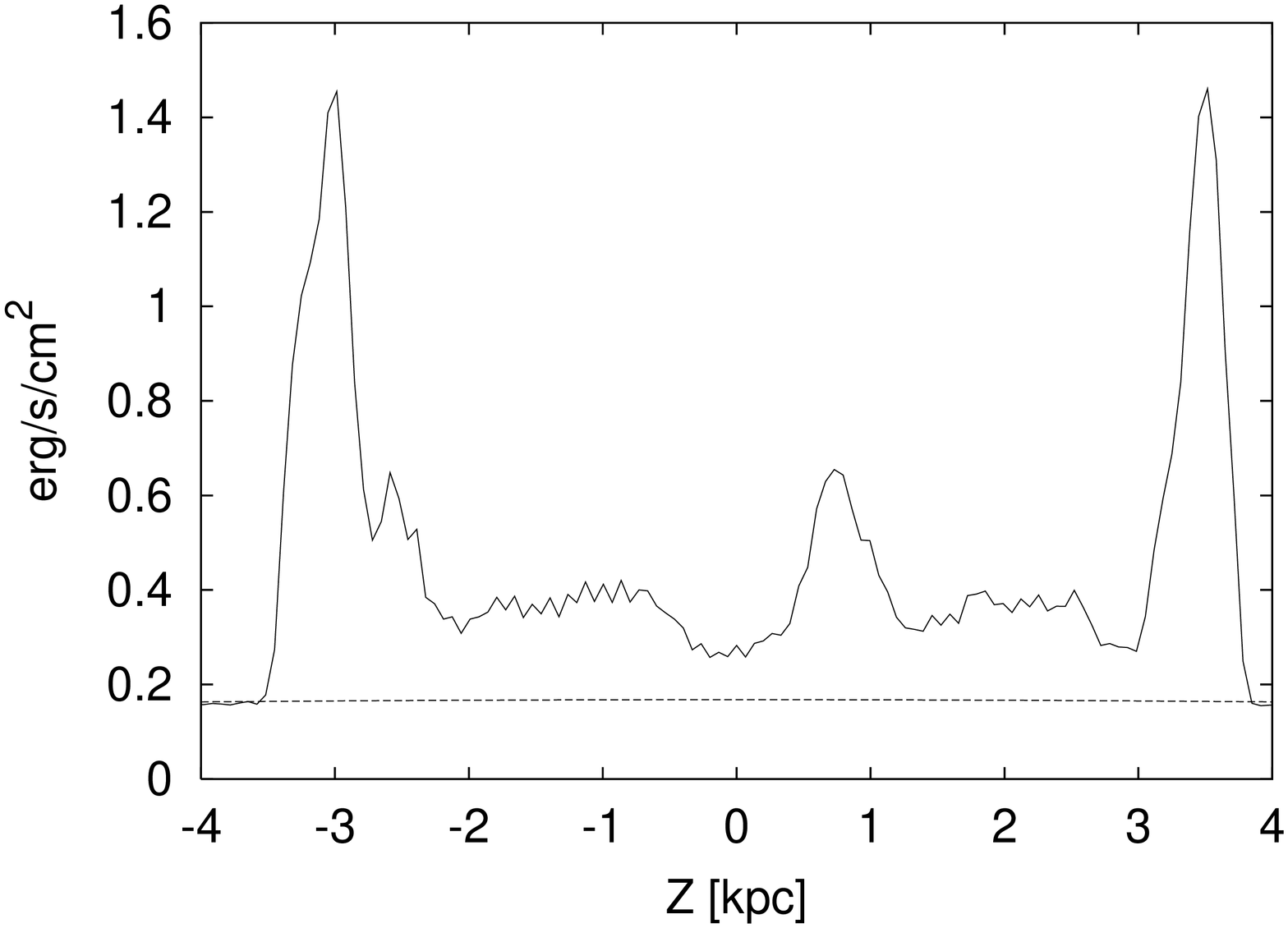}}\\
\rotatebox{-90}{\includegraphics[height=.38\textwidth]{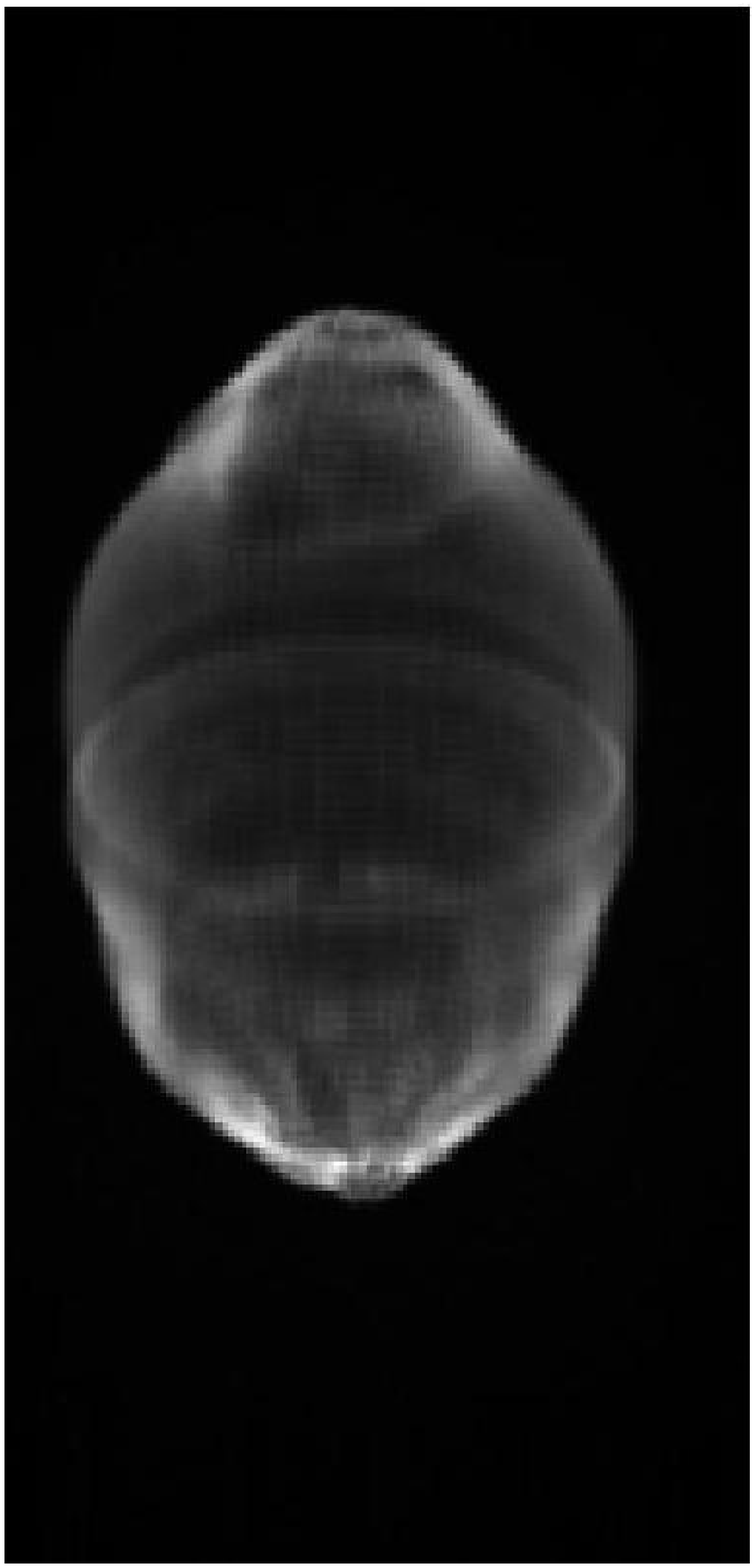}}
\rotatebox{-90}{\includegraphics[height=.30\textwidth]{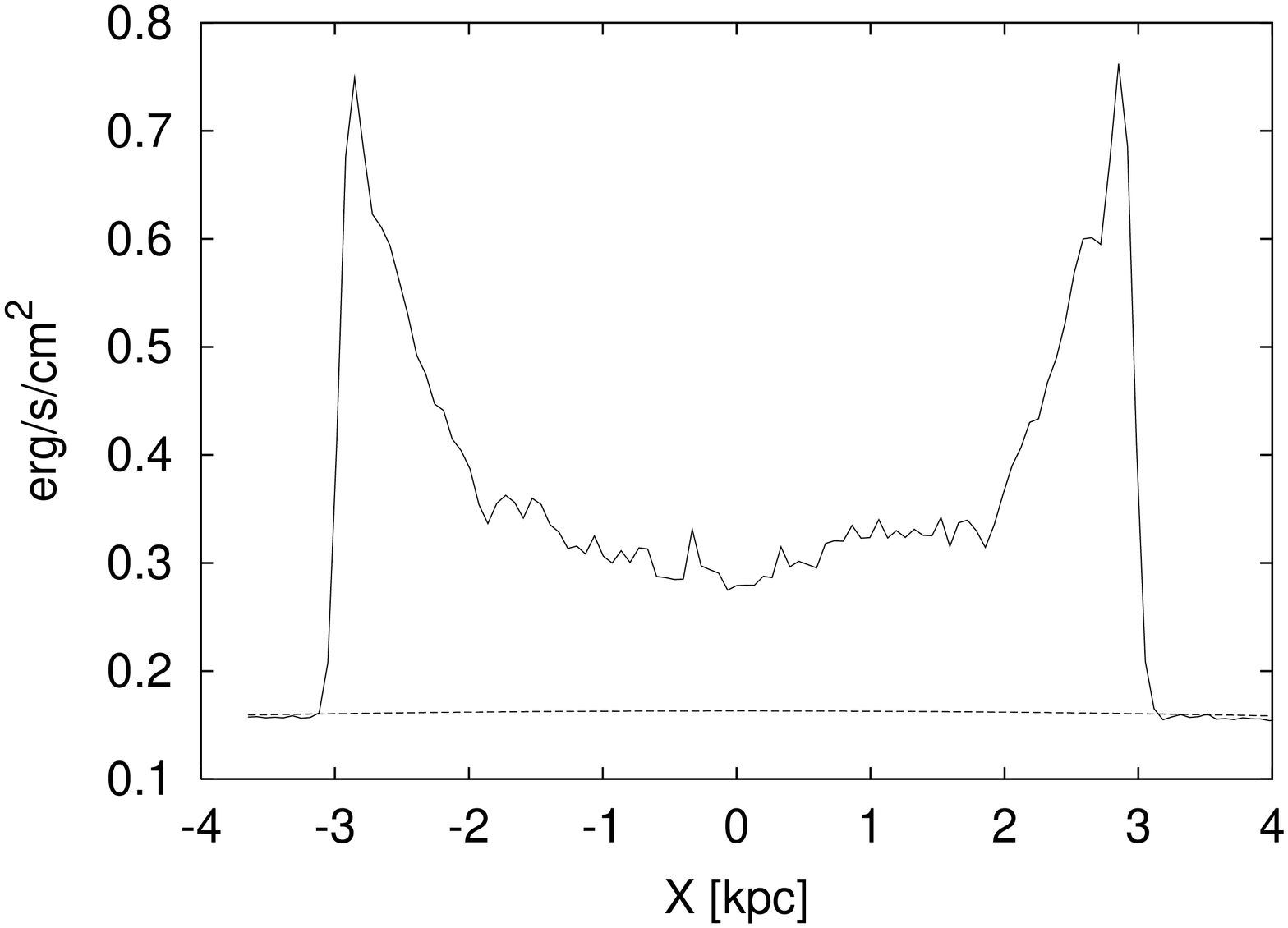}}
\rotatebox{-90}{\includegraphics[height=.30\textwidth]{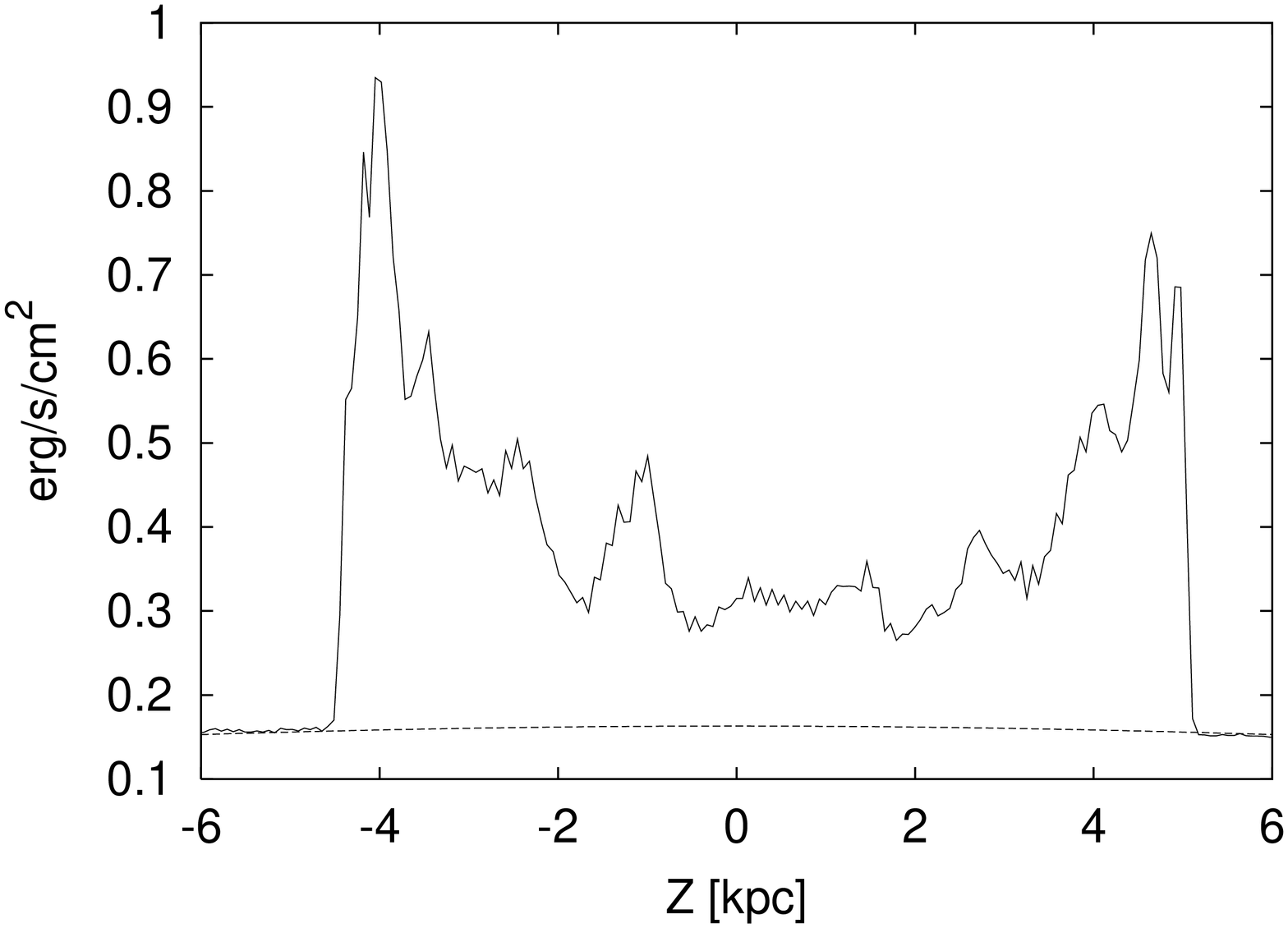}}\\
\rotatebox{-90}{\includegraphics[height=.38\textwidth]{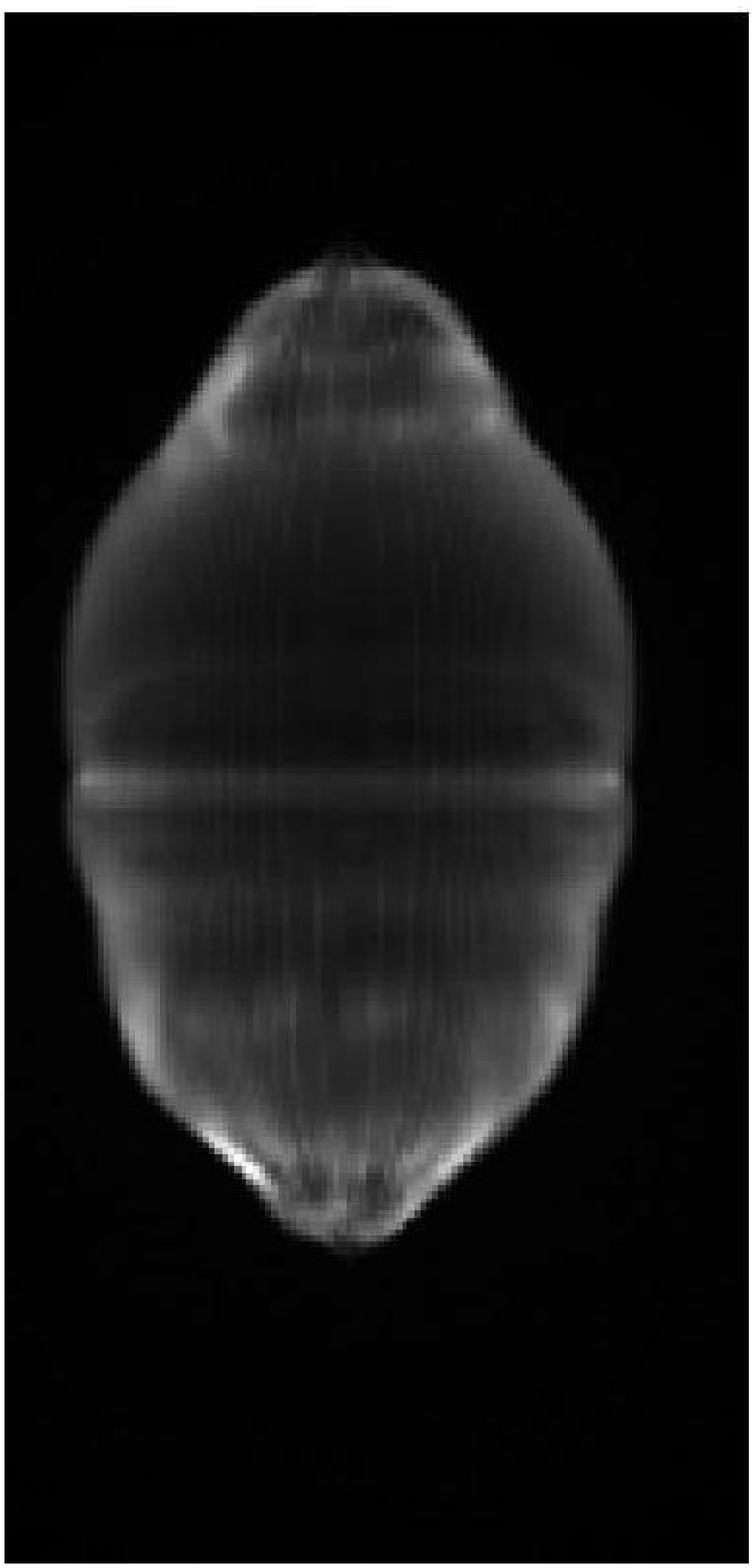}}
\rotatebox{-90}{\includegraphics[height=.30\textwidth]{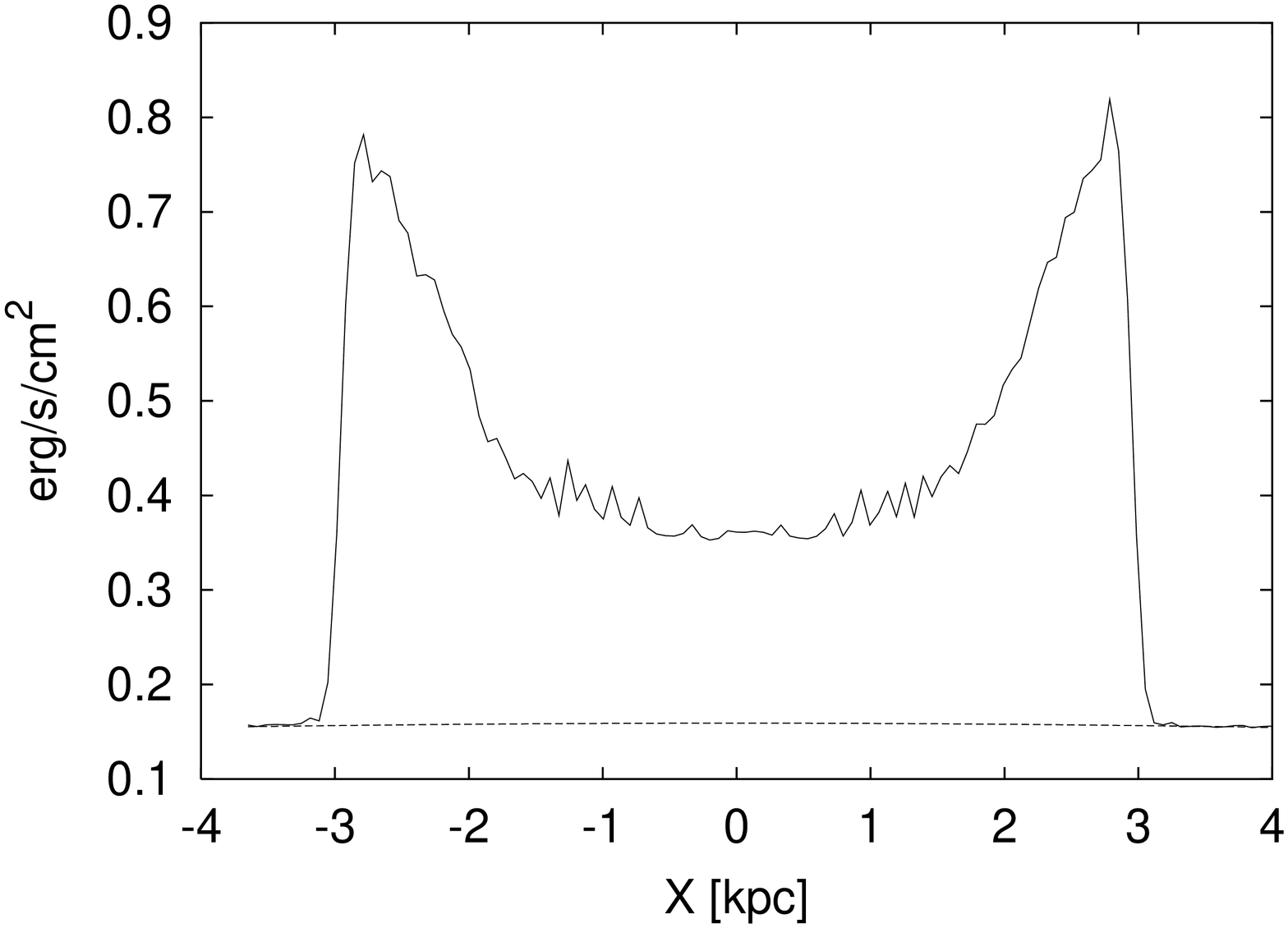}}
\rotatebox{-90}{\includegraphics[height=.30\textwidth]{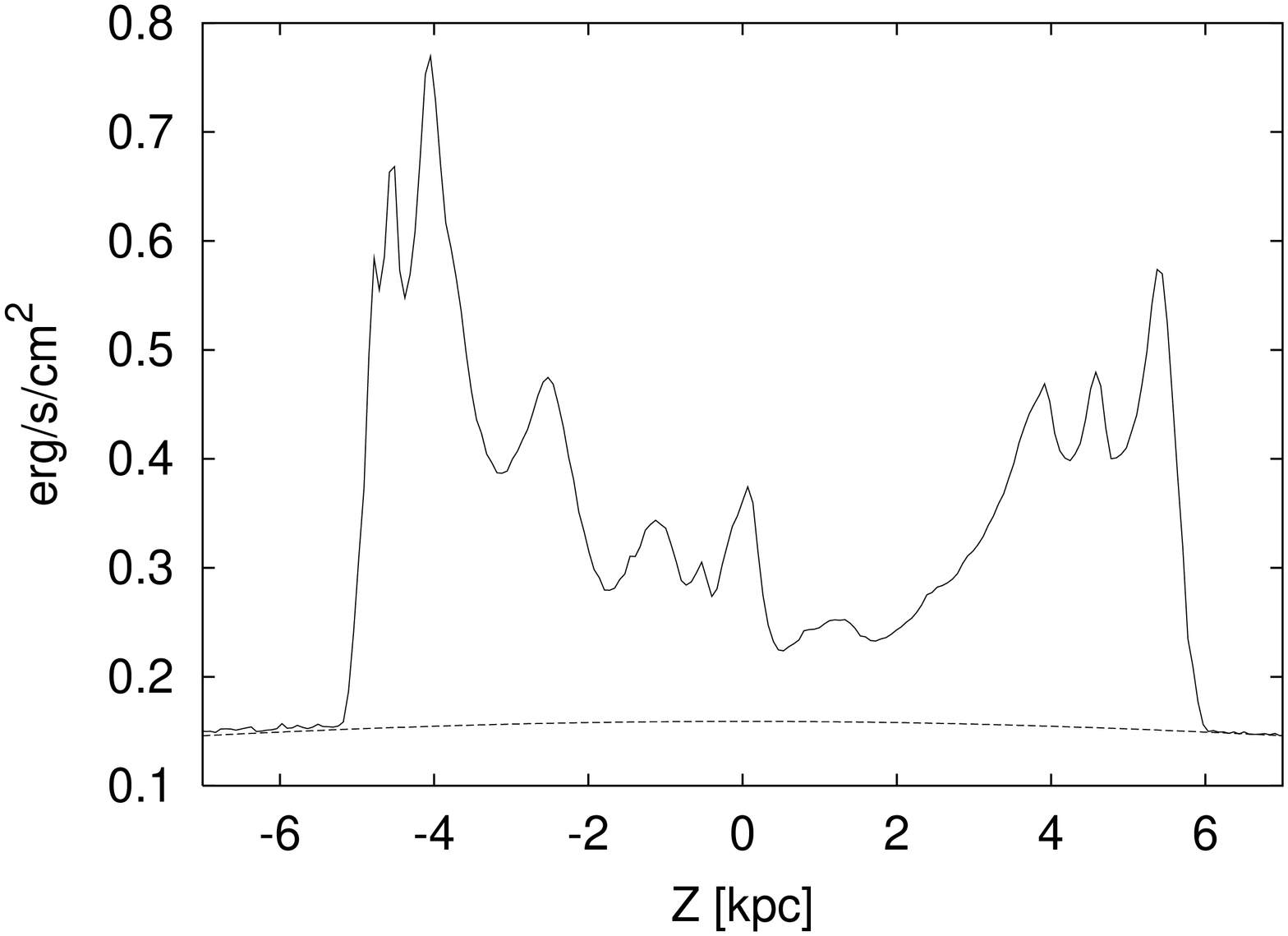}}\\
 \caption{\small Bremsstrahlung emission maps for the 3D run at $t=0.32$~Myr. 
        From top to bottom, the viewing angle is $0^\circ$, $10^\circ$,
        $30^\circ$, $60^\circ$, and $90^\circ$.
        The left column shows the emission map, the middle one a vertical slice
        through the center, and the right one a horizontal slice through the center 
        of the emission map. For comparison, the undisturbed cluster emission is given.
        The bow shock morphology is reflected in the ring like structures of differing
        sizes and surface brightness profiles. 
        \label{emimapa}}
\end{center}
\end{figure*}

\begin{figure*}[bth]
\begin{center}
\rotatebox{-90}{\includegraphics[height=.44\textwidth]{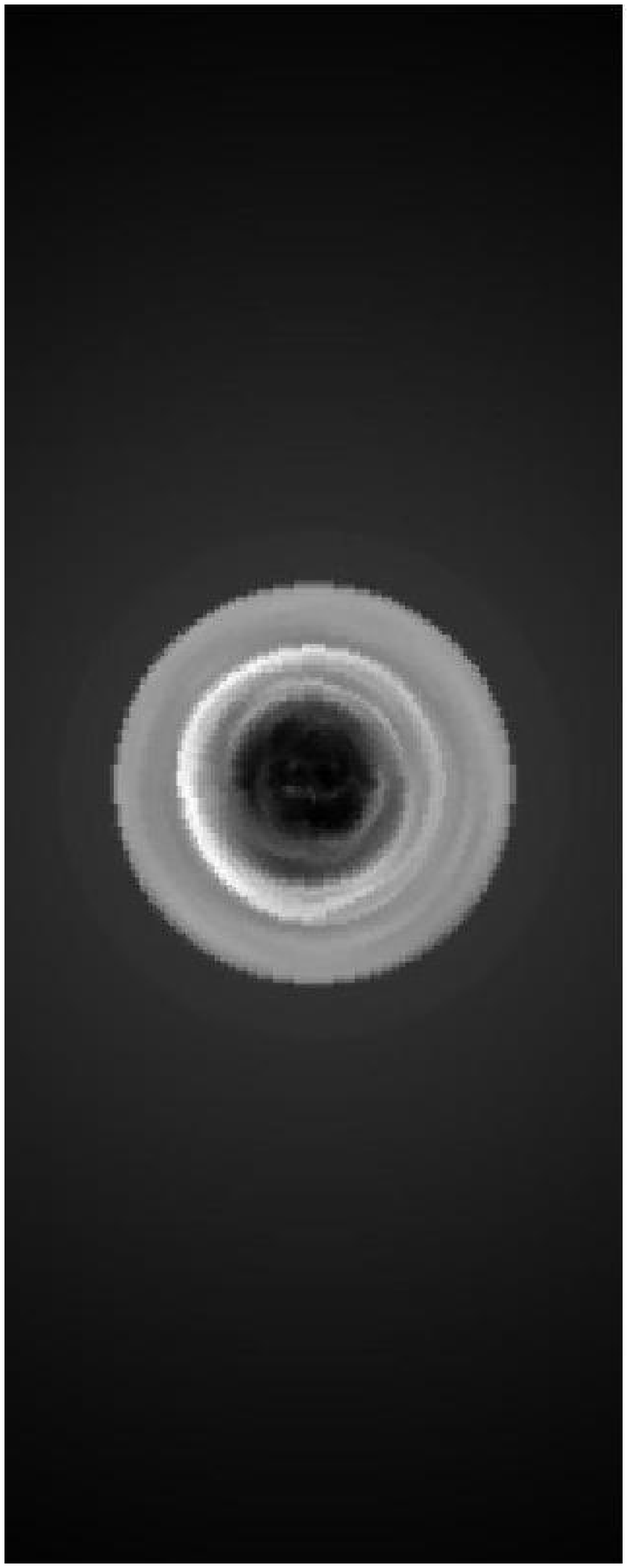}}
\rotatebox{-90}{\includegraphics[height=.27\textwidth]{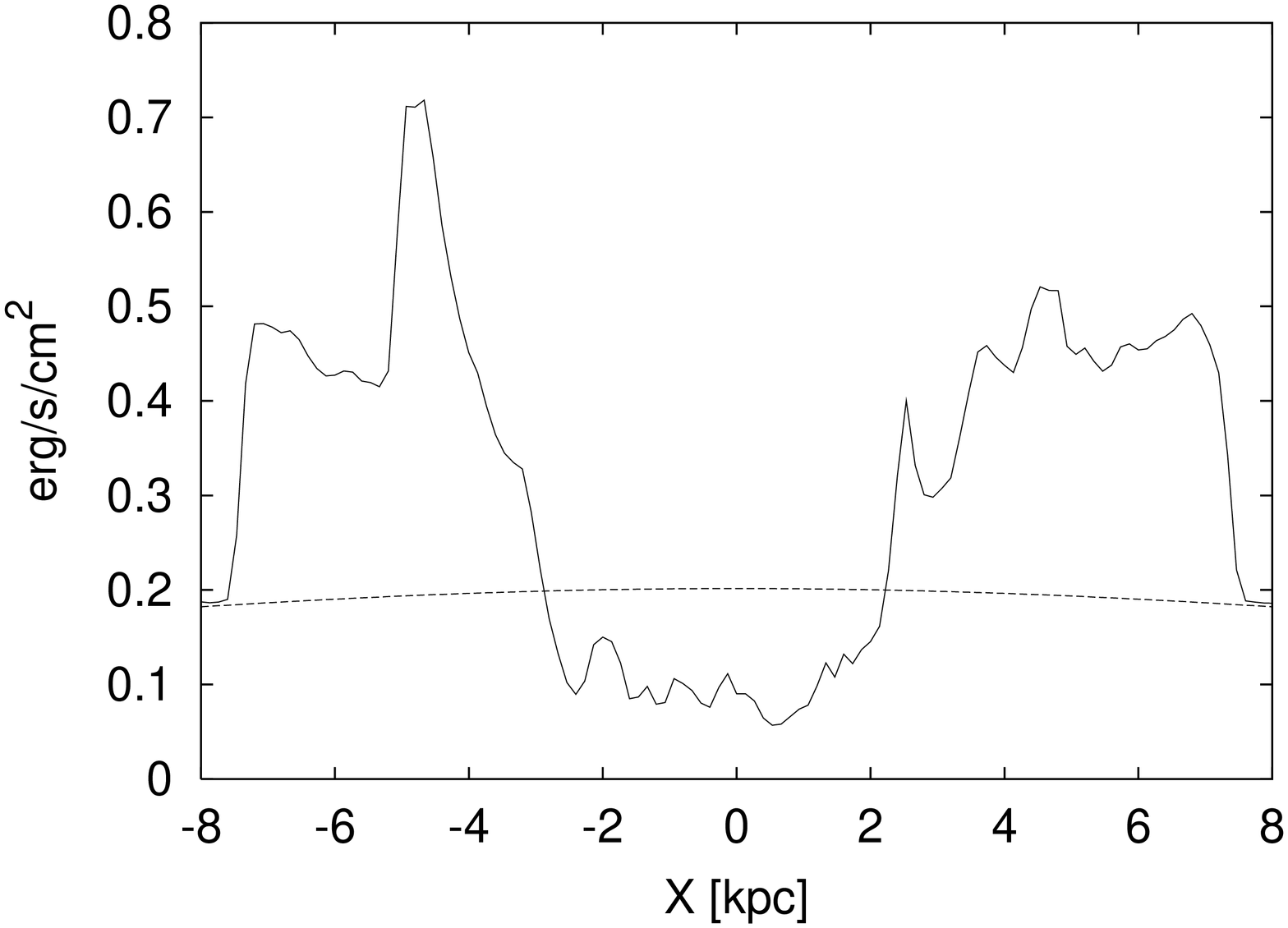}}
\rotatebox{-90}{\includegraphics[height=.27\textwidth]{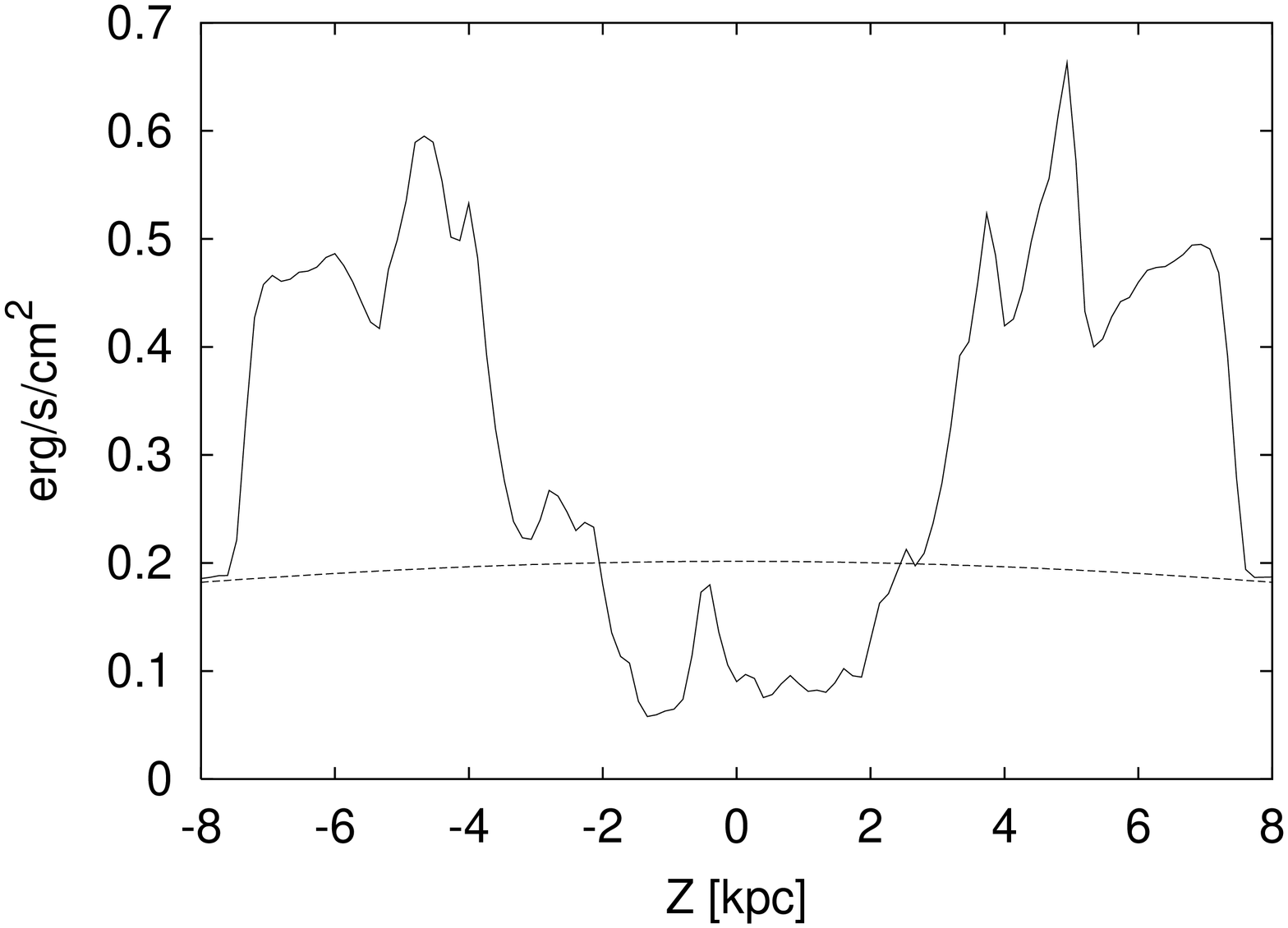}}\\
\rotatebox{-90}{\includegraphics[height=.44\textwidth]{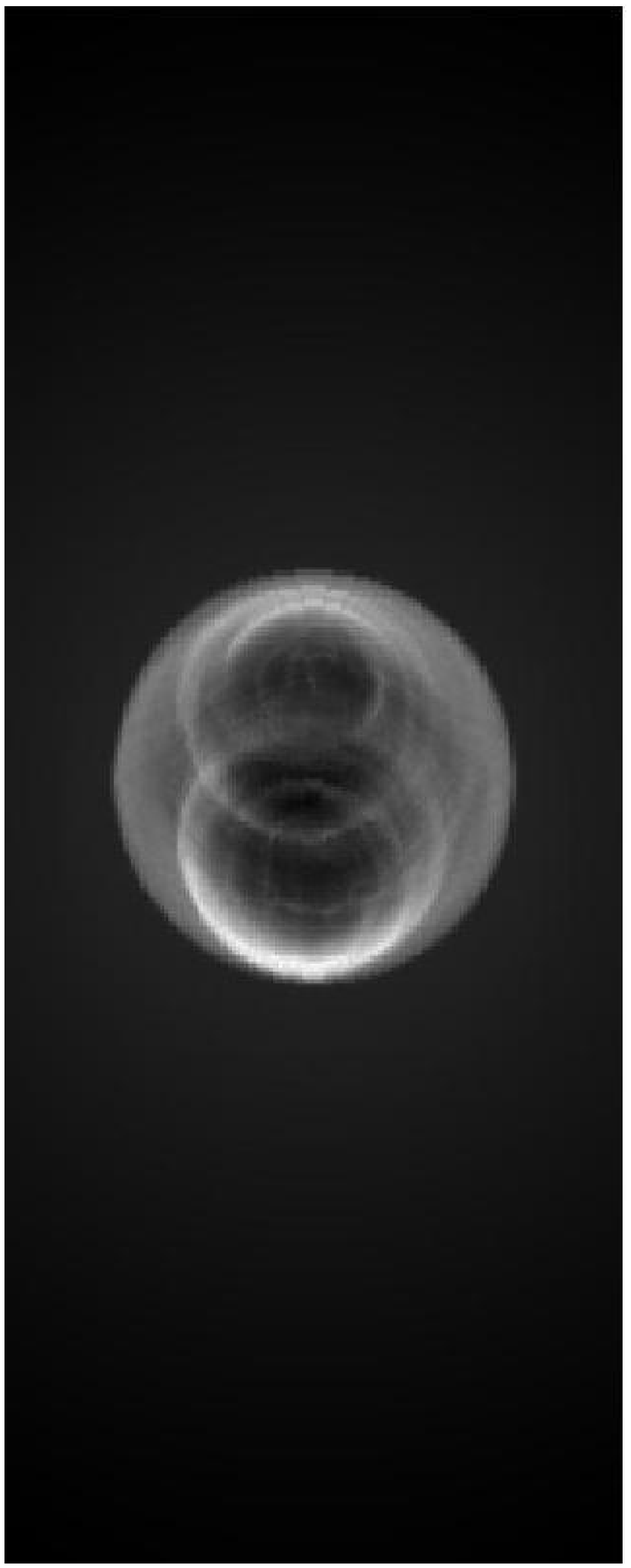}}
\rotatebox{-90}{\includegraphics[height=.27\textwidth]{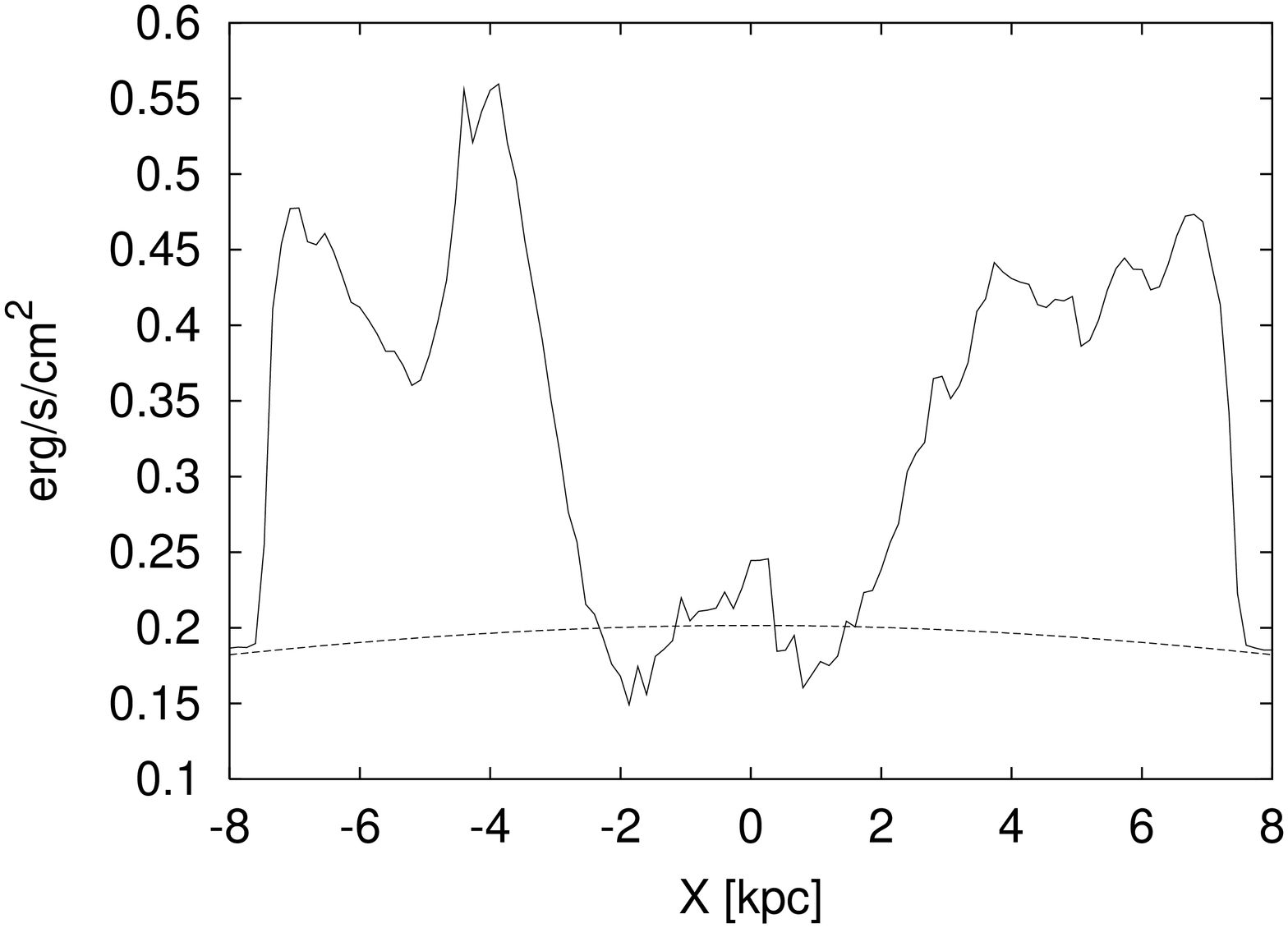}}
\rotatebox{-90}{\includegraphics[height=.27\textwidth]{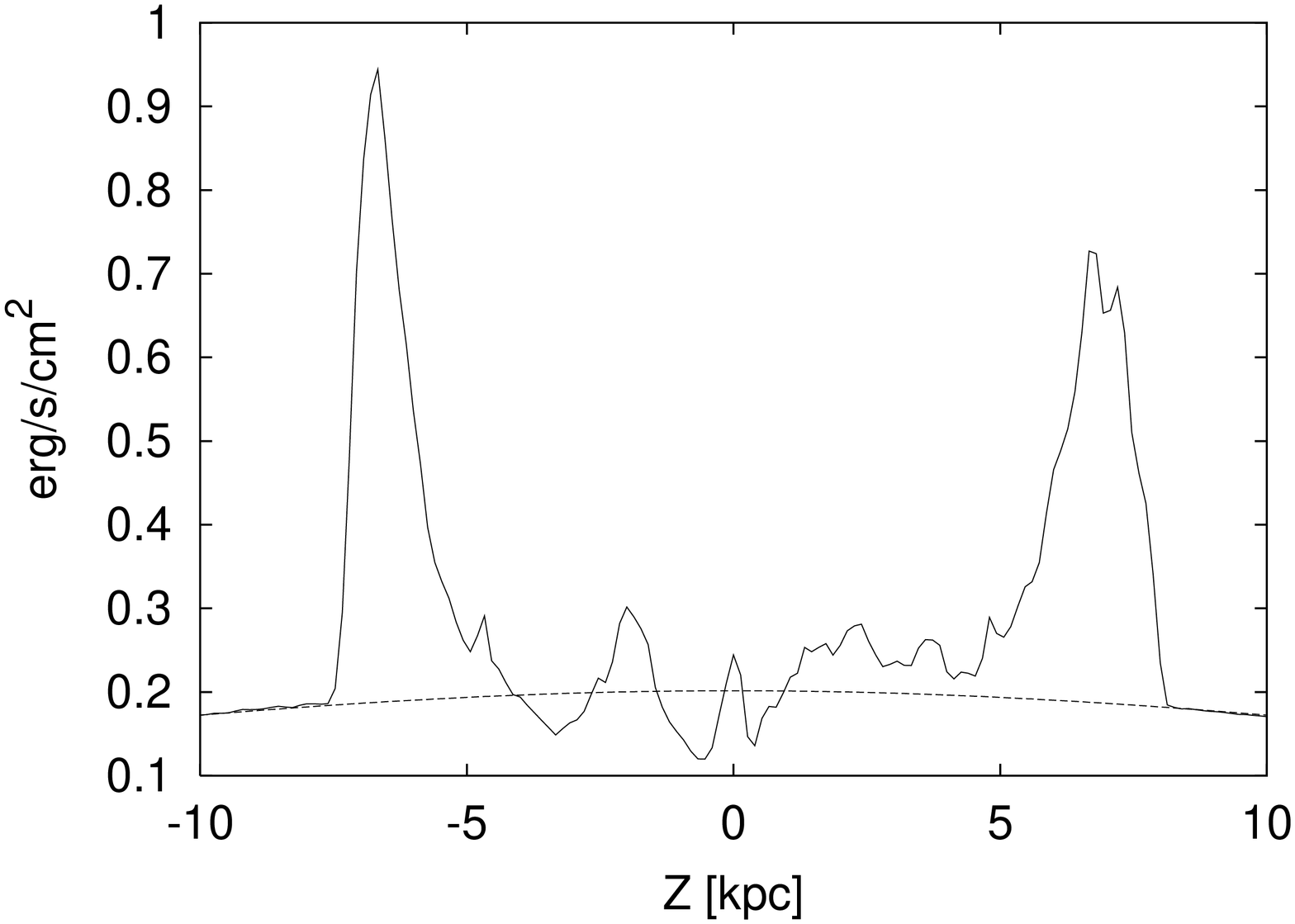}}\\
\rotatebox{-90}{\includegraphics[height=.44\textwidth]{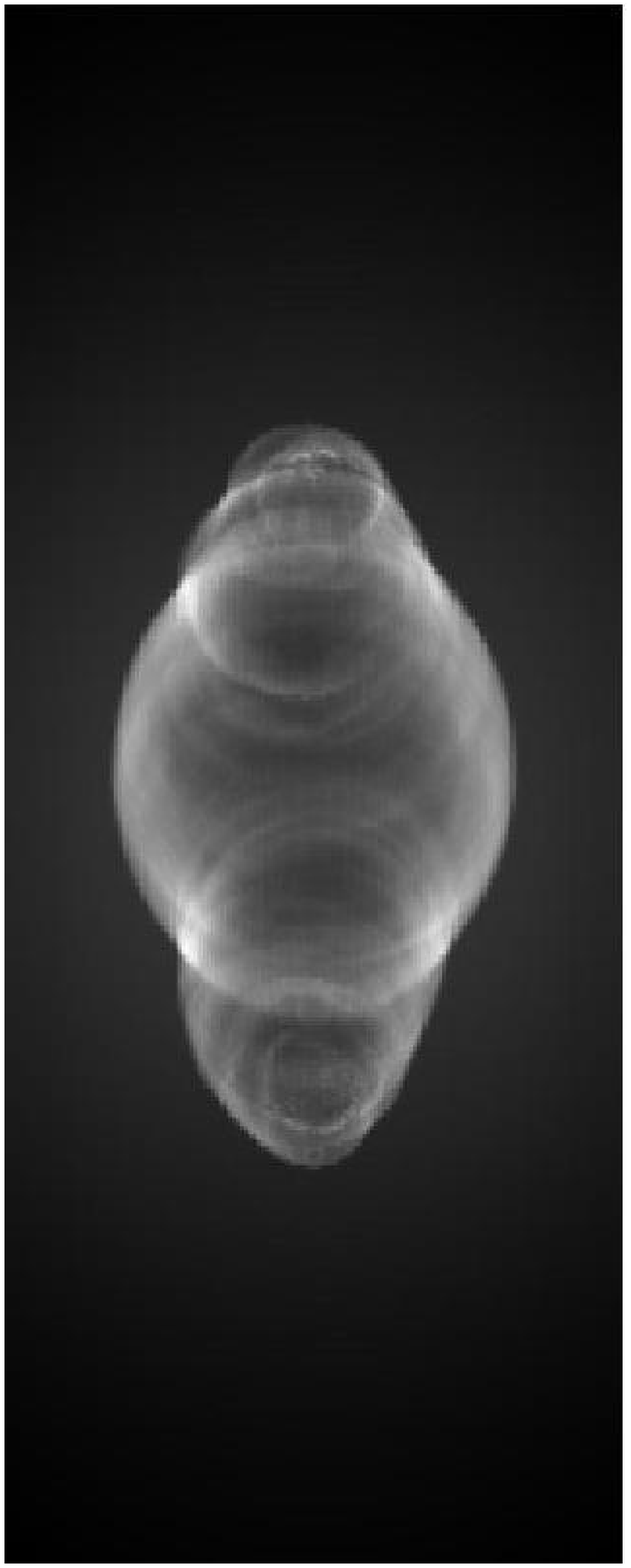}}
\rotatebox{-90}{\includegraphics[height=.27\textwidth]{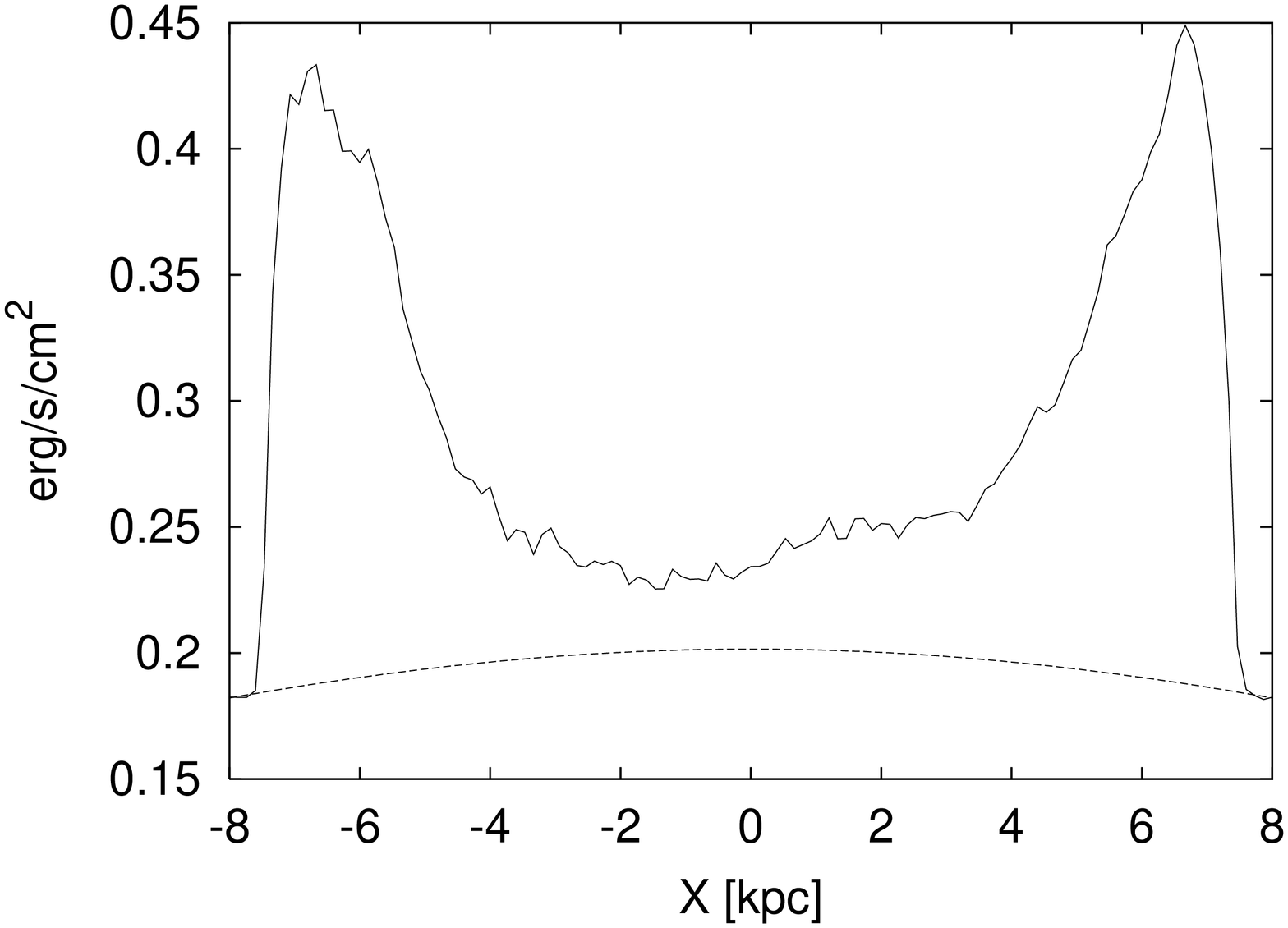}}
\rotatebox{-90}{\includegraphics[height=.27\textwidth]{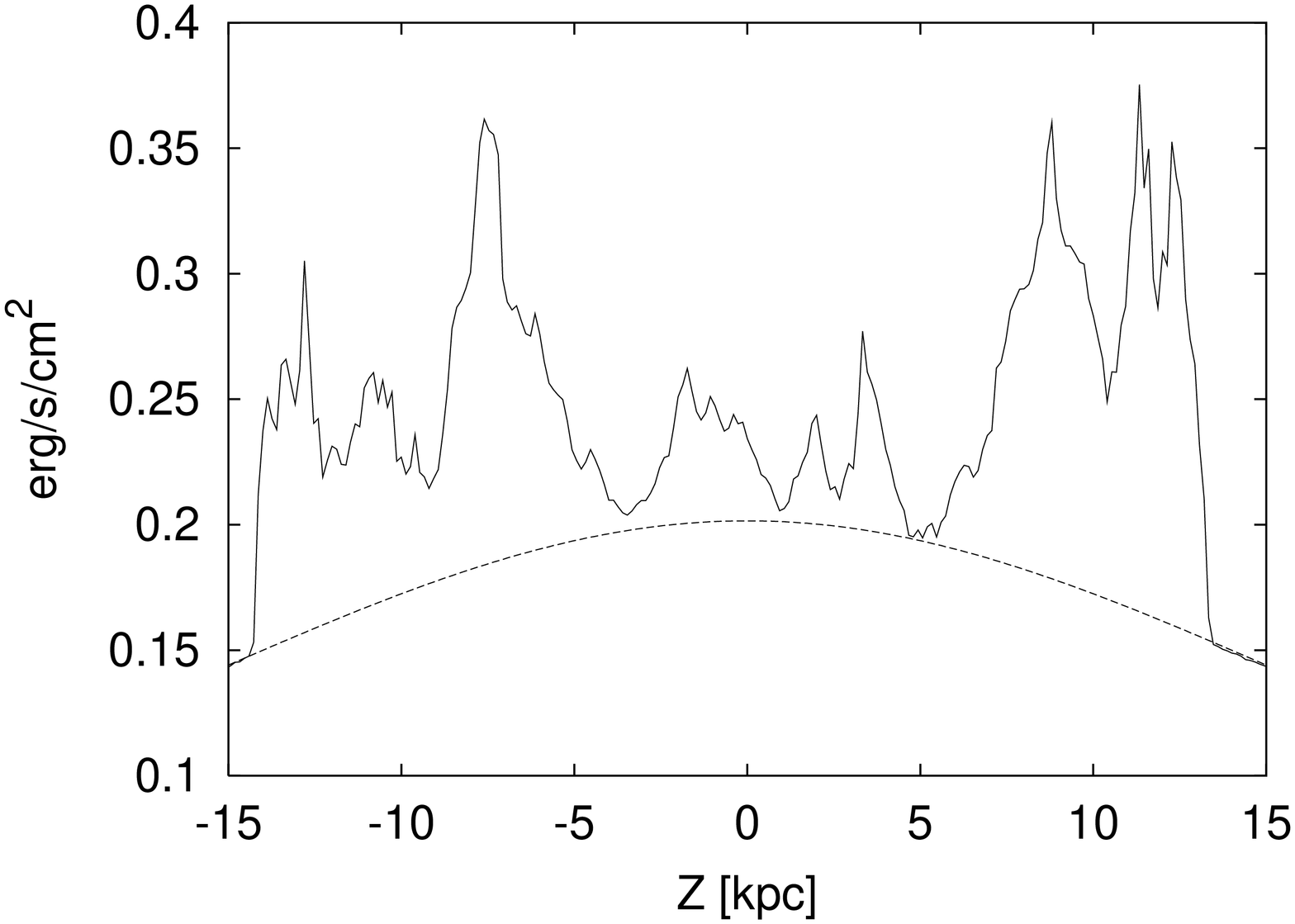}}\\
\rotatebox{-90}{\includegraphics[height=.44\textwidth]{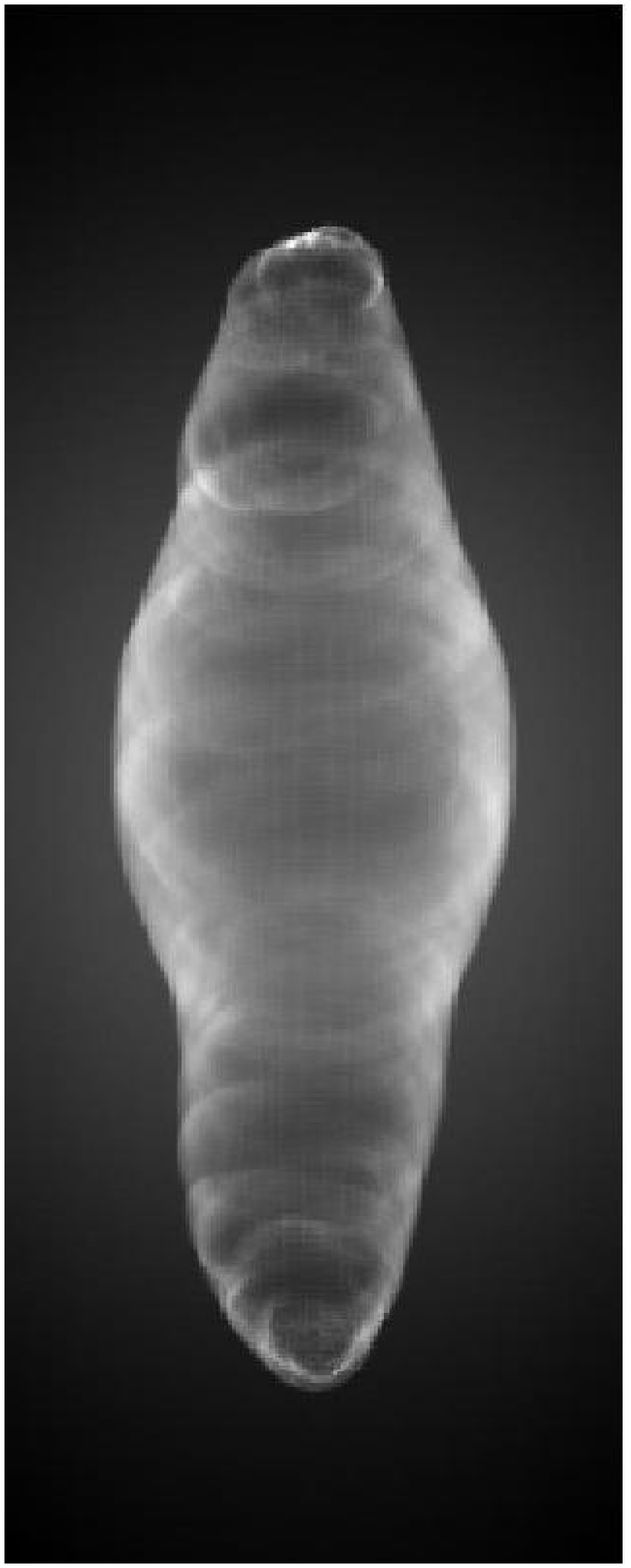}}
\rotatebox{-90}{\includegraphics[height=.27\textwidth]{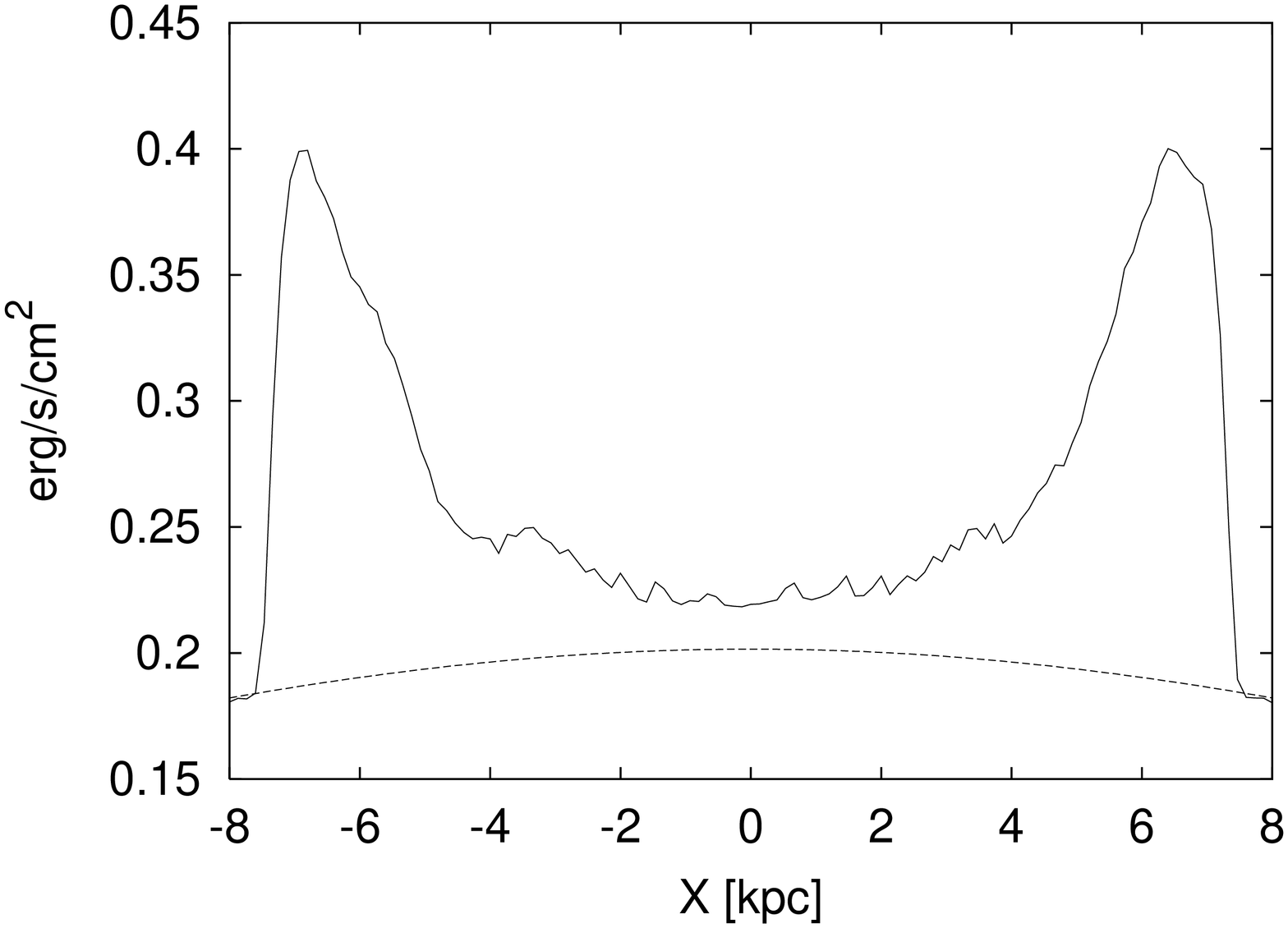}}
\rotatebox{-90}{\includegraphics[height=.27\textwidth]{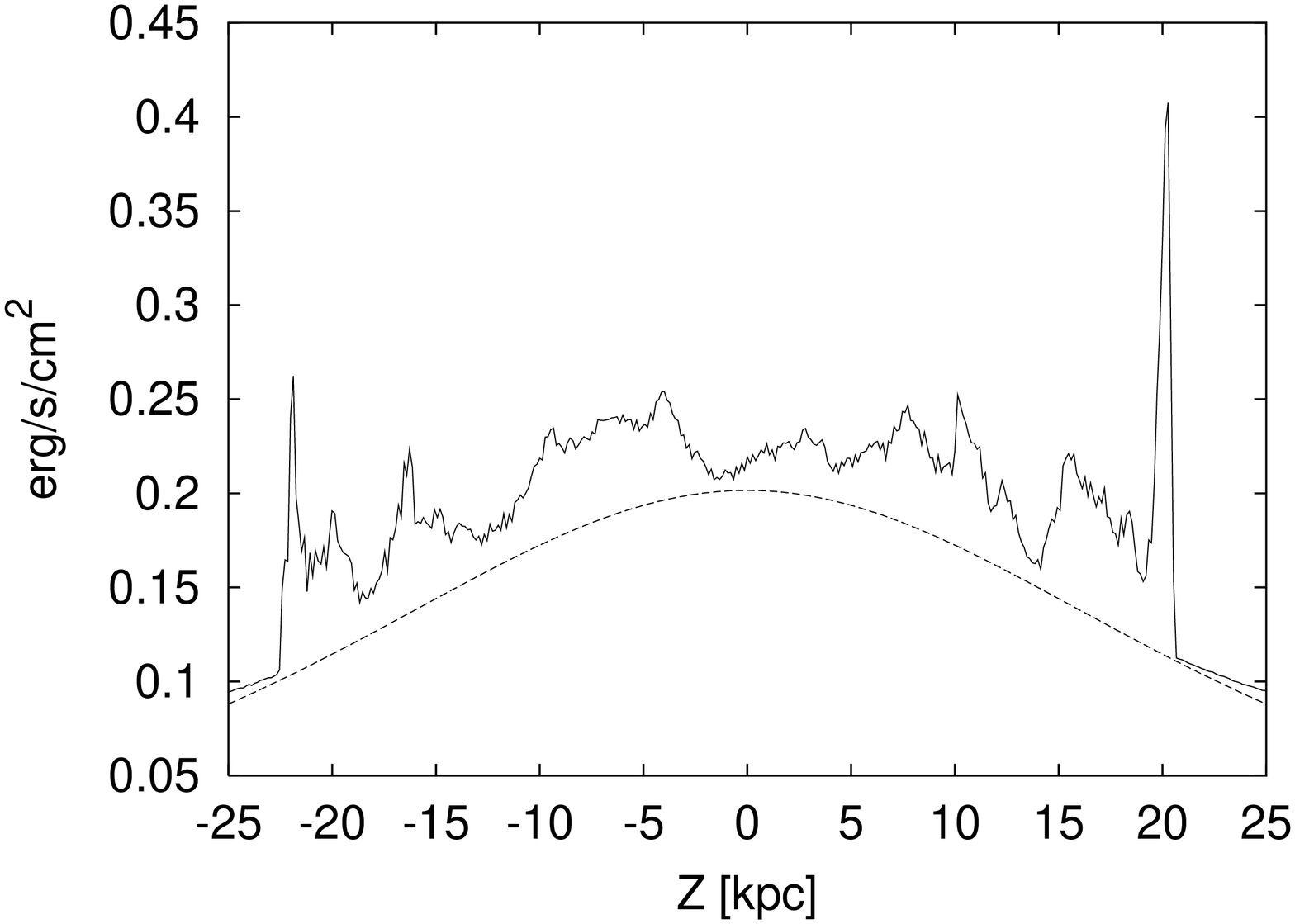}}\\
\rotatebox{-90}{\includegraphics[height=.44\textwidth]{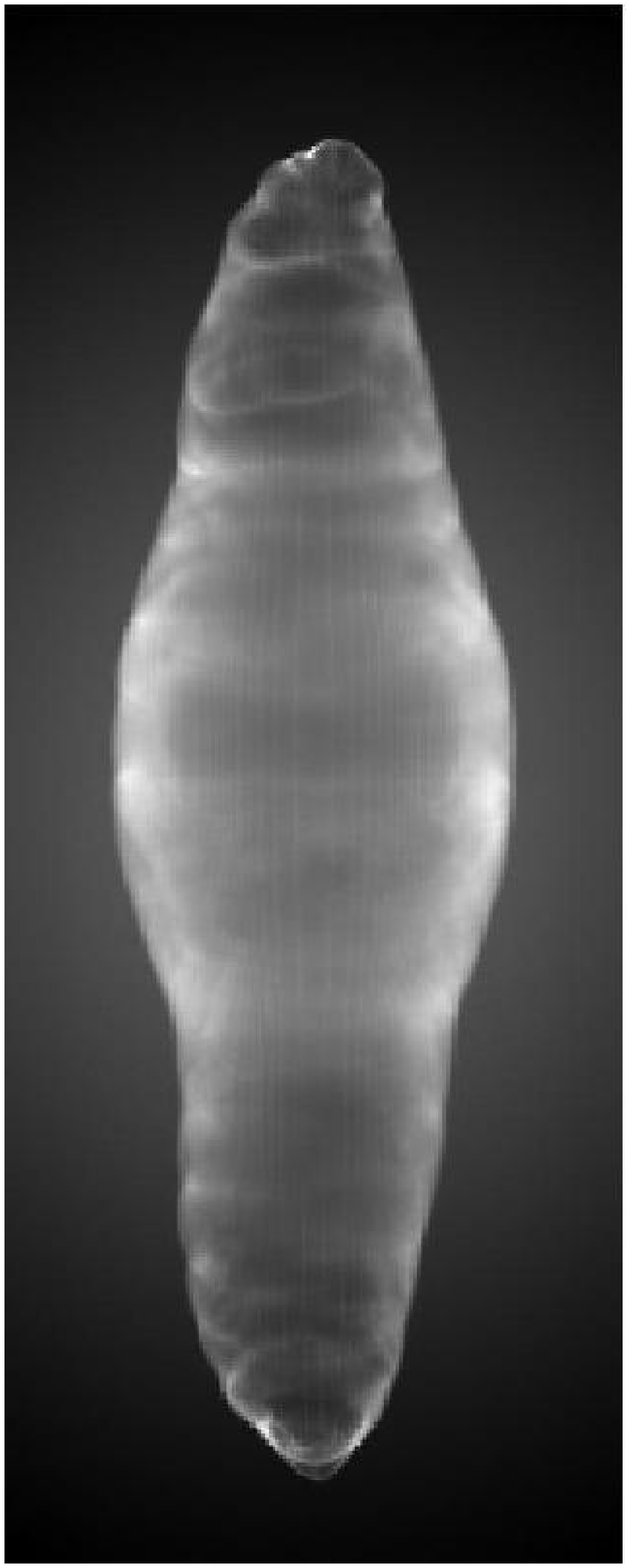}}
\rotatebox{-90}{\includegraphics[height=.27\textwidth]{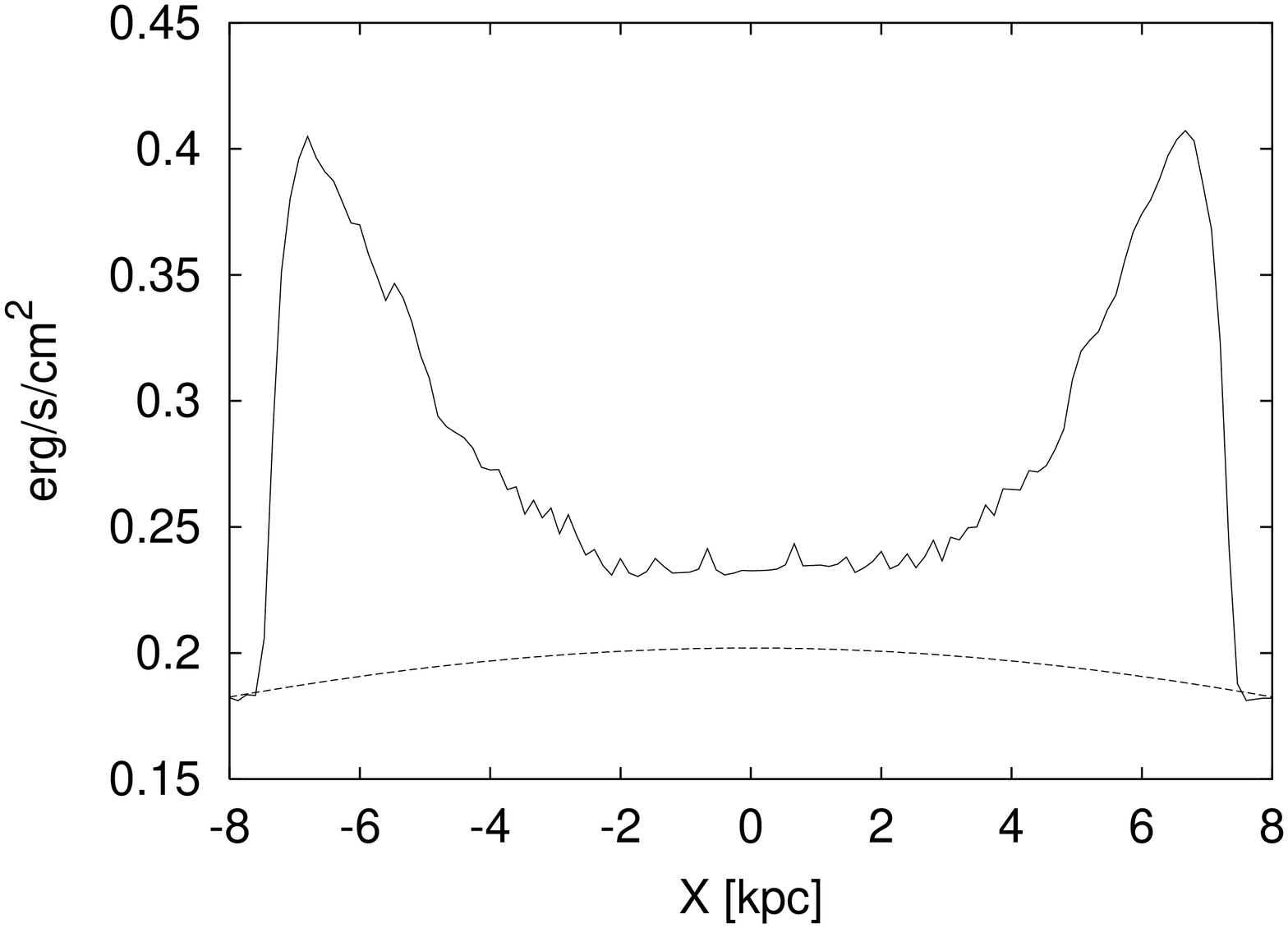}}
\rotatebox{-90}{\includegraphics[height=.27\textwidth]{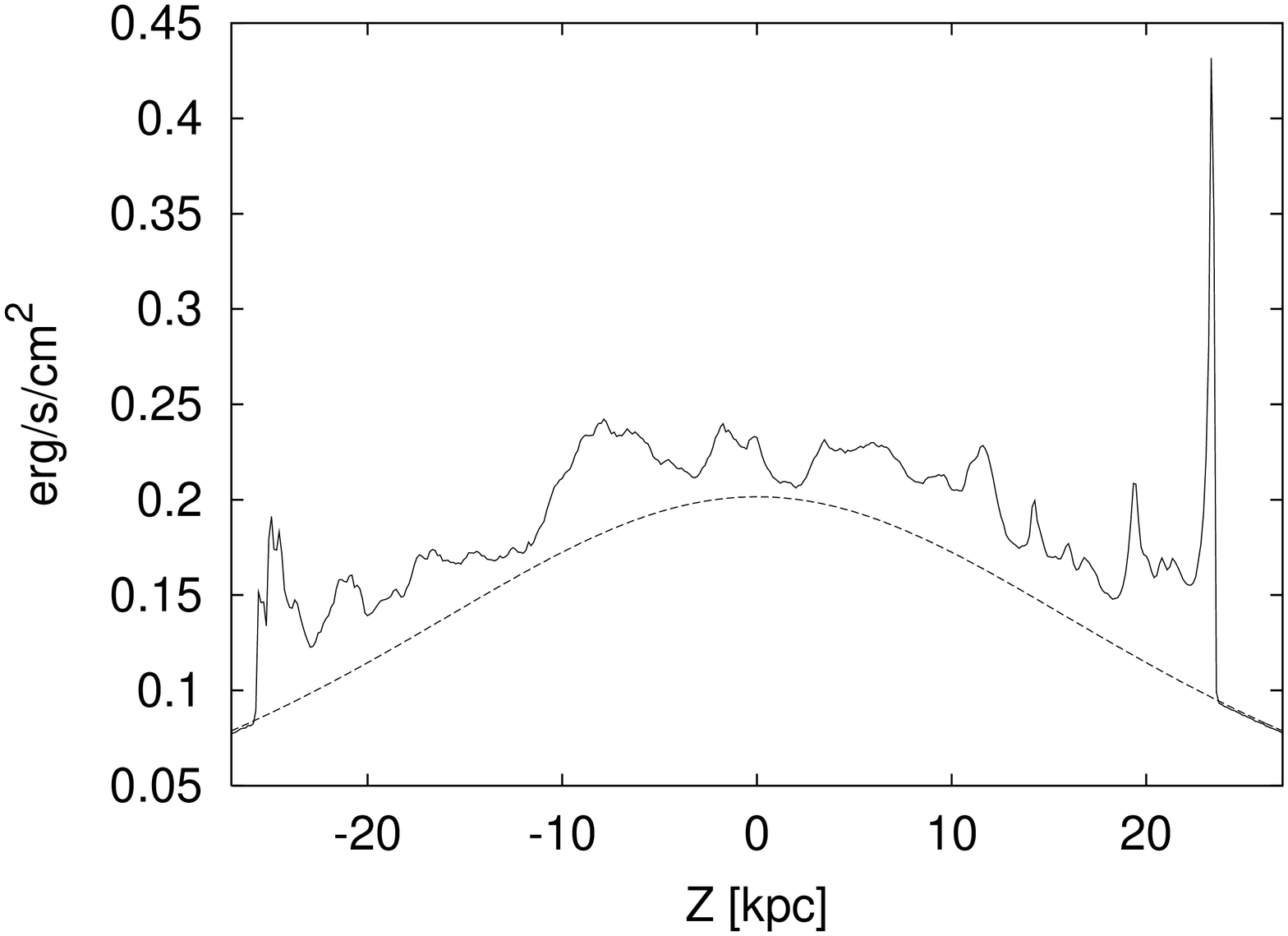}}\\
 \caption{\small The same as Fig.~\ref{emimapa} for $t=2.04$~Myr. Some artefacts of the 
        rotation procedure are visible at large~Z~values.
        \label{emimapb}}
\end{center}
\end{figure*}

\subsubsection{Emission maps}\label{emisec}
The emission due to optically thin bremsstrahlung \citep[e.g.][]{Shu92i} was computed,
mapped onto a Cartesian grid, and
integrated for different viewing angles (Figs.~\ref{emimapa}~and~\ref{emimapb}).

The general X-ray emission properties of shocked ambient gas in jet simulations 
have been discussed by \citet{CHC97}, which has been updated recently by \citet{Zanea03}.
The idea is that the gas is pushed aside by the jet cocoon. Depending on its compression,
it will or will not form X-ray deficits at the location of the cocoon, and bright shells
at the edges. Averaging the gas distribution and neglecting flows in the shocked ambient gas,
the critical parameter is the relative shell thickness
$\xi$, defined as the width of the shocked ambient gas region divided by the local
bow shock radius. Then, the ratio of observed flux to the flux of the initial condition is 
given by:
\begin{equation}
f^\prime/f= {\cal F} \xi^{-1} (2-\xi)^{-2}\enspace,
\end{equation}
where ${\cal F}$ depends on the pre-shock and post-shock temperatures, 
and is close to unity. Therefore, X-ray deficits could be observed for a shell thickness
$\xi$ above 38\%, for sources located at $90^\circ$ inclination.
Here, the shell thickness is comparatively low for the whole simulation. 
Therefore, at high inclinations the X-ray surface brightness never falls below that of the 
undisturbed King atmosphere. This picture implies that the deficit is pronounced for
low inclinations, since most of the gas is shifted aside. 
Indeed, we find these deficit regions
in Fig.~\ref{emimapb} for angles of $10^\circ$ and lower. The deficit is much
more evident at the later time. This is due to the different morphology:
At $t=0.32$~Myr (Fig.~\ref{emimapa}) the bow shock shape is essentially elliptical with
only a small cigar-like extension. In that phase, the gas is compressed more isotropically
(compare Fig.~\ref{jet1}).
The thickness of the shocked ambient gas shell is 
roughly the same in all directions. At the later time, the prominent cigar has 
displaced the gas sideways, producing the deficit when viewed pole on.
Another reason is the decreasing density. The region in front of the jet head 
is furthermost from the center and the gas is therefore thinnest.

The two phases of the bow shock, cigar and elliptical (comp. sect.~\ref{bowshap}), 
show up prominently in the emission maps. They form circular and elliptical rings,
respectively, depending on the viewing angle.
Where the rings partially overlap, they are brighter, producing the impression
of ring segments (e.g. Fig.~\ref{emimapa}, $10^\circ$ and $30^\circ$., Fig.~\ref{emimapb},
$10^\circ$). The structures can also intersect on the line of sight, 
producing bright spots (Fig.~\ref{emimapb},~$30^0$). The pole-on figures show at least 
two rings: one from the cigar phase and one from the inner elliptical bow shock part.

Initially, the edges are brightened for any viewing angle. This changes at late times
for high inclination. Here, the edges are prominently brightened only for the direction
perpendicular to the jet axis. The reason is the declining density profile that is 
superimposed
on the jet features.
Edges and rings are typically enhanced by a factor of a few in surface brightness.
They should therefore be detectable for some sources in this phase.


\section{2.5D simulation}
\label{2p5d}

\subsection{Simulation Setup}
In order to study the jet evolution on larger scale,
a 2.5D simulation was performed with similar initial conditions to the 3D simulation 
in the previous section. The simulation was run for 20~Myrs of simulation time;
during that time the jet reached an extent of 110~kpc which corresponds to
220 jet radii. In the radial direction, the jet reached 31~kpc.
The jet radius was set to 0.5~kpc and the resolution was set to 20 cells per jet radius.
This corresponds to $[4400 \times 1240]$ cells in total.
With that resolution
global parameters like the bow shock velocity on the Z-axis or energy and
momentum conservation are accurate to $\approx$~10\% \citep{mypap01a}.
$\rho_\mathrm{e,0}$, $a$ and $\beta$ were chosen slightly different to the
previous model, because the sideways expansion of the bow shock should be able 
to escape the constant density part during the simulation time. 
The values are: $\rho_\mathrm{e,0} = m_\mathrm{p}/\ccm$, $a=10\kpc$ and $\beta=0.75$.
The temperature was again set to $T=3\times10^7$~K.
The jet is injected with a density of 
$\rho_\mathrm{jet}= 10^{-4} \times \rho_\mathrm{e,0}$,
the sound speed in the jet was set to
$20\% c$, 
\ignore{$c$ being the the speed of light,} 
and the jet's Mach number to $M=3$.
The high internal sound speed and the low Mach number was chosen 
because a parameter study \citep{mypap03a} has shown that very light jets
adjust their pressure 
quickly to the cocoon pressure by oblique shocks (no expansion or contraction)
and that high Mach numbers cannot be sustained in a non-relativistic very light 
jet. 
No cooling was taken into account, and hence the simulation is scalable 
as outlined in \citet{Zanea03}.
\add{With these parameters, the simulation took about a month on an SX-5
supercomputer. It was therefore not possible to do this in 3D, for now.}

\subsection{Results}
\label{res}
\begin{figure*}
\centering
\begin{minipage}{\textwidth}
\begin{center}
\includegraphics[width=.48\textwidth]{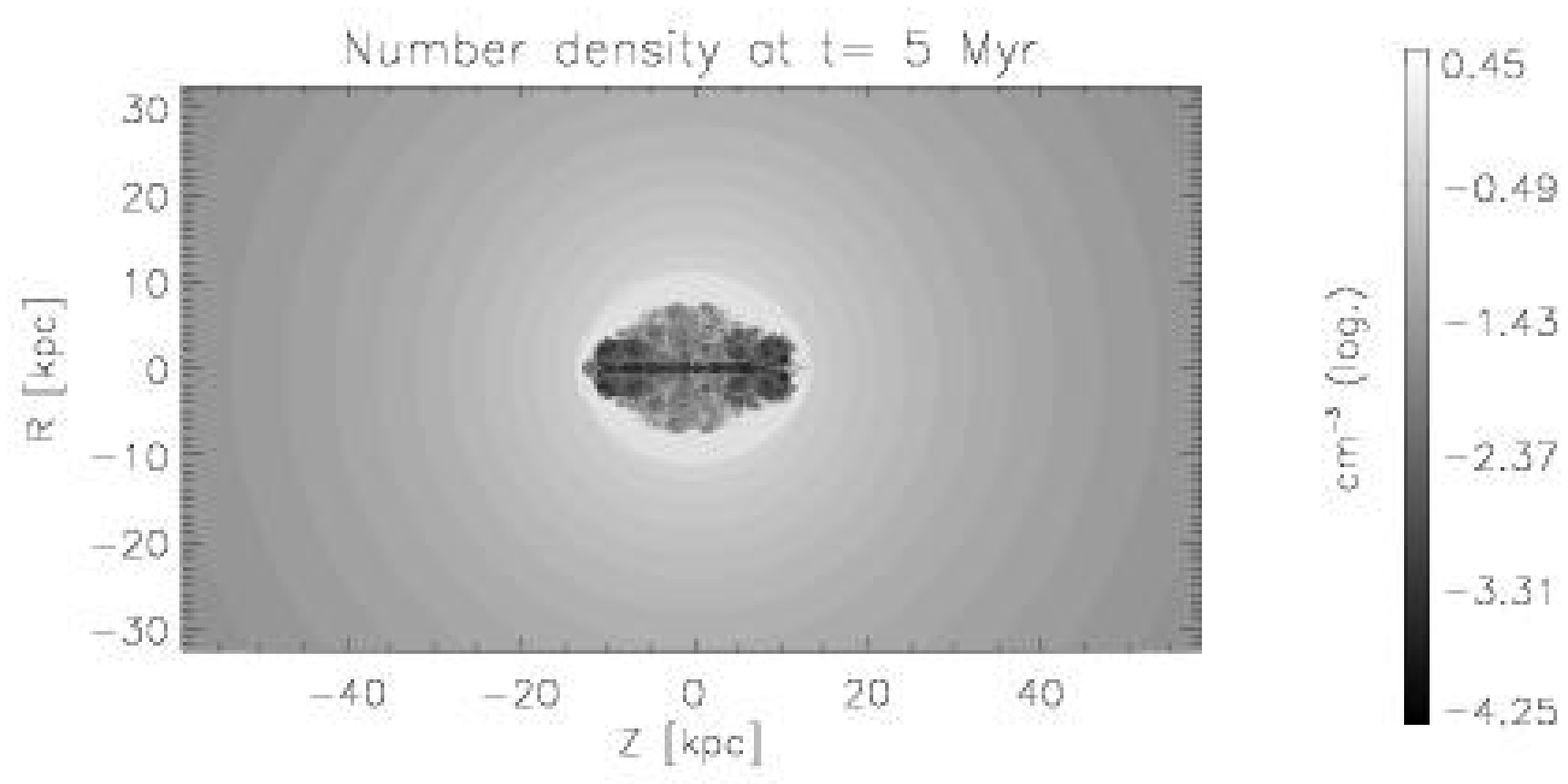}
\includegraphics[width=.48\textwidth]{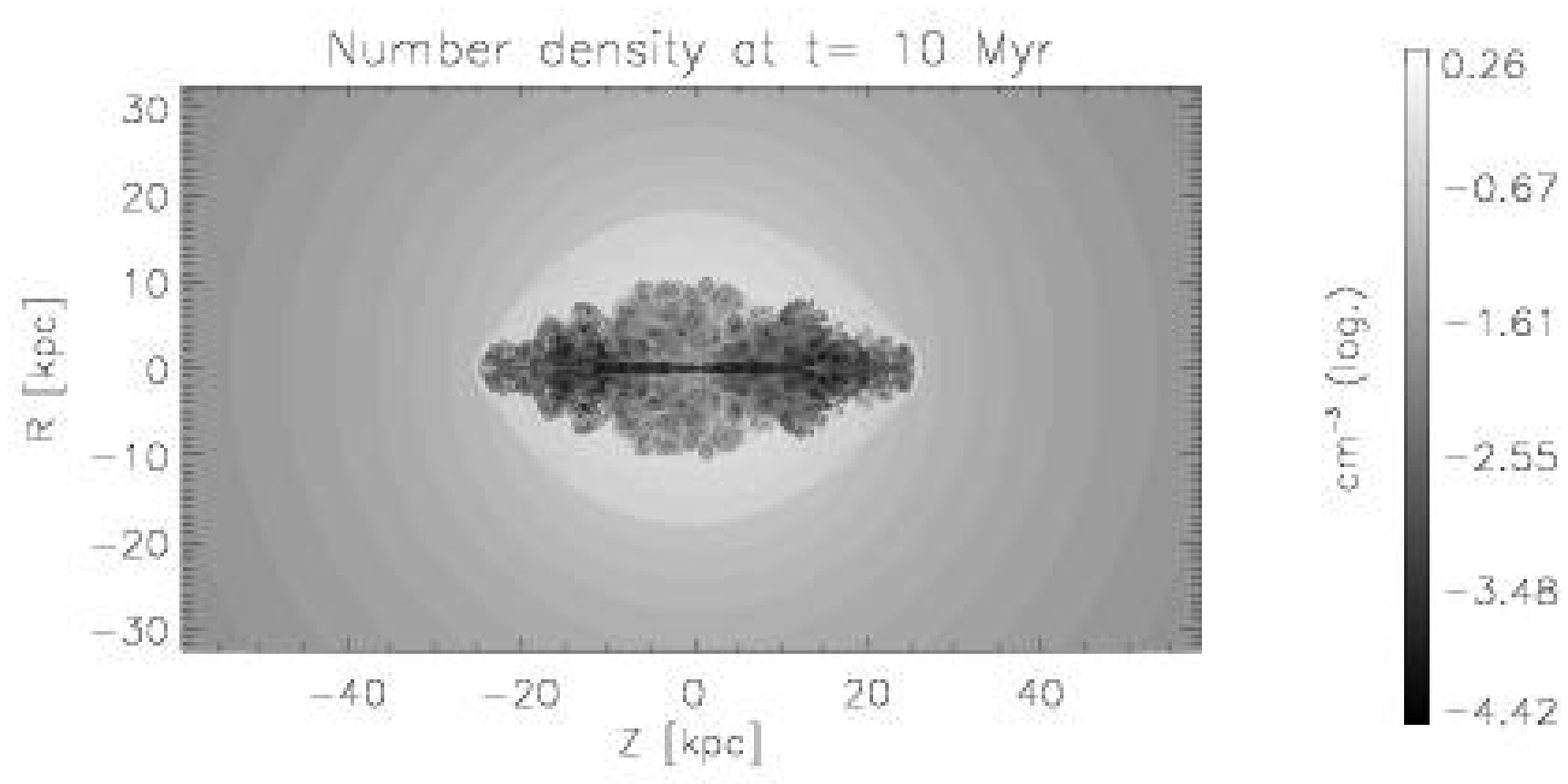}\\
\includegraphics[width=.48\textwidth]{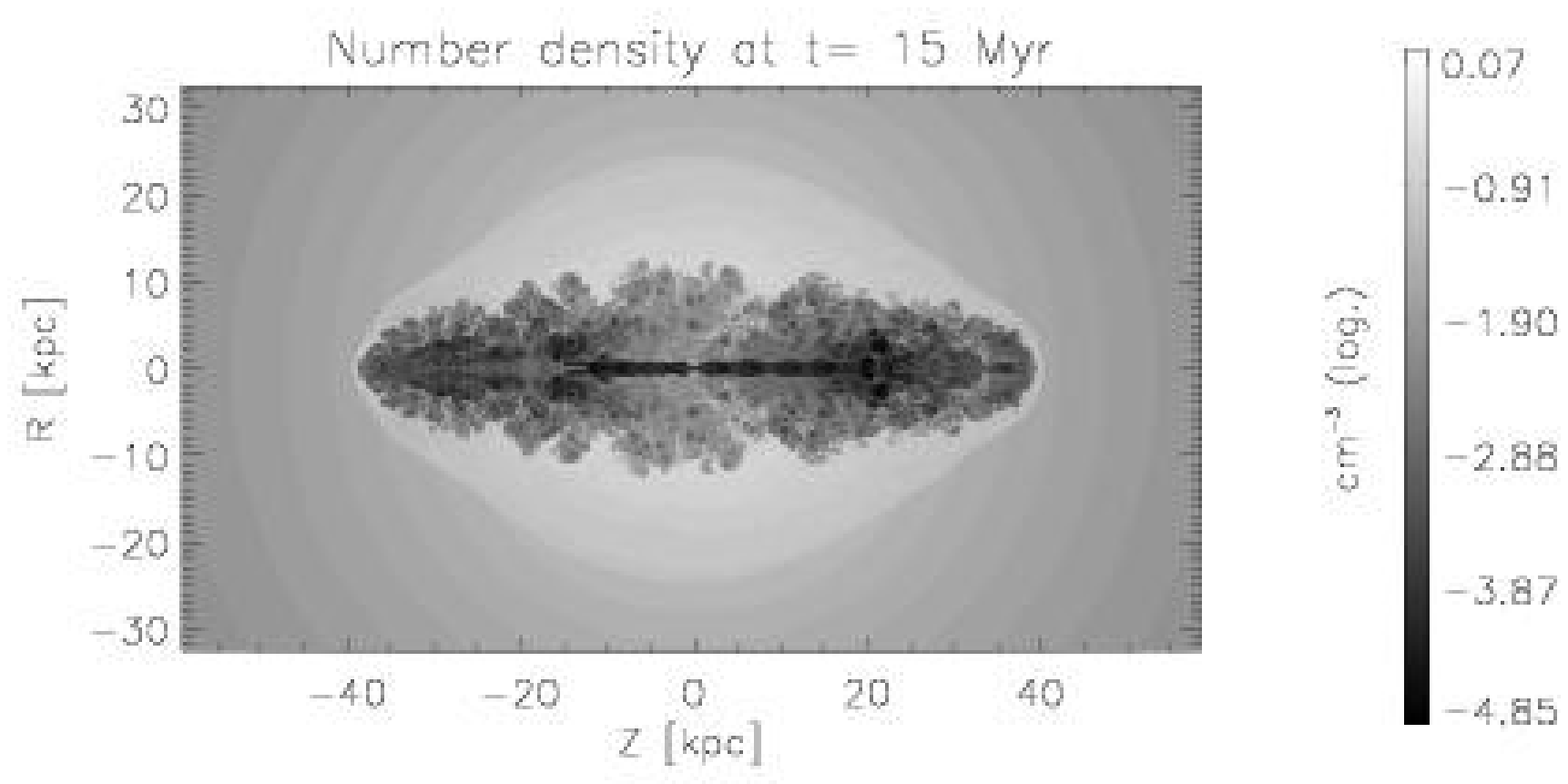}
\includegraphics[width=.48\textwidth]{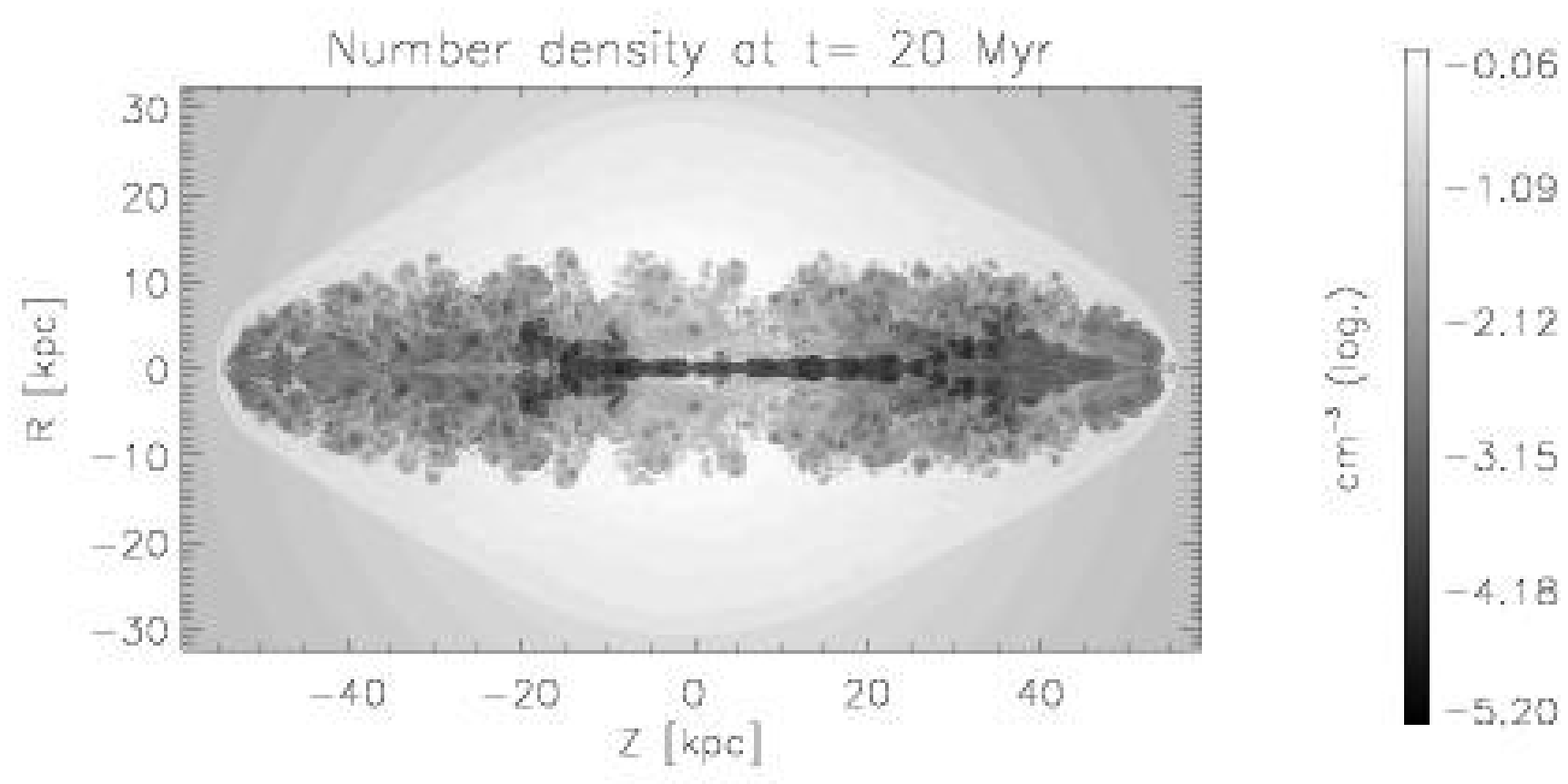}\\
\end{center}
\end{minipage}
%
%
\caption{\small Four snapshots of the 2.5D simulation. 
The logarithm of the number density is shown. The times for the snapshots are
indicated on top of the individual figures.
The same part of the grid is shown in each case.
The jet forms first a spherical bow shock, associated with a spherical cocoon.
The cocoon then transforms via a conical state to a cylindrical one.
The bow shock's aspect ratio (length/width) also grows, up to 1.8.}
\label{runAsnaps}       
\end{figure*}

We present logarithmic density plots of the simulation results for four
different simulation times ($5,10,15,20$~Myr) in Fig.~\ref{runAsnaps}.
The morphology that appears in these figures is a continuation of previous
simulations that could not reach the size shown here.
The 5~Myr figure shows the state that was reached by \citet{mypap03a},
and extensively discussed therein. In this early phase, the bow shock is spherical,
its radius following an expansion law given by the force balance equation 
(\ref{globeqmot}).
At later times the cocoon transforms via a conical phase towards 
a cylindrical one. This is a remarkable result because classical double
radio galaxies also often have cylindrical cocoons.

\begin{figure}
\centering
\rotatebox{0}{\includegraphics[width=.5\textwidth]{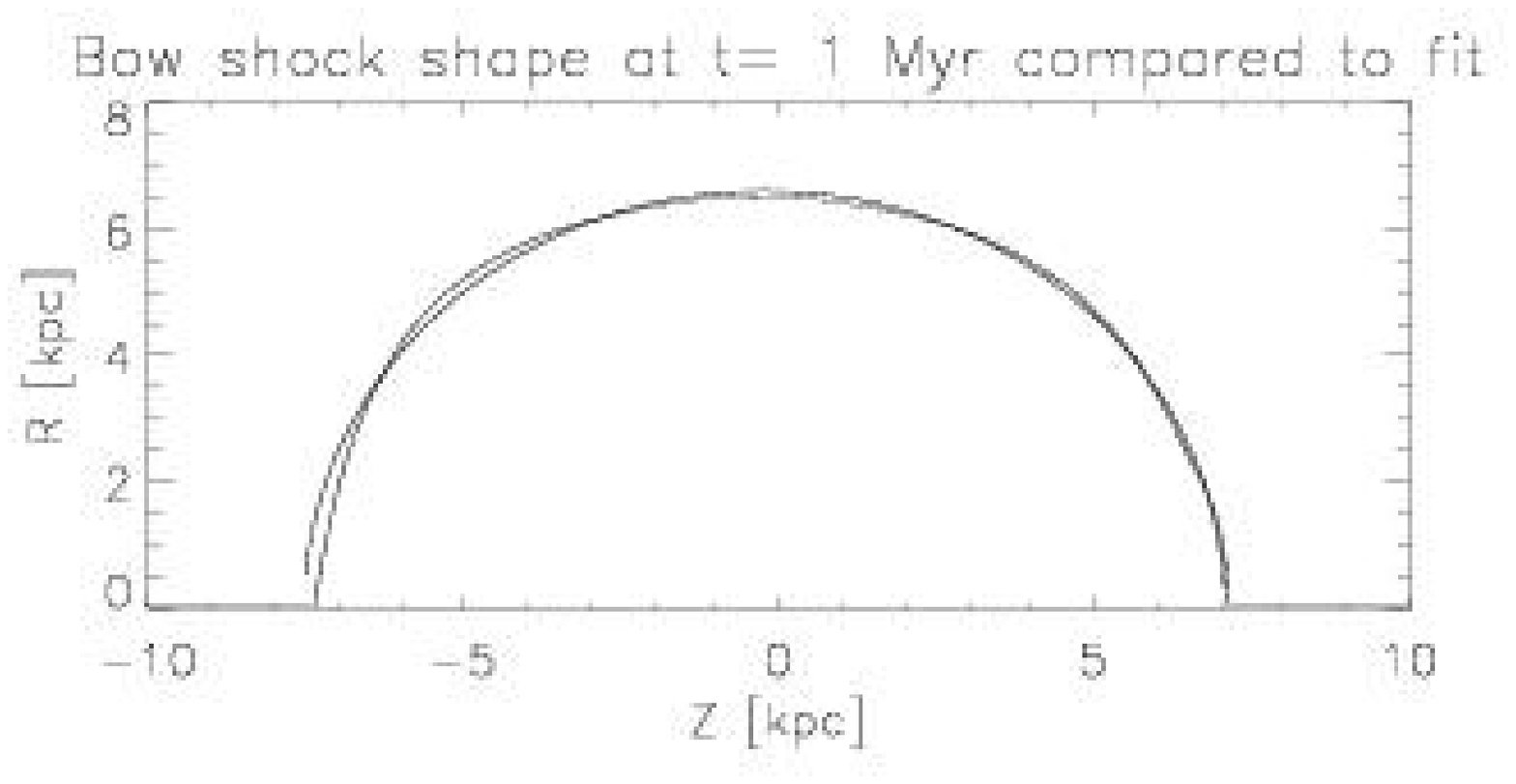}}\\
\rotatebox{0}{\includegraphics[width=.48\textwidth]{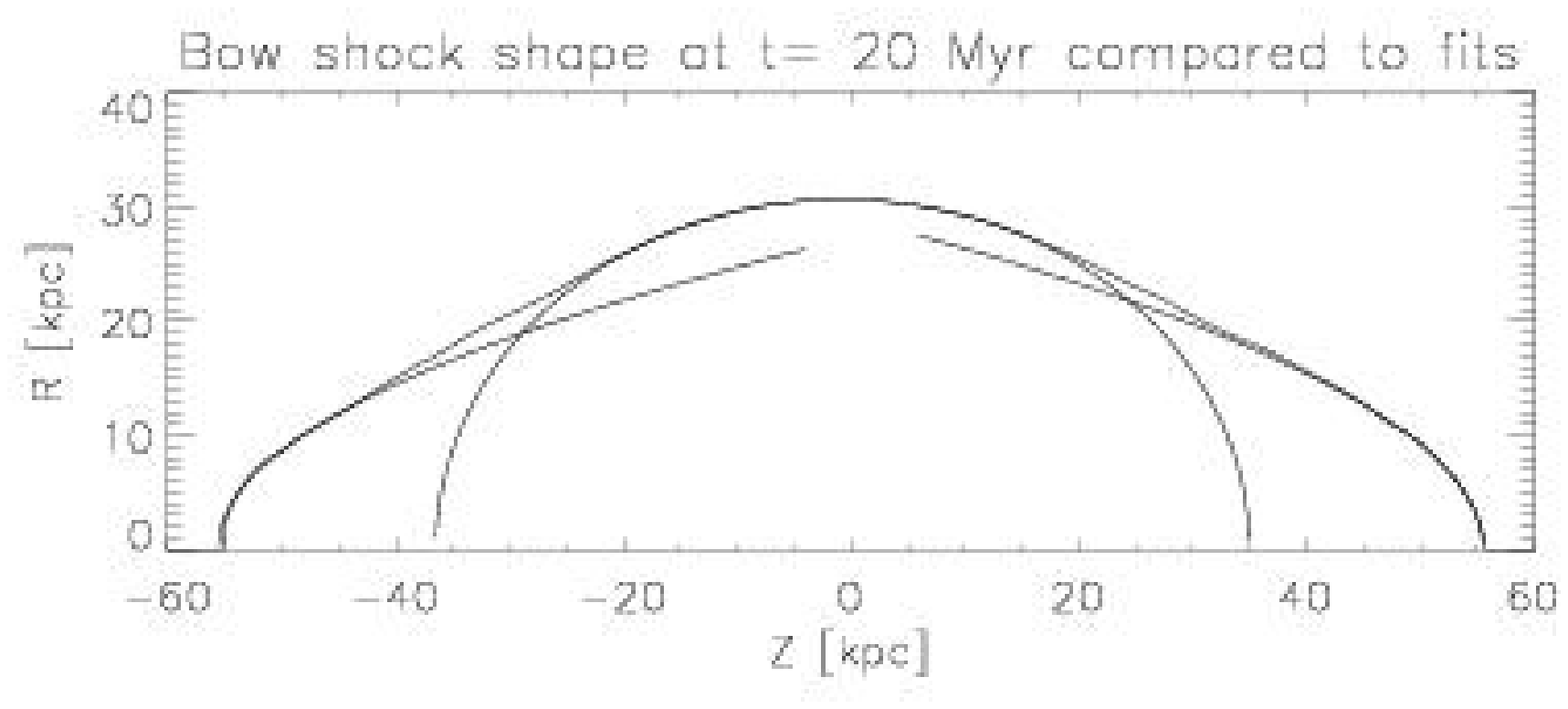}}\\
\caption{\small Shape of the bow shock at 1~Myr (top) and at 20~Myr (bottom)
for the 2.5D simulation. The
fit functions are: $R=6.61\times\sqrt{1-(Z+0.23)^2/53.28}$ (1~Myr), and 
$R=3.66\times\sqrt{Z+55.6}$, $R=3.91\times\sqrt{55.5-Z}$, and 
$R=30.7\times \sqrt{1-(Z+0.74)^2./1277}$ (20~Myr). This means that the bow shock evolves
from almost elliptical to elliptical + parabola extensions.
\label{bowshape2p5} }      
\end{figure}

\subsubsection{The shape of the bow shock}
The bow shock surface was extracted and fitted by elliptical functions 
as in the previous section. The result is shown in Fig.\ref{bowshape2p5}.
As in the 3D case, the bow shock can be represented by an ellipse in the center
and by a parabola near the tip. This parabola now has rank two while it was a rank three
parabola in the 3D case. The difference
is due to the different jet nature. The beam in the 3D simulation stays intact
all the way to the tip of the bow shock. Therefore it can deliver its momentum
to a smaller area. In the 2.5D simulation the beam becomes turbulent, entraining 
cocoon and shocked ambient gas, delivering the momentum to a larger area.
Also here, the center of the ellipse is shifted to the right, by 1.3~jet radii.
This \add{corresponds} to the armlength asymmetry, i.e. the right jet is longer for $t>3$~Myr,
by a few percent. This confirms the findings of the 3D simulation.

\subsubsection{Beam stability}
\begin{figure*}[bt]
\centering
\includegraphics[width=.74\textwidth]{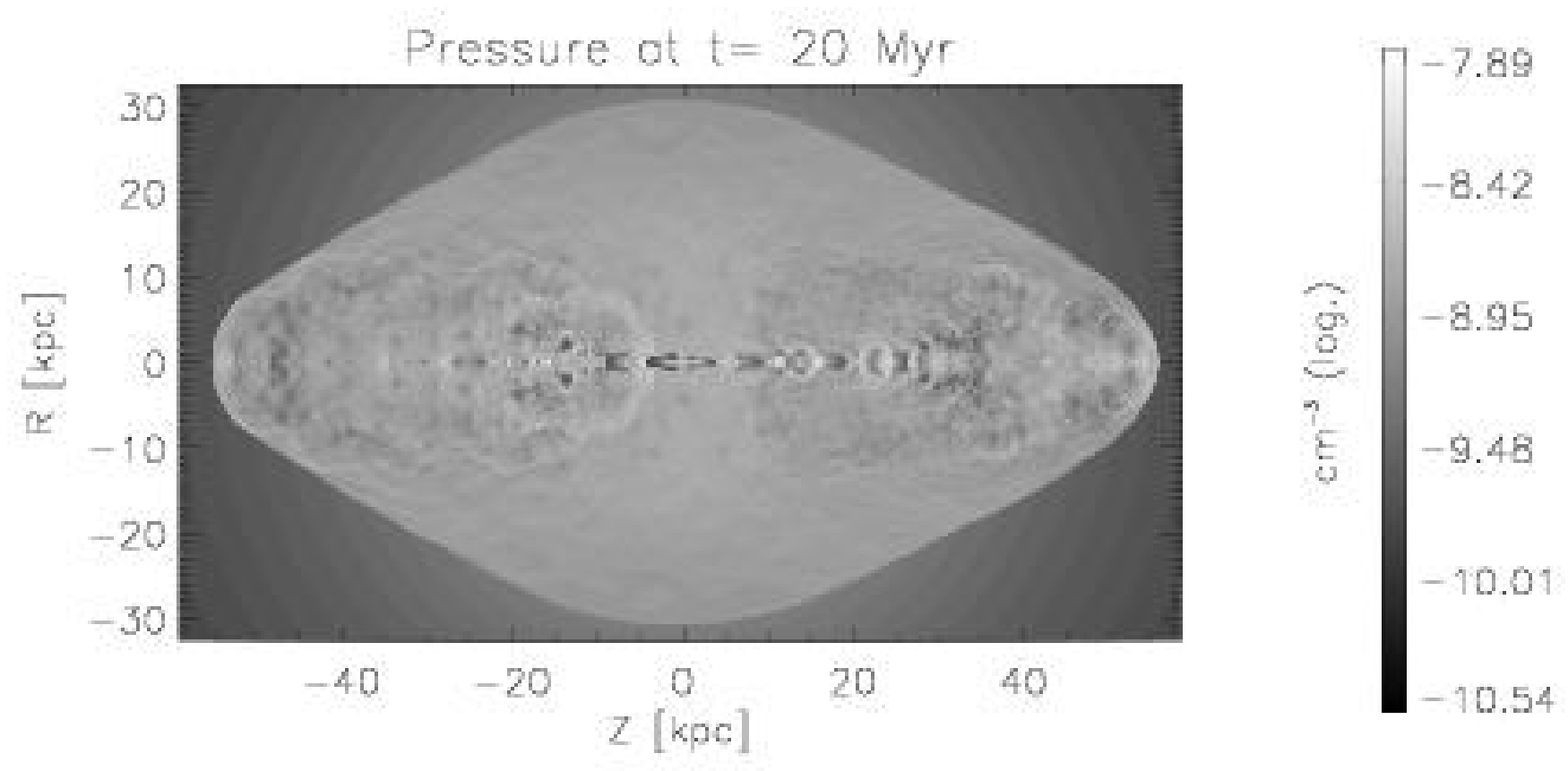}
\begin{minipage}[t]{.25\textwidth}
{
\vspace{-6.2cm}
\includegraphics[width=\textwidth]{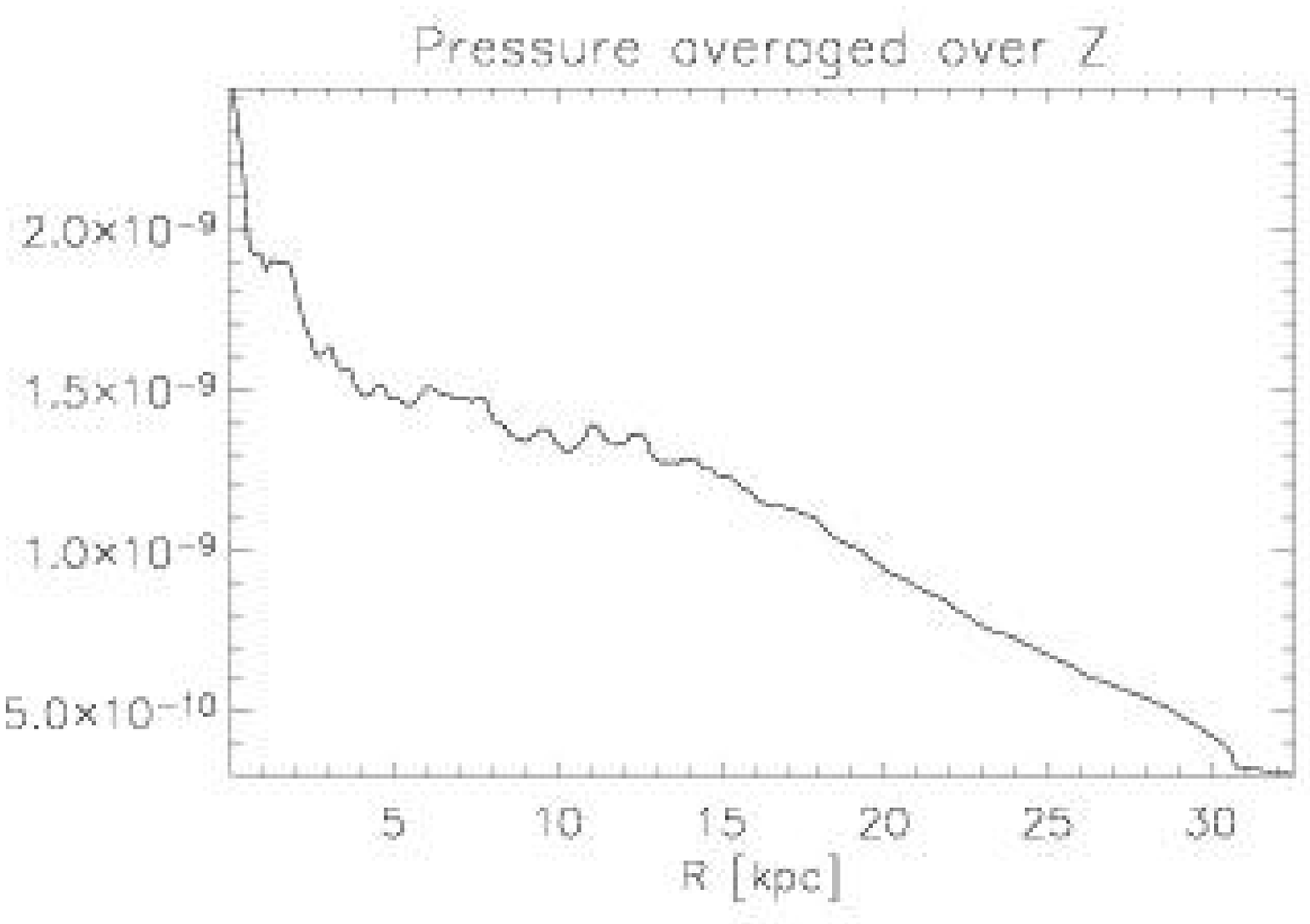} \\
\rotatebox{0}{\includegraphics[width=\textwidth]{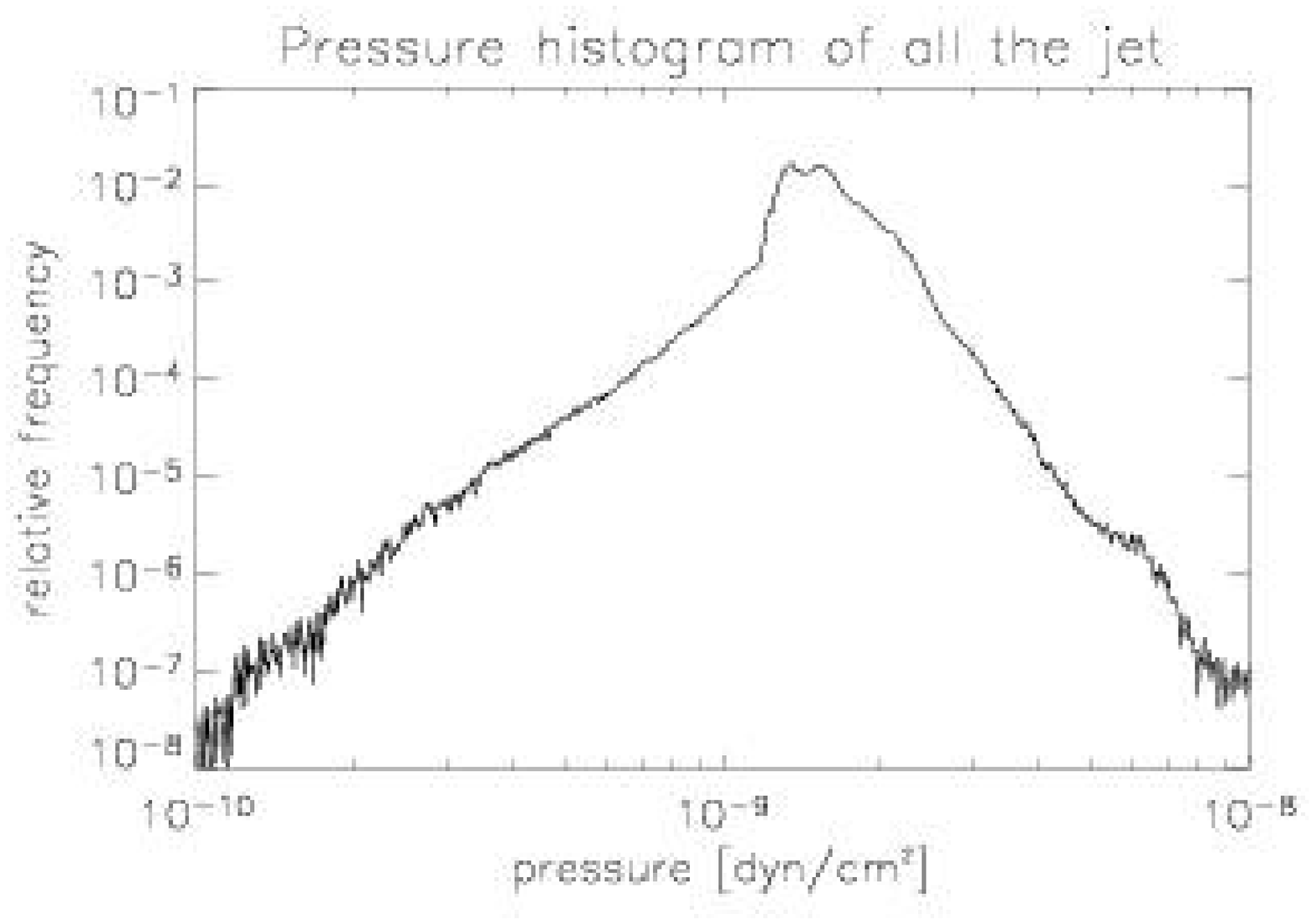}}}
\end{minipage}
\caption{\small 
Left: Pressure distribution for the 2.5D simulation after 20~Myr.
Top right: Axial pressure average over the radius. 
Bottom right: Pressure histogram within the jet (i.e. beam, cocoon, and 
shocked ambient gas). 
It can be represented  by a broken power law with humps
around the most frequent value. The power law indices are $5$ and $-8$, respectively.
90\% of the volume has a pressure in the range $[1-2]\times 10^{-9}$~dyn/cm$^2$.
The noise represents the statistical error.
\label{lgp1000} }      
\end{figure*}

The beam in the 3D simulation was stable, whereas the beam in the 2.5D simulation
becomes turbulent.
The critical factor for this behaviour is not so much the dimensionality
(that should work in the opposite direction, although the stability is probably 
somewhat enhanced by the choice of the cylindrical coordinates)
but the density contrast. Lighter jets automatically moderate their Mach number
\citep{mypap03a}.
This can be understood from the pressure equilibrium within the whole jet.
Because in a very light jet, the sound speed is high and the expansion speed is low,
the pressure is almost equal anywhere in the jet (compare Fig.~\ref{lgp1000}).
The jet beam adjusts to that pressure by varying the strength of its oblique shocks.
At the actual location of the shocks, the pressure is higher. This sums up to a 
decline according to \add{a} power law with exponent $-8$ in the histogram and can be seen
in the pressure distribution (Fig.~\ref{lgp1000}). 
The sound speed in the beam is therefore above $10\%$ the speed of light,
which makes it hard for the jet to be highly supersonic in a non-relativistic simulation
(the typical Mach number in the beam is between one and two).
The problem is even more severe at earlier times because the average pressure 
decreases with time (see below). It is well-known that jet beams at low Mach number
are disrupted quickly \citep[e.g. ][]{Bodea94,Bodea95,Bodea98}.

\begin{figure}[tb]
\centering
\rotatebox{-90}{\includegraphics[height=.5\textwidth]{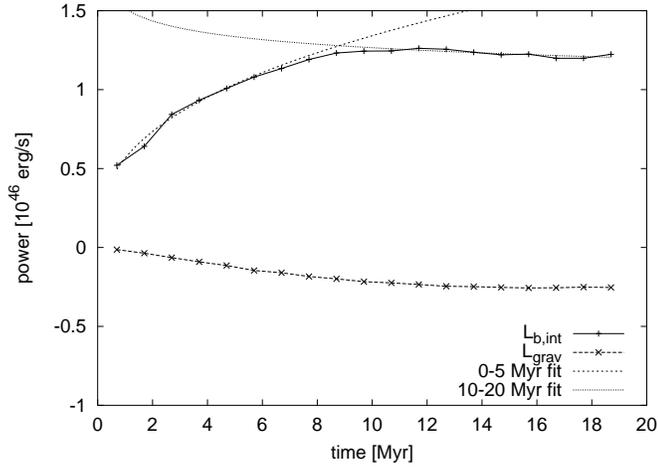}}
\caption{\small Approximate internal energy flux 
through the bow shock ($L_\mathrm{b,int}$) and energy loss rate due to 
gravitational uplifting ($L_\mathrm{grav}$)
over simulation time. $L_\mathrm{int,e}= 4\pi r^2 p_\mathrm{e} v / (\gamma-1)$
was calculated using the bow shock's radius and sideways velocity from the simulation.
The fitted functions are: $0.57\,t^{0.37}$ (0-5~Myr), and 
$1.51\,t^{-0.08}$ (10-20~Myr).
The gravitational energy loss rate levels off at 30\% of the beam energy flux.
\label{ll} }      
\end{figure}

\begin{figure}[tb]
\centering
\rotatebox{-90}{\includegraphics[height=.5\textwidth]{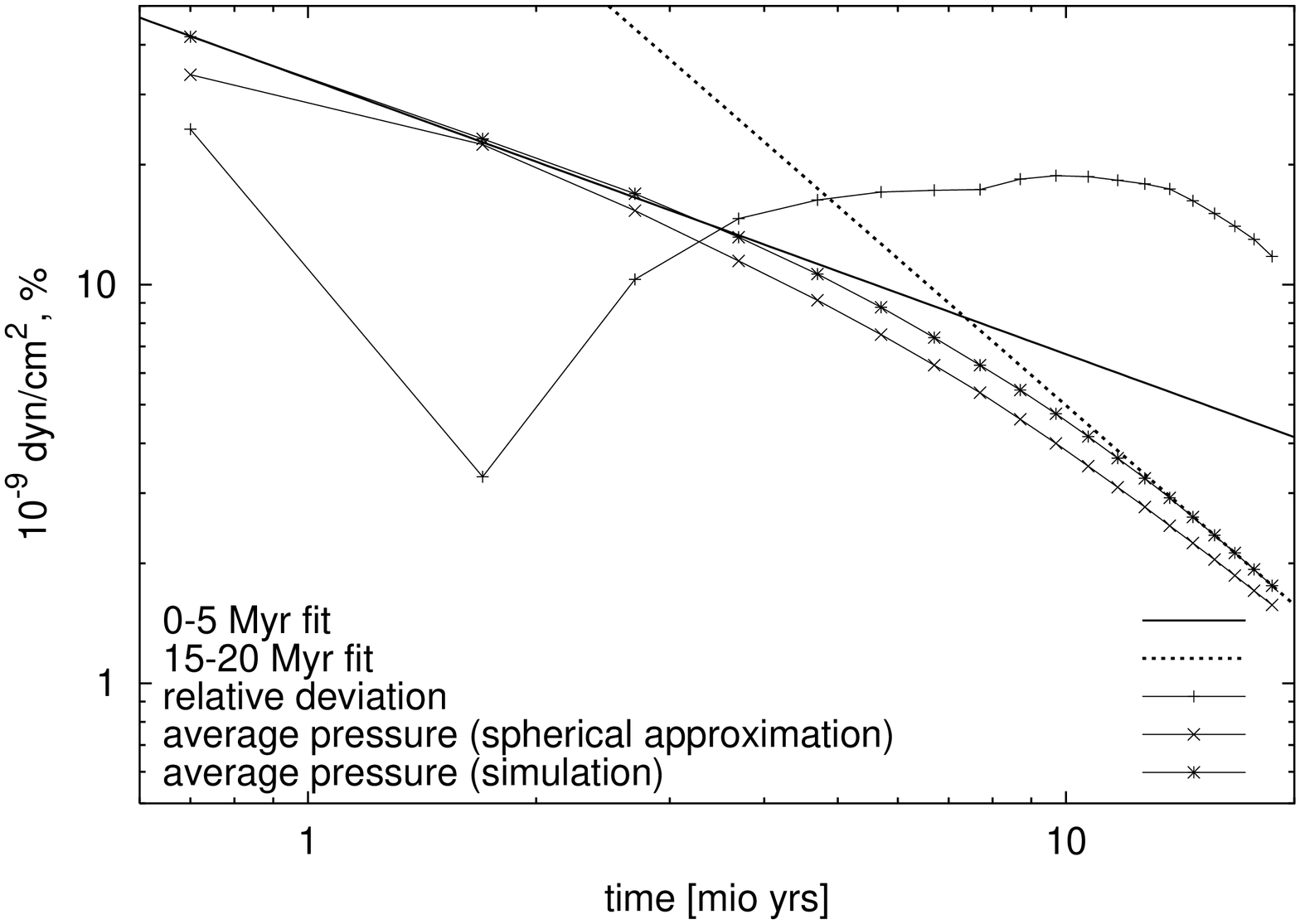}}
\caption{\small Average jet pressure over simulation time.
        The stars show the values measured in the simulation,
        crosses mark the pressure according to a spherical approximation,
        plusses show the relative difference between the former (in \%). 
        Corresponding symbols are connected with solid lines.
        The lines show 
        fits to the pressure measured from the simulation.
        The fits are:  $32.84\,t^{-0.69}$ (0-5 Myr, solid line), and
        $227.95\,t^{-1.66}$ (15-20 Myr, dashed line). The best fit for the spherical 
        approximation in the range 15-20~Myr is: $126.05\,t^{-1.50}$ (not shown).
\label{por} }      
\end{figure}

\subsubsection{Pressure evolution}

The average 
jet pressure is the driving force of the inner elliptically shaped part of the bow shock.
Figure~\ref{lgp1000} shows that the pressure in the jet 
system monotonically decreases with radius. Close to the axis, the pressure is higher 
because of shocks in the beam region. In the shocked ambient gas region, 
a new equilibrium of gravity and pressure appears. The 
smallest pressure values are located at the bow shock, roughly 20\% below the average.

In the previous section, it was shown that the inner part of the bow shock roughly follows
the spherical propagation law (\ref{globeqmot}) inside the core radius.  
The accuracy of this law will be
checked in the following also for the larger 2.5D simulation. In this case,
the bow shock has propagated more than three core radii in the sideways direction.
Notice that the \add{formulae} derived for blastwaves with strong shocks are still applicable here,
as shown in the appendix.
In the spherical approximation, 
the average pressure inside the whole jet is given by
\footnote{The formula neglects the increasing but small part of kinetic 
energy stored in the beam.}:
\begin{equation}\label{peq}
p_\mathrm{j}=(\gamma-1)\frac{Lt-{\cal M}v^2/2}{V_\mathrm{j}},
\end{equation} 
\add{Here, $\gamma=5/3$ is the adiabatic index, and}
$V_\mathrm{j}$ is the jet volume
(including beam, cocoon and shocked ambient gas). The power L includes all sources
of energy, i.e. the flux of kinetic and internal energy through the jet channel,
the flux of internal energy entering through the surface of the bow shock, and the
energy lost by work against the gravitational field.
The simulation takes these effects properly into account.
In order to test the pressure evolution against deviations from the 
simple spherical law these power sources are calculated as follows:

Via the jet beam, a constant power is injected, given by:
$L_\mathrm{j}=
L_\mathrm{j,kin}+L_\mathrm{j,int}=\pi \rj^2 \rho \vj^3(1+(\gamma(\gamma-1)M^2/2)^{-1})
=9\times10^{45}$~erg/s.

Through the jet's bow shock, the power 
$L_\mathrm{b,int}=4\pi r_\mathrm{b}^2 v_\mathrm{b} p_\mathrm{King}(r_\mathrm{b}) / (\gamma-1)$
enters the jet system in the spherical approximation. $r_\mathrm{b}$ and $v_\mathrm{b}$
are the bow shock's position and velocity, which is measured at the position of the
greatest cylindrical radius. $p_\mathrm{King}$ denotes the initial pressure distribution 
of the isothermal King profile.
In the spherical approximation, $L_\mathrm{b,int}$ should be proportional to
$t^{\delta(\kappa+3)-1}$, where $\kappa$ would be the exponent of an isothermal
power law distribution in pressure and density ($p,\rho \propto r^\kappa$), 
and $\delta$ the exponent of a 
local power law approximation to the sideways bow shock propagation ($r\propto t^\delta$,
$\delta=3/(\kappa+5)$). But for a
King profile, it takes some time until $\delta$ adjusts to the value it should
have due to the local power law approximation.
The latency is caused by the memory of the flow of its history due to the fixed amount 
of energy and mass at a given time.
$L_\mathrm{b,int}$ is shown in Fig~\ref{ll} together with fits at early and late times.
It should first rise proportional to $t^{4/5}$.
The fit gives an exponent of $0.37$ which again reflects the effect of the 
initial conditions.
 The fitted 
exponent for $L_\mathrm{b,int}$ of $-0.08$ after 15~Myr is in agreement with 
a local power law exponent for the pressure of $\kappa=-1.84$, where $\delta=0.8$ 
has been adopted from the measured bow shock propagation for that time span.
The local $\kappa$ in the fitted region is in the range -1.7 to -2.0. 

The rate of change of gravitational energy, $L_\mathrm{grav}$, is computed from the 
simulation data, averaging over one million years and is plotted in Fig~\ref{ll}.
$L_\mathrm{grav}$ rises from one to thirty percent of the jet beam's energy flux,
where it seems to converge towards the end of the simulation.
The gravitational energy loss rate and the power entering as internal energy through the 
bow shock make up a similar contribution to the energy within the jet system as 
the energy flux through the beam. 

Using $L=L_\mathrm{j}+L_\mathrm{b,int}-L_\mathrm{grav}$,
where $L_\mathrm{b,int}$ and $L_\mathrm{grav}$ now denote the time-averaged value 
at a given time, (\ref{peq}) can be evaluated, where $v$ and $t$ 
are given by (\ref{globeqmot}):
\begin{eqnarray}
v&=&\frac{Lt^2}{{\cal M} r}\\
t&=&\left( \frac{3}{L} \int_0^r {\cal M}(r^\prime) \mathrm{d}r^\prime \right)^{1/3}\enspace,
\label{tofr}
\end{eqnarray}  
and $r$ denotes the sideways extent of the bow shock.
Figure~\ref{por} shows this analytical estimate together with the data from the 
simulation. Here, $r$ was related to time via measurement from the simulation.
The agreement is quite good, in general. The analytical formula 
follows the slope of the simulation data, but underestimates it by up to $\approx 20$\%.

\ignore{After relaxation of the initial condition, the agreement is better than 3\%. 
Then the simulation data first falls more slowly and then steeper as 
the spherical approximation. Interestingly, the two curves come closer to each other again
for late times.}

Towards the end of the simulation, the pressure evolution can be approximated by a 
power law with exponent $-1.66$. For the same time range, the spherical approximation
is well approximated by a power law with exponent $-1.50$. 
Note that these are only local approximations to curved functions that have not yet 
reached the asymptotic power law regime.

\ignore{One should actually expect
the faster pressure drop in the simulation due to the excess volume created by the 
increase of the aspect ratio. Using elliptical deformation as first order approximation,
the volume increase should be proportional to the aspect ratio of the bow shock and 
therefore lower the exponent by 0.3 to 0.4. An even greater deceleration is expected 
due to the gravitational force of the dark matter halo, which is not taken into account 
in the
spherical approximation. The dynamics of the system compensates these effects,
and in the end, the evolution is close to the spherical approximation.}

\subsubsection{Sideways motion of the inner bow shock part}

\begin{figure}[tb]
\centering
\rotatebox{-90}{\includegraphics[height=.5\textwidth]{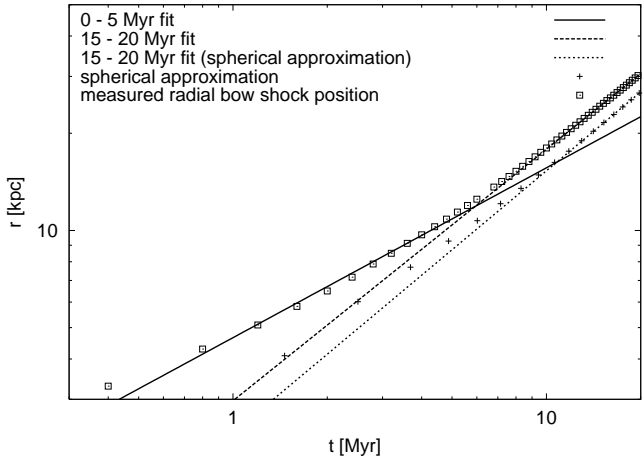}}\\
%
%
\caption{\small Bow shock radius at Z=0 versus time (squares) with fits and compared to 
spherical approximation, including all power sources (see text, plus-signs). 
The three fits are:
$4.66\, t^{0.53}$ (0-5~Myr),
$2.99\, t^{0.78}$ (15-20~Myr), $2.36\, t^{0.81}$ (15-20~Myr, spherical approximation).
}
\label{bowprop}       
\end{figure}

From the pressure evolution (previous section) one should expect that the sideways 
motion of the inner bow shock part roughly follows the spherical approximation for 
the whole simulation time.
The density profile used here has the asymptotic power law approximations:
$\lim_{r \mapsto 0}(\rho)=\rho_0$, and $\lim_{r \mapsto \infty}(\rho) \propto r^{-9/4}$.
Therefore, the bow shock should expand with $r\propto t^{0.6}$ at the beginning,
steepening towards $r\propto t^{1.09}$, at least as long as it remains spherical.
The radial bow shock position was determined every 0.4~Myr (Fig.~\ref{bowprop}).
The resulting function has a curved nature.
Following the arguments of sect.~\ref{bowprop3d}, the bow shock propagation was 
locally fitted 
by a function of type $a+b\,t^c$. The resulting parameters for the different regions are shown in Table~\ref{bowfits2dtab}. Usually, $a=0$, since only
the very late evolution of the jet is studied. 
Only for the time span up to 5~Myr a fit with
$a\not=0$ has been included because here it is possible that effects from the 
initial condition still dominate the propagation.
\begin{table}[bht]
\caption{\label{bowfits2dtab}Fit parameters for the bow shock position in the 2.5D simulation.
The star denotes a fit with fixed $a=0$. Two stars denote fits to the spherical
approximation with fixed $a=0$.}\nopagebreak
\begin{center} \noindent
\begin{tabular}{lccc}  \hline \hline
\multicolumn{4}{c}{\vspace{-3mm}}\\
time range [Myr]  & a       & b & c     \\ \hline
  {$[0:5]$}        & 2.00 & 2.67 & 0.76   \\
  {$[0:5]^*$}        & 0 & 4.66 & 0.35   \\
  {$[0:5]^{**}$}        & 0 & 3.22 & 0.67   \\
  {$[15:20]^*$}        & 0 & 2.99 & 0.78   \\ 
  {$[15:20]^{**}$}        & 0 & 2.36 & 0.81   \\ \hline \hline
\end{tabular}
\end{center}
\end{table}
For comparison, also fits to the detailed spherical approximation are given, computed 
by application of (\ref{tofr}). According to that, the exponent for the first five
million years should be $0.67$. Using the pure power law, an exponent of 0.35 is 
achieved in the simulation data. Allowing for the radial offset gives a best fit exponent
of 0.76. Since the exponent of 0.35 is much below any expectation, it follows that 
the initial condition is still important in that phase, and the fit with offset
is more appropriate.
The concurrence of the curves increases with time and for the last five million years,
the exponent for the power law fit of the simulation data ($0.78$) differs from that of 
the spherical approximation by only $0.03$. From the increasing aspect ratio, an exponent
lower than the one of the spherical approximation should be expected. 
The simulation shows that the effect is small.

\subsubsection{Axial bow shock propagation} 

\begin{figure}[t]
\centering
\rotatebox{-90}{\includegraphics[height=.5\textwidth]{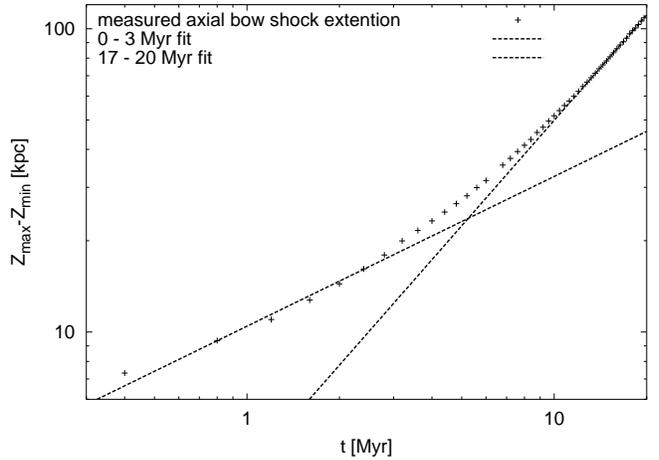}}\\
%
%
\caption{\small Bow shock extention on the axis versus time. The two fits are:
$10.45\,t^{0.49}$ (0-3~Myr),
and $3.49\,t^{1.15}$ (17-20~Myr).}
\label{bowpropax}       
\end{figure}
According to self-similar hydrodynamic jet models 
\citep[compare e.g.][ and references therein]{Alex02},
the velocity of the jet head should
be proportional to $v_\mathrm{bow,ax} =t^{(\kappa+2)/(\kappa+5)}$, 
where $\kappa$ is the exponent of a local 
power law for the external density ($\rho \propto r^\kappa$).
For a jet head with constant area, the velocity should be given by the 
one-dimensional estimate: 
$v_\mathrm{bow,ax} =\sqrt{\eta} v_\mathrm{beam} \propto t^{2/(\kappa+2)}$.
The last proportionality holds for $\kappa>-2$, only.

The bow shock propagation in axial direction is shown in Fig~\ref{bowpropax}.
Its evolution can be represented by two power laws that are much closer
to the expectation from the self-similar models (compare
Table~\ref{vexp}) than to the one-dimensional estimate.
\begin{table}[h]
\caption{\label{vexp} Comparison between the fitted exponent
$e$ for the axial bow shock velocity ($v_\mathrm{bow,ax} \propto t^e$) 
and the exponent expected in the self-similar and one-dimensional (1D) estimate. 
$\uparrow$ denotes denotes velocity growth faster than any power law.}\nopagebreak
\begin{center} \noindent
\begin{tabular}{lccc}  \hline \hline
\multicolumn{4}{c}{\vspace{-3mm}}\\
time range [Myr] & measured       & self-similar        & 1D    \\ \hline
  {[0:3]}        & $-0.51$ & $-0.4$ & 0   \\
  {[17:20]}        & $0.15$ & $0.09$ & $\uparrow$   \\ \hline \hline
\end{tabular}
\end{center}
\end{table}
For the time range up to three million years, the bow shock is still spherical.
The exponent should therfore be exactly $-0.4$. But the initial conditions 
set up the jet with a finite amount of energy from the beginning which would cause
an exponent of $-0.6$, analogous to the supernova case. The measured exponent of $-0.51$
shows that the jet is just in transition between these two phases.
Towards the end of the simulation the jet head velocity grows slightly faster than 
self-similar.
The good agreement with the self-similar models is probably due to the unstable beam,
at least in the later phases, where the momentum is distributed over a
large and increasing area. Jet models with intact beams (at higher jet density)
have been shown to have head advance speeds more consistent with the 
one-dimensional estimate than with the self-similar models \citep{CO02a}.

\subsubsection{Aspect ratio}
Defining the aspect ratio to be the length divided by the width of the bow shock, it grows 
proportionally to $t^{0.34}$ towards the end of the simulation. This is clearly a non-self-similar feature,
and may be partially due to the squeezing of the cocoon by the gravity of the swept up gas, as discussed recently by 
\citet{Alex02}.

\subsubsection{Evolution of the contact discontinuity.}
\begin{figure*}[t]
\centering
\rotatebox{-90}{\includegraphics[height=.48\textwidth]{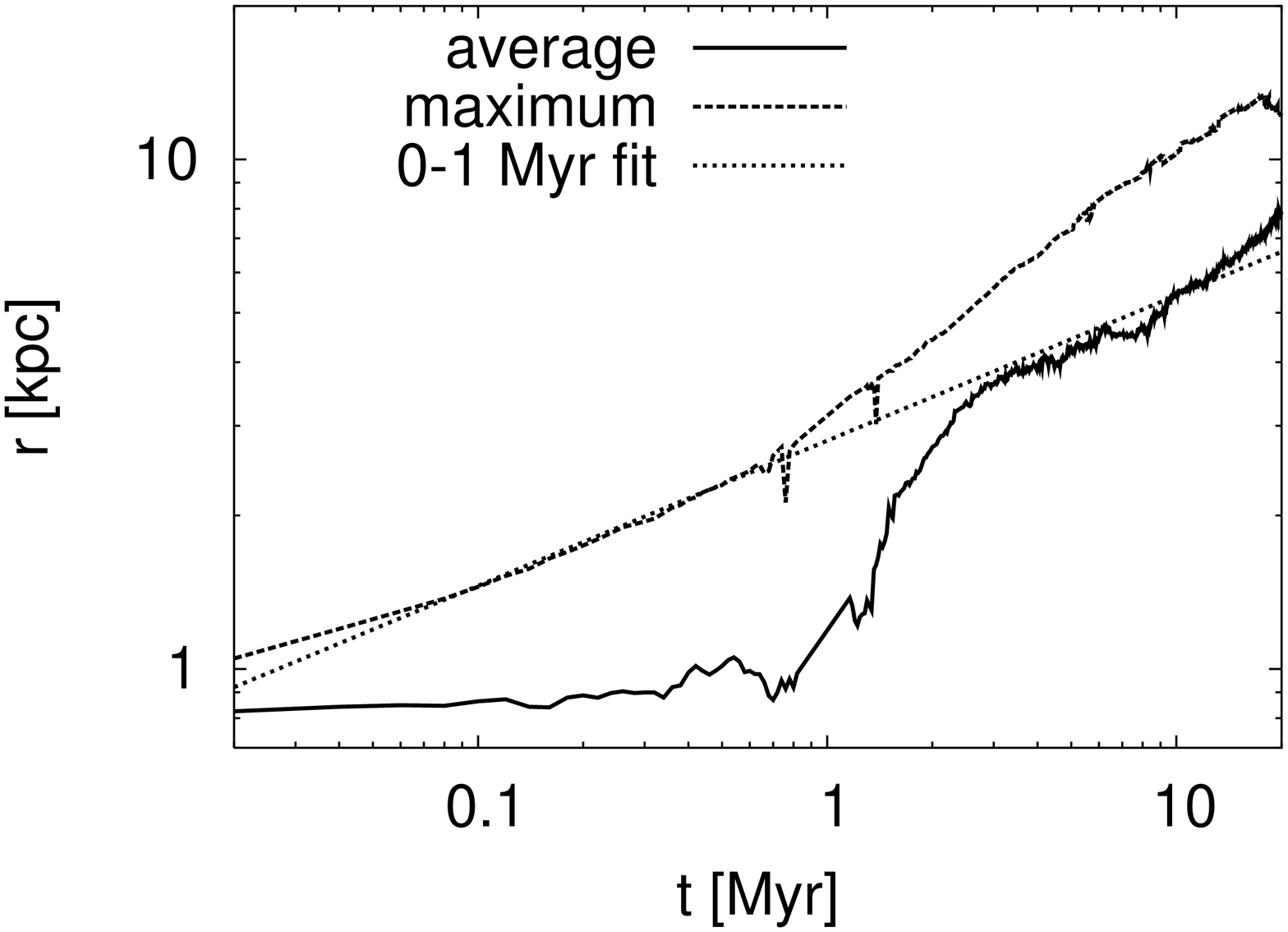}}
\rotatebox{-90}{\includegraphics[height=.48\textwidth]{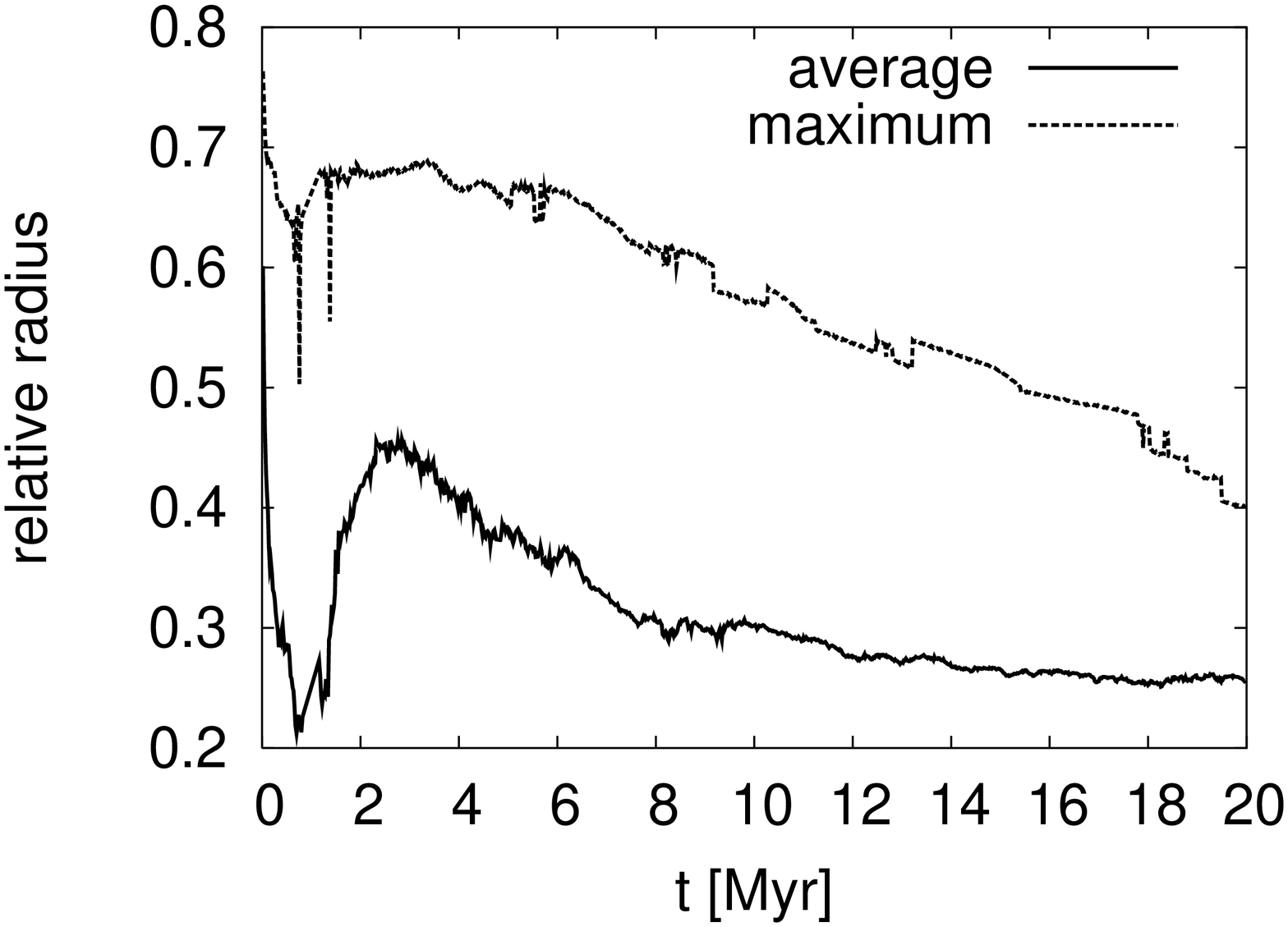}}\\
%
%
\caption{\small Position of the contact discontinuity over time. The contact discontinuity is determined by a passive tracer.
The maximum and the average value is shown. The left figure shows absolute values, the right one shows
the value relative to the maximum sideways bow shock radius. The fit in the left figure is a power law with exponent
0.28.
}
\label{condi}       
\end{figure*}

The temporal evolution of the contact discontinuity is shown in Fig.~\ref{condi}.
Both the maximum and the average radius are shown.
The position of the contact discontinuity was determined by a passive tracer variable that is advected with the flow.
Due to mixing, the cocoon plasma does not keep its initial value of zero. The contact surface is therfore
defined to be at the largest radius with the tracer having less than 10\% of the ambient medium value.
The average and maximum values of the contact discontinuity evolve in a dissimilar way,
which is due to the geometry changes discussed above.

\begin{figure*}[t]
\centering
\rotatebox{-90}{\includegraphics[height=.47\textwidth]{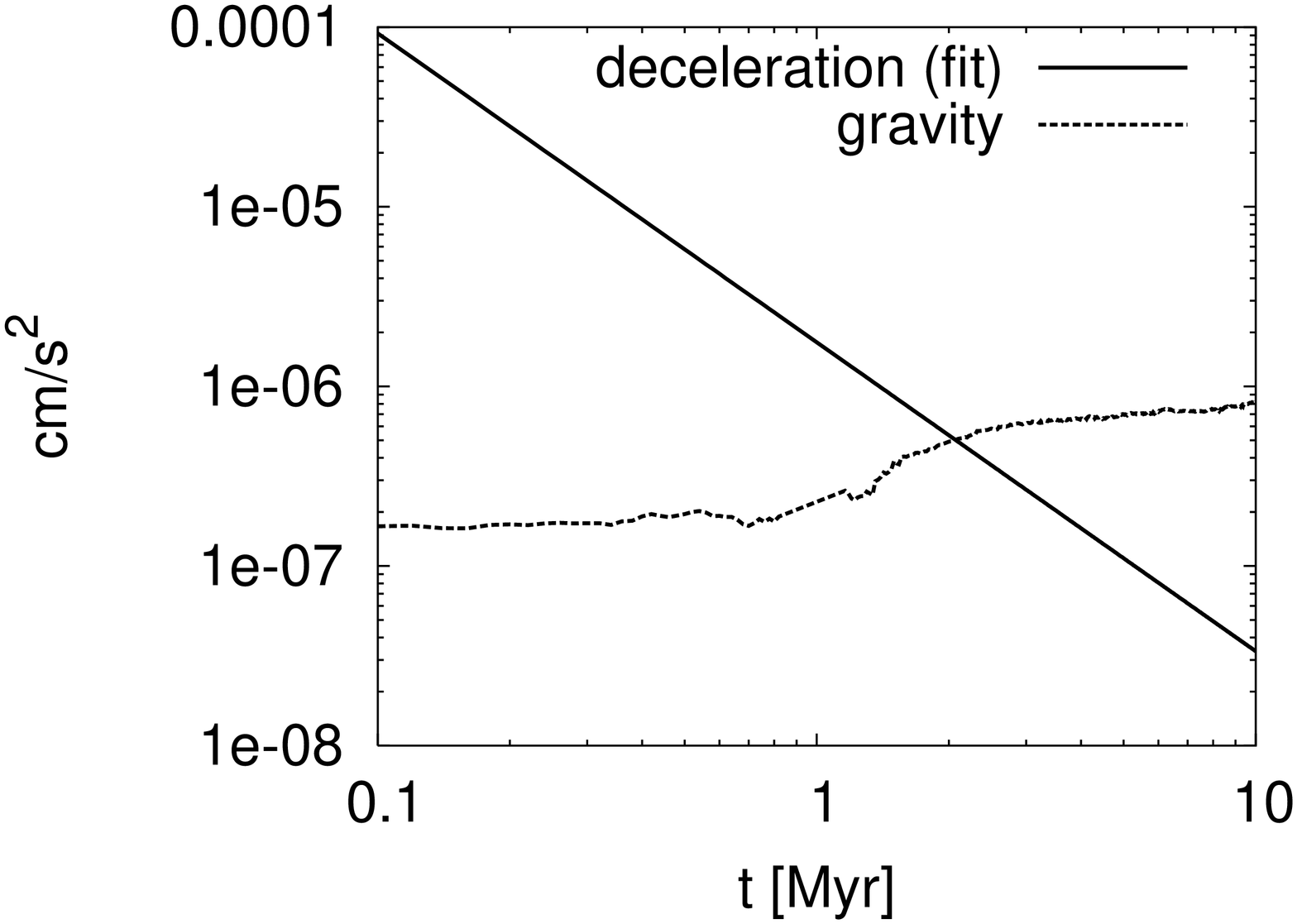}}
\rotatebox{-90}{\includegraphics[height=.47\textwidth]{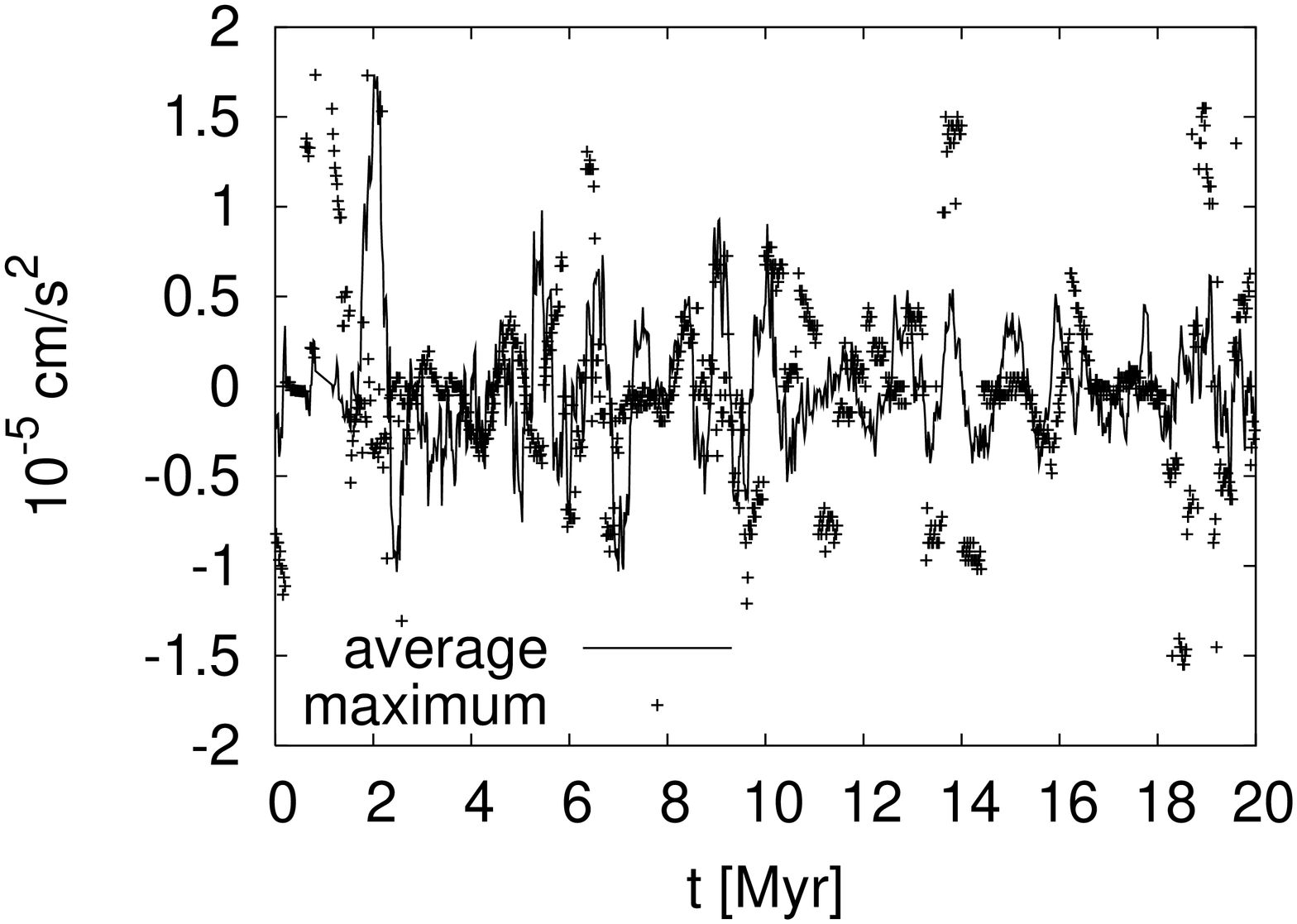}}
%
%
\caption{\small Left: Comparison of the deceleration (determined by a global fit for the time span up to one Myr)
of the contact surface against the local gravity (at the average position of the contact surface) of the dark matter halo. Right: Detailed acceleration averaged over 
0.2 Myr (because of the spatial resolution) for both, the maximum and the average position of the contact surface.}
\label{gcomp}       
\end{figure*}

Other than at the earliest times, the contact discontinuity is always decelerating.
For some time (up to $t=1$~Myr) this is sufficient to stabilise against 
the Rayleigh-Taylor instability. 
To be Rayleigh-Taylor stable, the deceleration of the contact
discontinuity must overcome the gravity of the assumed dark matter halo, which is given by 
\begin{equation}
g_\mathrm{D}=\frac{3\beta k T}{\mu m_\mathrm{H}  a} \frac{r/a}{1+r^2/a^2} 
=1.94 \, 10^{-6} \frac{r/a}{1+r^2/a^2} \, \mathrm{cm}/\mathrm{s}^2
\end{equation}
 The central region is represented by the maximum value, which shows the strongest deceleration at the beginning.
We determine a characteristic deceleration from a fit to the maximum value of the contact discontinuity up to 1~Myr.
Note that this is also an upper limit for the whole simulation.
The result is compared to gravity in Fig.~\ref{gcomp}.
This shows that the contact surface should be Rayleigh-Taylor unstable for $t>2$~Myr, which is consistent with the density plots.
The linear growth time 
($\approx \sqrt{l/g} = 2.5 \,\mathrm{Myr}\, \sqrt{(l/25\,\mathrm{pc}) (10^{-6}\,\mathrm{cm}\,\mathrm{s}^{-2}/g)} $, 
where $l$ is the wavelength of the instability and g the total acceleration) is typically below the simulation time
for wavelengths greater than the resolution limit. 

The details of the acceleration of the contact discontinuity are more complex than in the global discussion 
of the previous paragraph. Figure~\ref{gcomp} shows that it is constantly shaken with values of the order
$10^{-5}$~cm~s$^{-2}$.  For stability, the global deceleration has to exceed this value, which is the case for a fraction 
of the first Myr only.
Another complication is that the contact surface is not smooth but Kelvin-Helmholtz fingers always penetrate through.
The conclusion is that, at least in the central region, swept up gas is constantly entrained into the cocoon. 

\ignore{The entrained gas mass within the average contact discontinuity is shown at the end of the simulation in Fig.~\ref{me}.
It can be seen that most of the entrained gas is located in the central region. For the central 40~kpc, this adds up to a total
of $3\times 10^9 M_\odot$, i.e. 3\% of the mass sitting there before the jet activity, 
which corresponds to an average rate of $137 M_\odot$/yr.}

\begin{figure*}[t]
\centering
\rotatebox{-90}{\includegraphics[height=.47\textwidth]{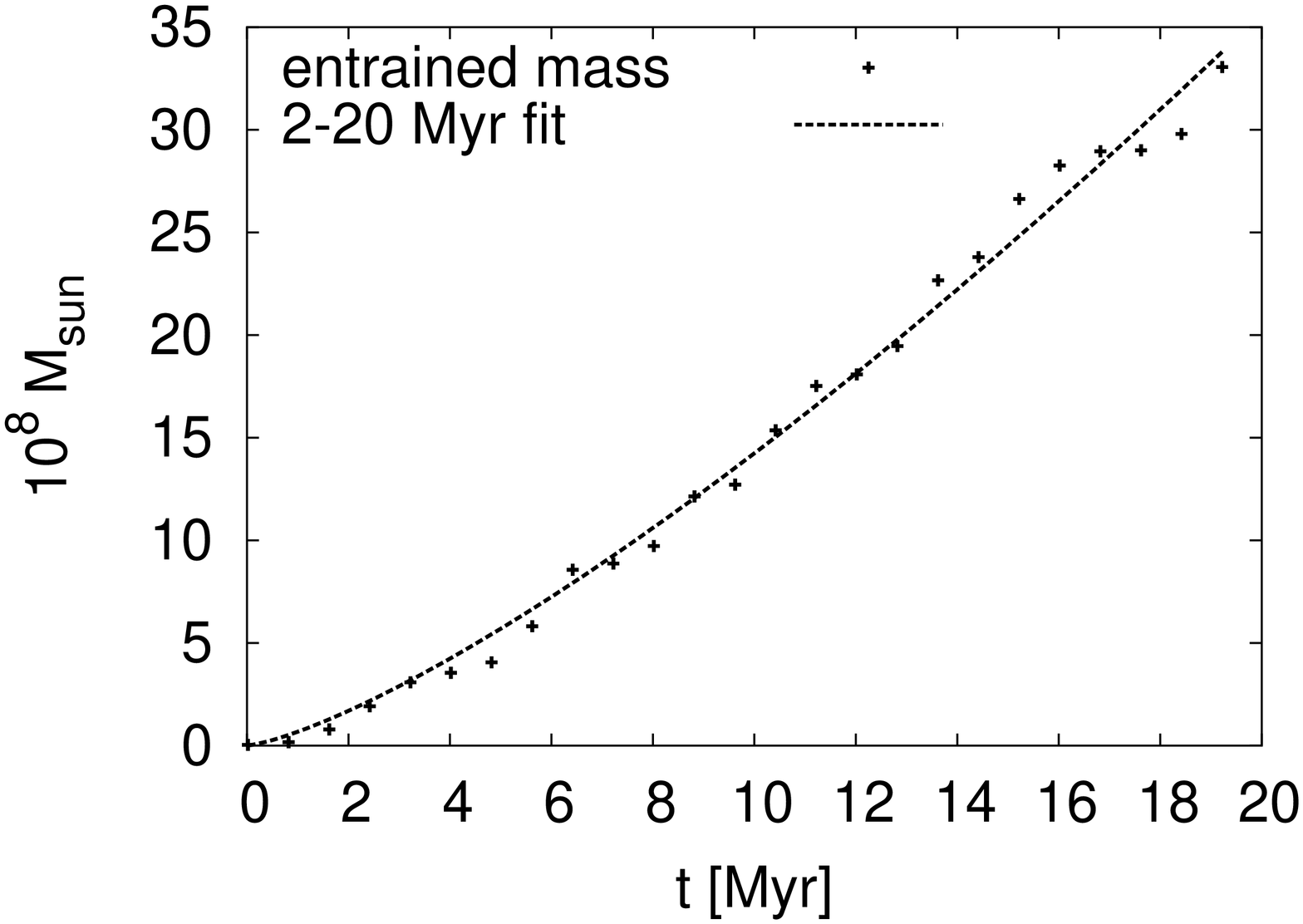}}
\rotatebox{-90}{\includegraphics[height=.47\textwidth]{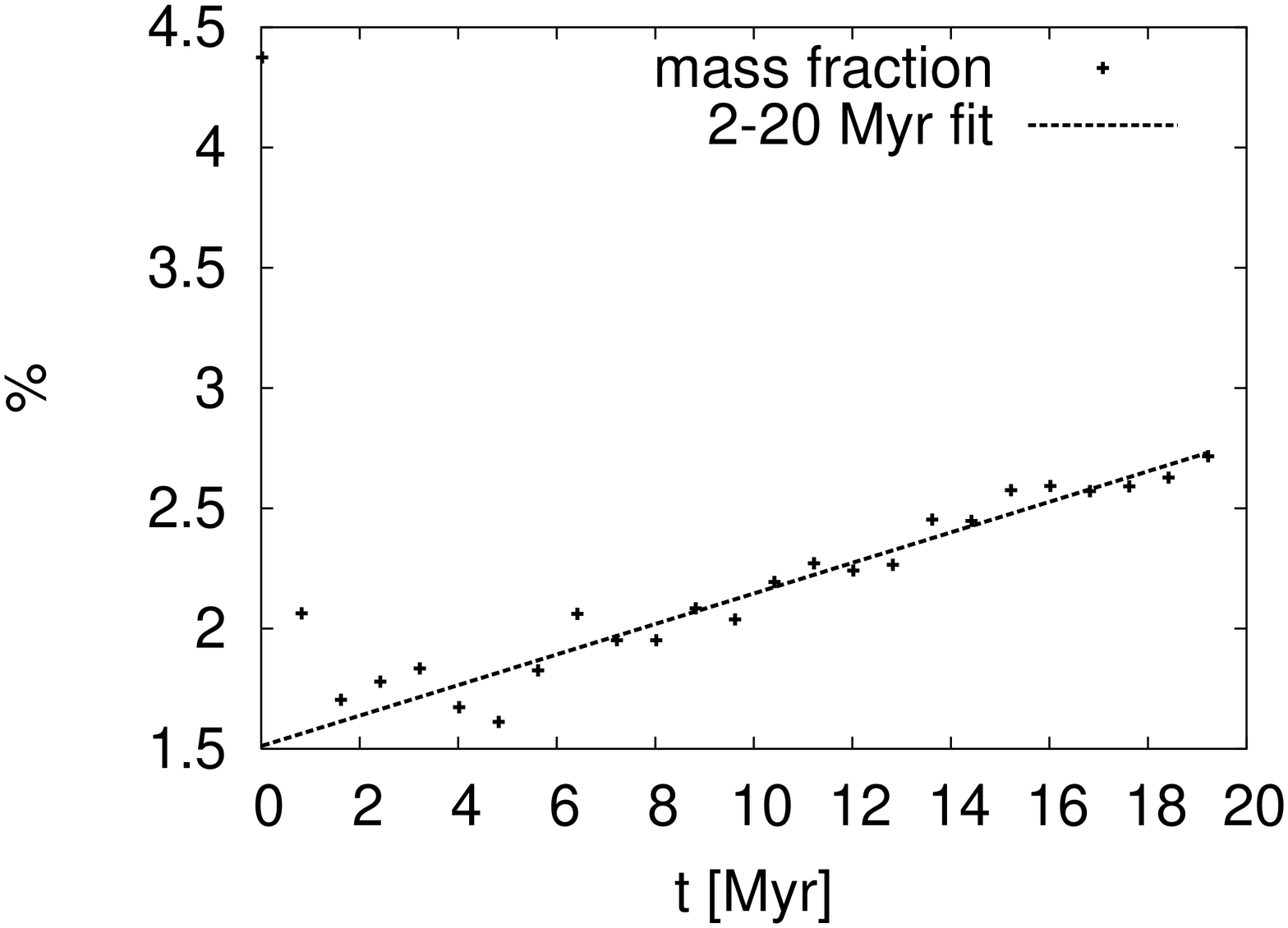}}
%
%
\caption{\small Entrained gas mass in the cocoon over time. Left: Absolute values, the fit function is: 
$0.677 (t/\mathrm{Myr})^{1.32}$.
Right: Percentage with respect to the mass of the initial condition in the same volume, the fit function is: 
$1.5+0.06(t/\mathrm{Myr})$.}
\label{met}       
\end{figure*}

\add{The entrainment rate is shown in}
\ignore{This gas is constantly entrained, as can be seen from} Fig.~\ref{met}. \ignore{The entrainment rate}
\add{It} is slightly
rising ($\propto t^{0.32}$). The entrained mass is compared to the mass of the gas that filled the cocoon volume
before the jet activity. This mass fraction is linearly rising, and would reach unity after 1.6~Gyr, if it would continue 
in the same way.

The width of the cocoon relative to that of the bow shock radius ($\lambda$) is also given in Fig.~\ref{condi}.
For self-similar evolution this value should be constant, and detailed self-similar models \citep{HRB98,KA99} place it 
in the range of 0.8 to 0.9. Such a high value is reached at no time. The relative width decreases 
with time to 0.4 (0.26) in the case of the maximum (average) value.
This result is in agreement with simulations by \citet{Zanea03}. 
For the late phase, the average value for $\lambda$ levels off. This may indicate a phase of nearly self-similar
behaviour, only disturbed by the gravity of the swept up gas, as discussed by \citet{Alex02}. 
The slowly growing aspect ratio also supports this view.
The small value of $\lambda$ may be due to the weak bow shock, as discussed in more detail by \citet{Zanea03}.

\begin{figure*}[t]
\centering
\rotatebox{0}{\includegraphics[width=.95\textwidth]{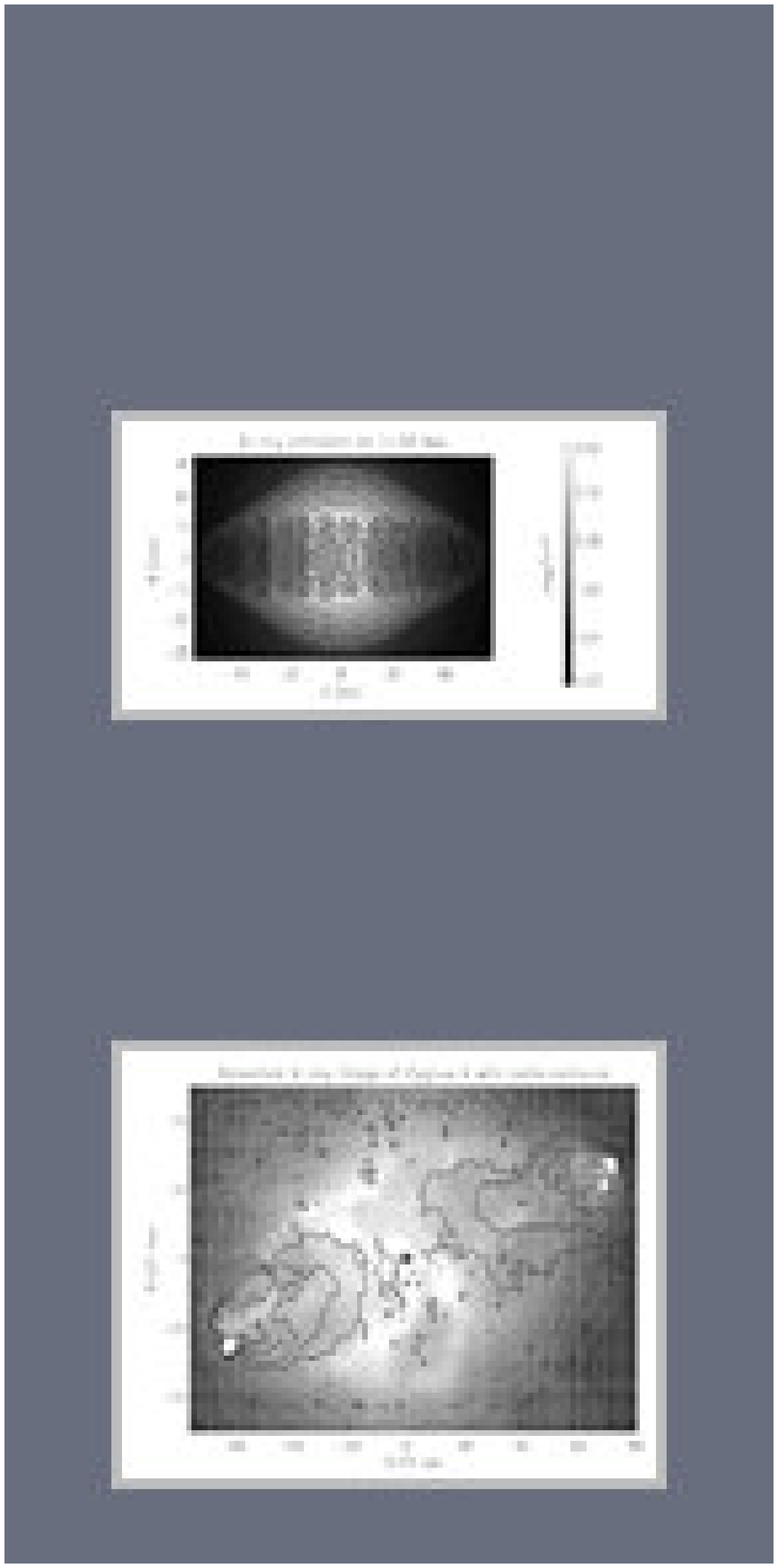}}
%
%
\caption{\small Top: Line-of-sight integrated X-ray emissivity for the 2.5D simulation. Bottom: smoothed X-ray map of Cygnus~A 
(Chandra archive, credits: NASA / UMD / A.Wilson et al.) with 6cm radio contours (VLA, credits: NRAO/AUI, Chris Carilli and Rick Perley, \citep[compare][]{CarBar96}) overlayed. $h$ denotes the Hubble constant in units of 100~km/s/Mpc.}
\label{cygA}       
\end{figure*}

\section{Comparison to observations} \label{obs}
\subsection{Cygnus~A}
This section refers to published data on Cygnus~A \citep{CarBar96,Sea01}.
For illustration, an overlay of the smoothed X-ray and radio data is shown in Fig.~\ref{cygA} below the
synthetic X-ray image of the 2.5D simulation.

The X-ray-radio comparison shows the jet-IGM interaction, impressively. Coincident with the radio cocoon,
there is a deficit of X-ray emission (disregarding for the moment the bright filaments in the center and the emission from the beam and the core).
As discussed above, and detailed in \citet{CHC97} and \citet{Zanea03}, the parameter $\xi$, i.e. the thickness of the shocked ambient gas region over the bow shock radius, should exceed 0.38.  Although the 2.5D simulation presented here differs from Cygnus~A because of the instable beam,
it should be appropriate for the overall morphology. 
In the simulation, the shocked ambient gas has elliptically shaped X-ray isophotes. \citet{Sea01} 
report elliptically shaped isophotes
from the cocoon boundary out to $66/(h/0.7)$~kpc (i.e. the outer boundary of their annulus four). 
Further out, the spherical fit is the better one.
Identifying the bow shock's radius with that position, together with a cocoon width of $\approx 32/(h/0.7)$~kpc,
results in $\xi=0.76$, in agreement with the discussed limit. Taking the average value
for the position of the contact discontinuity, the 2.5D simulation produces $\xi = 0.74$ at $t=20$~Myr, 
showing that such a high value is indeed possible for sources of that size. 

Using the knowledge of the sideways bow shock position, its aspect ratio (length to width) may be derived, which turns out to be
1.1 (using the hot spot separation of $147/(h/0.7)$~kpc from \citet{CarBar96}, it is 1.1 for the eastern and 1.2 for the western jet). 
From the low aspect ratio, \citet{mypap03a} has concluded a density ratio of $\eta<10^{-3}$. However, also the
$\eta=10^{-4}$ 2.5D~simulation presented here reaches an aspect ratio of  1.8 at the end of the simulation.
Since aspect ratios can only grow \citep{mypap03a}, and the simulated jets are smaller than the observed jet in Cygnus~A,
it follows that the real jet still encounters more mass than the simulated ones. This is certainly due to a shallower 
density profile in the atmosphere of Cygnus~A of $\beta=0.51$ \citep{Sea01} compared to $\beta=0.75$ employed for
the simulations (the latter was the old value from the ROSAT data). At the largest extent of the bow shock in the 2.5D~simulation
this would increase the ambient density by a factor of three, which already might result in the desired reduction of the aspect ratio.
A still lower central value for the density contrast cannot be precluded. However, a density contrast of $\eta\approx 10^{-4}$
would also be supported by the large cocoon width \citep{Rosea99}. Another constraint for the density contrast was presented 
by \citet{mypap03a}, who showed that $\eta>3\times10^{-6}$, because the jet is no longer in the spherical bow shock phase.

It has been shown above that the sideways, pressure-driven part of the bow shock satisfies (\ref{globeqmot})
with good accuracy even at an aspect ratio of~1.8. Given the low aspect ratio of Cygnus~A, this
equation should be very accurate here, and can be used to derive the jet's power and age.
In order to do that, one needs information on the gas distribution before the jet activity.
One requirement is that the distribution should join the density distribution far from the radio lobes, smoothly. Assuming 
it was a King distribution (\ref{kprof}), as found for other clusters of galaxies, the following equation has to be satisfied:
\eq{\label{arholink} \rho_\mathrm{e,0} =1.64 \,(a/10\,\mathrm{kpc})^{-1.53} m_\mathrm{p} /\mathrm{cm}^3}
The core radius $a$ (for the density profile before the jet activity) cannot have been much greater than 
$\approx 20/(h/0.7)$~kpc.
This follows from the enhanced X-ray emission next to the radio cocoon. Such enhancements have been shown 
analytically to appear
in atmospheres steeper than $\rho\propto r^{-1}$ \citep[][and references therein]{Alex02}. In the 2.5D~simulation,
the line-of-sight integrated X-ray emission is strongest next to the radio lobes for positions of the contact discontinuity
in excess of 60\% of the core radius. There is no obvious lower limit to the core radius. However, it will be shown in the
following that the exact value of $a$ is not needed for the present discussion.

Approximating the gas density $\rho_\mathrm{e,0}$ with a constant, 
one infers the following conditions 
for the jet's power and age:
\eq{L=\frac{100\pi}{27} \rho_\mathrm{e,0} r_\mathrm{bow}^2 v_\mathrm{bow}^3 \label{lsimp}}
\eq{t=\frac{3r_\mathrm{bow}}{5v_\mathrm{bow}}\enspace . \label{tsimp}}
An interesting feature of these equations is that the source age is independent of the external density, which remains 
true for the normalisation of arbitrary density distributions.
Applied to the simulation data, this method yields quite accurate values, as demonstrated in sect.~\ref{pow3d}.
In order to apply it to Cygnus~A, the bow shock's sideways 
radius is taken to be 66~kpc (see above), and the velocity from the temperature
jump measured from the X-ray data at that position. According to \citet{Sea01}, the temperature of the preshock
gas is at about 7-8~keV at large radii, falls to 5.4~keV immediately before and then rises significantly to 9.2~keV at the position
where the bow shock was proposed to be situated above. It is therefore a weak shock. 
The shock conditions for a weak shock yield \citep[e.g.][]{Shu92ii}:
\begin{eqnarray}
\nonumber v_\mathrm{bow} & = & \zeta \sqrt{\frac{2\gamma k_\mathrm{B}T_0}{m_\mathrm{p}}} 
= 1313 \zeta \sqrt{\frac{T_0}{5.4\, \mathrm{keV}}} \, \frac{\mathrm{km}}{\mathrm{s}}\\
\nonumber \zeta^2 &=& \frac{8}{5} 
        \left(\frac{T_1}{T_0} -\frac{7}{8} + \sqrt{\left(\frac{T_1}{T_0}-\frac{7}{8} \right)^2 +\frac{15}{64}}\right)  \enspace .
\end{eqnarray}
The measurement indicates a temperature jump $T_1/T_0$ of 1.7, and errors limit it to smaller than 4.
Also, the bow shock velocity should exceed the speed of sound in the preshock gas: 
$c_\mathrm{s}=1313 \sqrt{T_0/5.4\, \mathrm{keV}}$~km/s.
This results in a sideways bow shock velocity of:
\dmath{v_\mathrm{bow}= 2217\pm^{1948}_{904} \, \mathrm{km/s} \enspace .}
Inserting this into (\ref{lsimp}) and (\ref{tsimp}) and assuming an average pre-jet density of  
$\rho_\mathrm{e,0}=0.05 m_\mathrm{p}$~/cm$^3$ yields
$L=4.4 \pm ^{25}_{3.5} \times 10^{47}$~ erg/s and $t=17 \mp ^{8}_{12}$~Myr.

\begin{figure}[t]
\centering
\rotatebox{-90}{\includegraphics[height=0.48\textwidth]{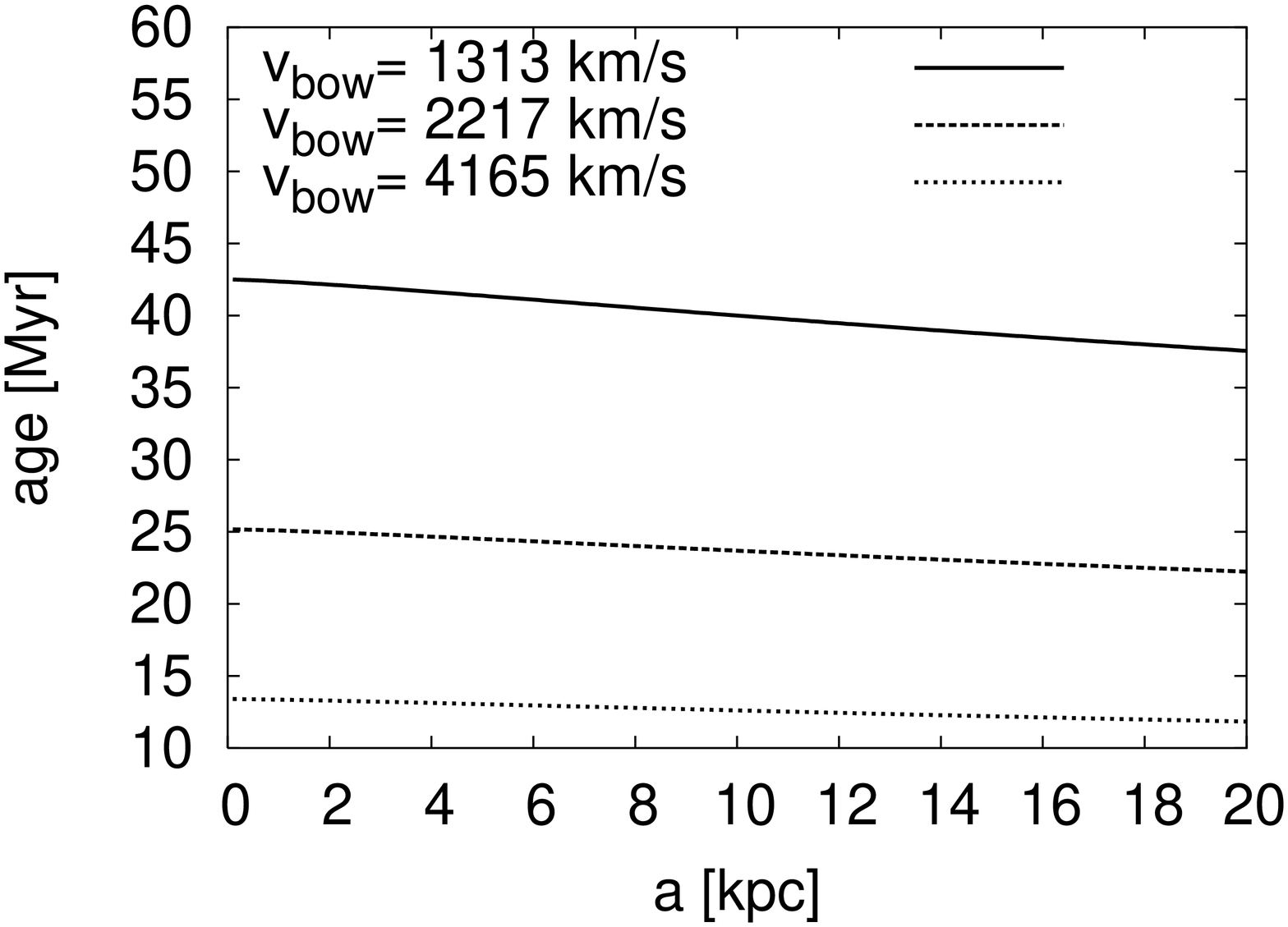}}\\
\rotatebox{-90}{\includegraphics[height=0.48\textwidth]{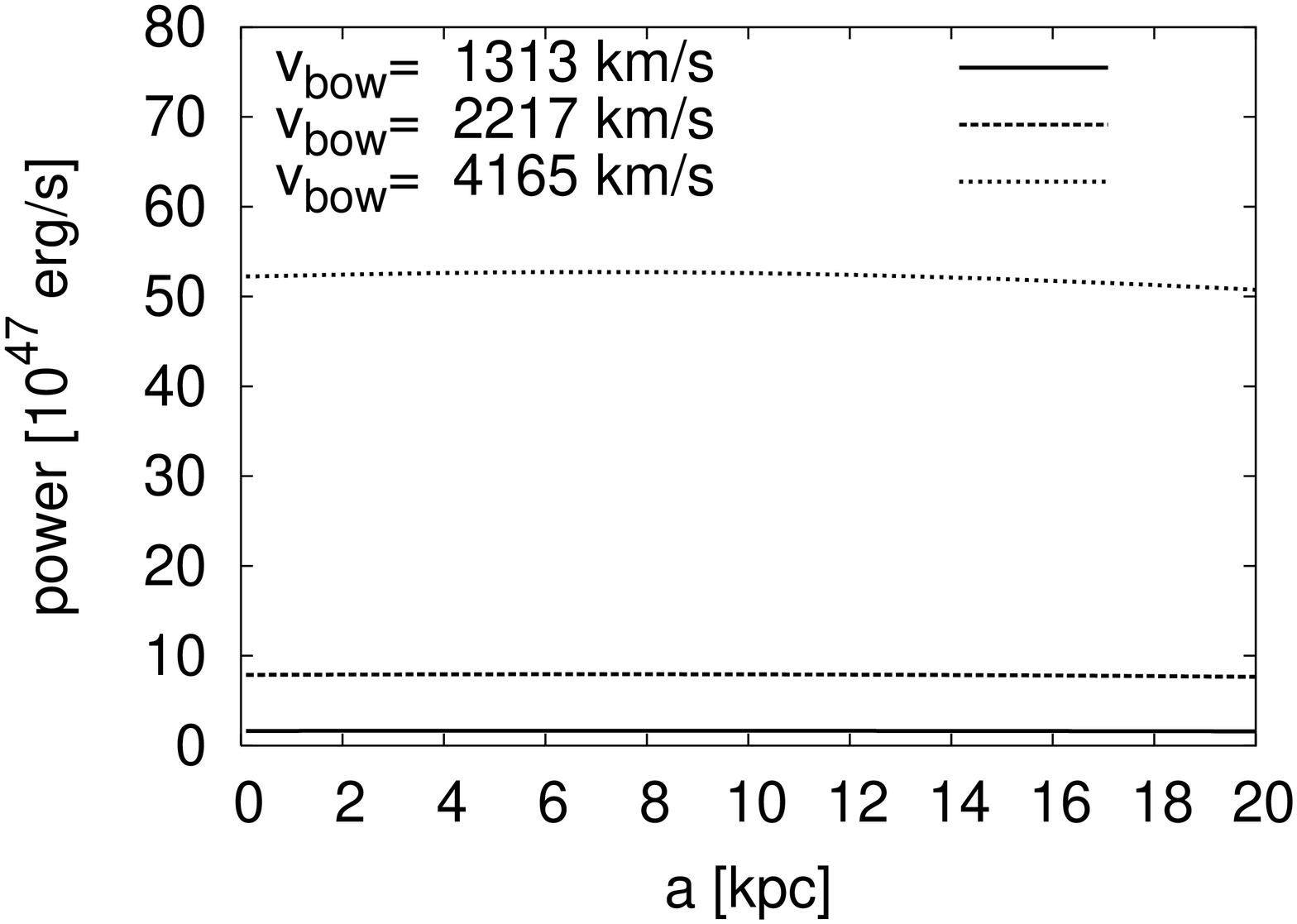}}\\
%
%
\caption{\small Detailed results for age (top), and power (bottom) for Cygnus~A's jets,
assuming $h=0.7$ and the sideways bow shock velocity indicated on the figures.}
\label{cygAdetpar}       
\end{figure}

The parameters can also be estimated assuming a King distribution in the pre-jet era.
It follows from (\ref{globeqmot}):
\begin{eqnarray}
L & = & \frac{4 \pi \rho_\mathrm{e,0} r_\mathrm{bow}^3 v_\mathrm{bow}^3}{9 a} 
\frac{I(r_\mathrm{bow}/a)^3}{J(r_\mathrm{bow}/a)^2} \\
t & = & \frac{3a^2}{r_\mathrm{bow} v_\mathrm{bow}} \frac{J(r_\mathrm{bow}/a)}{I(r_\mathrm{bow}/a)}\\
I(y)& = & \int_0^y x^2 (1+x^2)^{0.765} \, \mathrm{d}x  \nonumber \\
J(z)& = & \int_0^z I(y) y \, \mathrm{d}y \enspace , \nonumber 
\end{eqnarray}
where $\rho_\mathrm{e,0}$ is given by (\ref{arholink}). The result is shown graphically in Fig.~\ref{cygAdetpar}.
It turns out that all the parameters are only weakly dependent on the core radius.  The average value for each curve
is given in the following summary:
\begin{eqnarray}
L & = & 7.9 \pm ^{44.3}_{6.3} \times 10^{47} \, \mathrm{erg/s} \nonumber \\
t & = & 24 \mp ^{11}_{16} \, \mathrm{Myr} \nonumber \\
E & = &  5.9 \pm ^{14.9}_{3.8} 10^{62} \mathrm{erg} \enspace , \nonumber
\end{eqnarray}  
where E is the total energy released by the jets.
As discussed above, in order to obtain the power of the beam, the flux of internal energy through the 
bow shock has to be subtracted and the gravitational energy increase has to be added.
A reasonable estimate of the internal energy contained within the bow 
shock region before the jet activity can be obtained by 
assuming an isothermal King atmosphere. Adopting a core radius of 10~kpc, the X-ray data far from the center 
is consistent with a central pressure of $p_0=10^{-9}$~erg/cm$^3$. This leads to a total of $10^{60}$~erg, which is
negligible here. The gravitational energy increase is not so easy to estimate. But for the present discussion it is
sufficient to say that the 2.5D~simulation indicates that up to 30\% of the jet power is used to lift up gas. 
Adopting a value of 20\%, the power through the jet channel is determined to $L>1.9\times 10^{47}$~erg/s.
Neglecting this rather uncertain contribution, the lower limit to Cygnus~A's jet power
is $1.6\times10^{47}$~erg/s.

This has to be compared to the power in the jet channel. For a non-relativistic jet:
\vbox{
\begin{eqnarray}
L_\mathrm{kin} & = & \pi \rj^2 \rho_\mathrm{j} v_\mathrm{j}^3 \nonumber \\
  & = & 5\times 10^{45} \,(\rj/0.57\, \kpc)^2 (10^4 \eta_0) (\rho_\mathrm{e,0}/m_\mathrm{p} \, \mathrm{cm}^{-3}) \nonumber \\
 & & (v_\mathrm{j}/(c/2))^3 \, \mathrm{erg/s} \enspace . \nonumber
\end{eqnarray}
}
The jet's radius was adopted from \citet{CarBar96} using the latest value for the Hubble constant, $h=0.72$ \citep{Spea03},
and the subscript zero denotes values in the center.
Hence, even when using most optimistic numbers, the kinetic jet power falls short of the lower limit derived above,
by a factor of ten or more. This means that the internal energy cannot be responsible for the rest, since this would yield 
a subsonic jet. Therefore the energy should be in the (toroidal) magnetic field:
\begin{eqnarray}
L_B &=& 2 \pi \rj^2 u_B \Gamma^2 v_\mathrm{j} \nonumber \\
      &=& 5 \times 10^{44} (\rj/0.57\, \kpc)^2 \beta \Gamma^2 (u_B/u_{B,\mathrm{HS}}) \enspace . \nonumber
\end{eqnarray} 
Here, $\Gamma$ is the bulk Lorentz factor $\beta=v_\mathrm{j}/c$, and $u_{B,\mathrm{HS}}=9\times 10^{-10}$~erg/cm$^3$ 
is the magnetic energy density measured in the hot spots by \citet{WYS00}. It is again an upper limit, since the 
magnetic field in the jet will be lower than in the hot spot, where it is shock compressed.
It is clear that the derived power can also not be raised by the magnetic energy flux, as long as we assume a non-relativistic jet.

\begin{figure}[t]
\centering
\rotatebox{-90}{\includegraphics[height=0.48\textwidth]{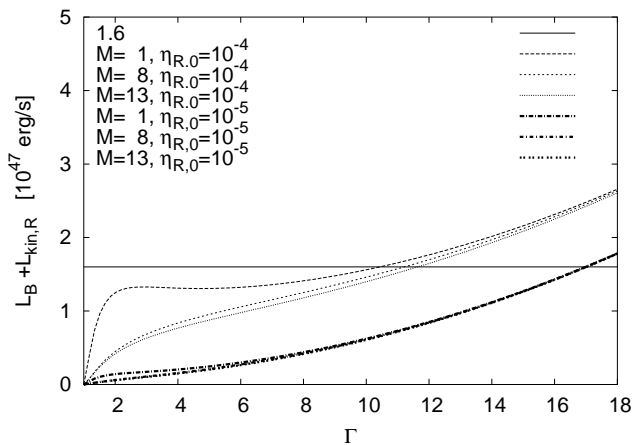}}\\
%
%
\caption{\small Power through the jet channel of Cygnus~A over the Lorentz factor compared to the lower limit
derived in the text. Curves for differing Mach numbers are almost identical.}
\label{powsum}       
\end{figure}

Therefore, the jets in Cygnus~A have to be relativistic.
The kinetic power for a relativistic jet may be written as follows.
\begin{eqnarray} \nonumber
L_\mathrm{kin,R} & = & 2 \pi \rj^2 \eta_\mathrm{R} \rho_\mathrm{e} (1- (\Gamma h)^{-1})\beta c^3 \\
                           & = & 8.8 \times 10^{46}  (\rj/0.55\, \kpc)^2 (10^4 \eta_\mathrm{R,0})  \nonumber \\
                                   &    & (\rho_\mathrm{e,0}/m_\mathrm{p} \, \mathrm{cm}^{-3})
                                        (1- (\Gamma h)^{-1}) \beta \, \mathrm{erg/s} \nonumber \\
\eta_\mathrm{R}  & = & \rho_\mathrm{j} h \Gamma^2 / \rho_\mathrm{e} \enspace , \nonumber
\end{eqnarray}
where $h=1+\gamma e/ \rho c^2$ is the specific enthalpy. $\eta_\mathrm{R}$ denotes the relativistic generalisation of the density 
contrast. This number should be similar to the value for the non-relativistic $\eta$ \citep[compare ][]{Rosea99}. 
The central cluster density $\rho_\mathrm{e,0}$ has a weak dependency on the core radius
which is subsumed in a variation of $\eta_\mathrm{R,0}$, which is regarded to have a value below $10^{-4}$.
This can be combined with constraints on the relativistic Mach number $M=\Gamma \beta/ \Gamma_\mathrm{s} \beta_\mathrm{s}$,
where the index~s denotes the values for the sound speed \citep[e.g.][]{Martea97}. The sound speed is given by
$c_\mathrm{s}^2 = \gamma p / h \rho = (\gamma-1)(1-1/h) c^2$,
which yields for the specific enthalpy:
\begin{equation} \label{hdet}
h=\frac{M^2+\Gamma^2\beta^2}{M^2+g \Gamma^2\beta^2}, \,\,\,  
g=\frac{\gamma-2}{\gamma-1} \enspace.
\end{equation}
Since Cygnus~A shows a stable and laminar jet, the Mach number has to exceed unity.
Most likely the Mach number is greater than that.
\citet{CarBar96} infer a Mach number of $M\approx 8$ (from the oblique shocks) and $M<13$ (from the opening angle).
The total power derived in this way is plotted against the Lorentz factor in Fig.~\ref{powsum} for different
values of the central density contrast and Mach number. The lower power limit derived above results in a Lorentz
factor around 10, the central value gives 37 to 39. 
An important result is that the jet's Lorentz factor should exceed ten for reasonable assumptions.
According to (\ref{hdet}) $M$ should then exceed 14, in order to get $h>1$, which might point to 
a significant Alfv\'en speed.

If one would regard a Lorentz factor above twenty as unreasonable, the jet power could be further constrained to less than
$3\times~10^{47}$~erg/s. 

%
%

%
%

The derived jet power exceeds the total emitted radio power by at least a factor of $\approx 100$ \citep{CarBar96}, and might be compatible 
with estimates by \citet{KA99} ($2\times10^{46}$~erg/s) and \citet{Zanea03} ($3\times10^{46}$~erg/s), who do not state 
their errors. 

The simulations have shown that the Kelvin-Helmholtz instability produces large fingers in the central regions of the cocoon.
These may be identified with the belts of cooler (4~keV) gas inside the cocoon of Cygnus~A.
Additionally, the contact discontinuity might be Rayleigh-Taylor unstable \citep{Alex02}. The 2.5D-simulation shows that the
deceleration is stronger at the beginning, and falls below the critical value for most of the simulation time.
Also, local shaking dominates over the global deceleration. Therefore the contact discontinuity is expected to be Rayleigh-Taylor unstable.

\subsection{Hercules~A}
Radio and ROSAT data on Hercules~A are given by \citet{GL03,GL04}. 
This source is an intermediate one,
showing huge radio lobes like FR~II sources, but no edge brightening. 
It seems to be in a similar state
to the 2.5D~simulation towards the end, presented above. 
The jet is unstable, probably because it cannot 
reach the required Mach number in order to stabilise it up to the tip of the cocoon.
The whole cocoon seems to be turbulent, and entraining external gas across the contact discontinuity.
This might be the reason for the lack of an X-ray deficit as observed in Cygnus~A.
However, the ROSAT data may lack the necessary resolution.
What it does show is that the X-ray gas is elongated in the direction of the radio source. 
This can be interpreted
as a wide region of shocked ambient gas, as in Cygnus~A.
 
\subsection{Abell 2052}
X-ray, radio, emission line, and other data on this central cluster galaxy can be found in 
\citet{Blanea01,Blanea03}. 
It is an old radio galaxy, the radio emission is diffuse, and no sign of hot spots or 
beams has been found.
The diffuse radio emission is devoid of X-rays, and is surrounded by bright X-ray shells. 
\citet{Blanea01,Blanea03}
argue that the shells consist of two oppositely situated bubbles, as found in other X-ray clusters. 
However, the study of
the dependency of the X-ray emission on the viewing angle suggests that the situation may be similar 
to 
the ten degree plot of Fig.~\ref{emimapb}: A low density jet may have blown an approximately 
spherical bubble.
After reaching the critical radius (see above), a jet with a narrow beam, i.e. high thrust
bored a cigar-like extension into the shell. 
Now, we view the cigar-like extension from such an angle that it partly overlaps
with the big shell, producing enhanced emission at those parts of the X-ray rings.

An interesting point in this interpretation is that when modeling the X-ray enhanced regions by 
two elliptical, overlapping rings,
these are elongated in the same direction, and deprojecting to a circular ring results in an 
inclination of roughly
$37^\circ$ for both. This inclination is compatible with the estimate of $\approx 45^\circ$ 
from the radio data by
\citet{Blanea03}. 

A question concerning this interpretation arises from the temperature map of the X-ray shells. 
The smaller shell seems to be colder 
than the larger one. While the uniform temperature of each of the 
elliptical shells is consistent with their proposed nature,
the cigar part, corresponding to the smaller elliptical ring, 
is always hotter in the simulations than the inner bow shock part.
This problem might be alleviated by the fact that the jet in this source is presently not active. 
Therefore, the radio plasma that 
was present in the cigar may have left it in the backflow, leaving behind an underpressured region. 
The gas in the shell
would then expand somewhat into the cigar cavity, thereby lowering its pressure. For that reason, 
the expansion velocity
of the leading bow shock would slow down. Assuming a cylindrical shell for the cigar part, 
the adiabatic expansion 
($T\propto V^{-8/3}$) would require a change of the inner radius of the shell from e.g. 80\% to 70\% 
of the outer one,
in order to explain a temperature change by a factor of two.

Following the smaller X-ray shell that has been interpreted as cigar part above, 
there are emission line filaments \citep{Blanea01}.
These might be due to weak radiative shock waves inside of the shell 
that might have been excited by the re-expansion described above.  

\subsection{Radio galaxies at higher redshift}
Radio galaxies with redshifts greater than $z \approx 0.6$ show associated 
large scale emission line regions that are aligned 
with the radio jets \citep{MC93}. The aligned emission line regions 
are often located where one would expect the radio cocoon,
and are stronger towards the center \citep[e.g. ][]{BLR96}.
It has been shown in this paper that the central cocoon regions can entrain 
several percent of the gas mass that was 
situated there before the jet event. The X-ray belts in Cygnus~A have been 
interpreted above as evidence for this entrainment process. These belts are 
the coldest X-ray gas found in the Cygnus~A system. 
It is therefore possible that such entrained gas
reaches thermal instability, if the cluster's gas density is higher, 
or the source has some more time to cool. The cooling time for these belts is some 100~Myr. 
One can therefore imagine a situation where the entrained gas 
cools down to emission line temperatures whereas the other shocked ambient 
gas does not cool. 
As suspected above for Abell~2052, the cooling could also proceed within 
X-ray filaments that contain weak, radiative shocks.
This would explain why the line ratios for some sources (preferentially the smaller) 
are consistent with shock excitation
\citep[e.g. ][]{Inskea02c},
without relying on the compression of pre-existing clouds by the bow shock.
Wether this picture can cope with other observational data has yet to be explored.

\section{Summary and conclusions}
Bipolar jet simulations have been presented in 2.5D~and~3D. For the 3D~simulation, 
a cylindrical grid was employed,
and the boundary conditions were given. It has been shown that the influence of the 
axial boundary remains
acceptably small. 
Because of small disturbances in the ambient medium, the backflows are located at 
different distances from the jet axis,
and permeate each other in the center. There, a turbulent mixing region emerges that 
entrains shocked ambient gas across
the contact discontinuity via Kelvin-Helmholtz and Rayleigh-Taylor instabilities. 
After a certain time, 
both simulations show two bow shock phases. The shape of the outer one is well fitted by a parabola. 
The inner part has an elliptical shape and stays
axi-symmetric, even in the 3D-simulation. It follows the blastwave equation of motion 
with good accuracy (a few percent
difference in the exponent of a local power law fit), even when the aspect ratio is high.
The aspect ratio keeps growing for the whole simulation time, i.e. a self-similar regime 
is not reached.
The viewing angle-dependent emission maps show that the different parts of the bow shock may 
form circular and 
elliptical rings, and produce elliptical isophotes when viewed at high inclination. 
Varying the inclination can also produce X-ray
deficits.
At late times of the 2.5D~simulation, the beam is unstable, barely reaching the tip of the bow shock.
This could be expected from stability analysis, because a non-relativistic very light jet cannot keep a high Mach number, 
which is necessary for stability.
Regarding the propagation of the tip of the bow shock, an armlength asymmetry of a few percent was measured in both simulations.
For late times, the jets are faster on the side with the on average higher density. 

The X-ray structure in Abell~2052 was shown to be explainable by two elliptical rings, one from the outer bow shock part
of a former jet and one from the inner one. From the ellipticity, an inclination of $37^\circ$ is derived consistent with other estimates in the literature. Based on the simulation results, the low inclination is one reason for the pronounced 
X-ray deficit inside of the rings.

The X-ray gas in Hercules~A and Cygnus~A shows elliptical isophotes elongated in the direction of the radio jets.
The ellipticity decreases at a certain distance from the center. The location where the ellipticity drops was identified
as being the bow shock position. For Cygnus~A, the quality of the data is sufficient to determine the sideways bow shock position 
to 66~kpc. At the same position, a temperature jump can be infered from the X-ray data, implying a sideways bow shock velocity
between 1300~km/s and 4200~km/s, i.e. a Mach number between one and three. The width of the radio cocoon is only a
quarter of the bow shock width, which is consistent with the observation of an X-ray deficit, and shown to be possible in the
2.5~D simulation presented here. However, in self-similar models, this fraction is usually above 80\%, which is a shortcoming
of these models. Applying the simulation results and using the width and velocity of the sideways bow shock,
a jet power of $>10^{47}$~erg/s and an age of $\approx 24$~Myr is derived. This result was found to be inconsistent
with the jet being non-relativistic, and a lower limit for the Lorentz factor of ten was infered.

Explaining the belts of low temperature X-ray 
gas within the radio cocoon of Cygnus~A by entrained gas over the contact discontinuity,
it was suggested that such gas could cool further to form emission line regions, if the density of the ambient gas 
would be somewhat higher. This scenario could be realised at higher redshift and explain the origin of the gas 
producing the alignment
effect. 

\begin{acknowledgements}
This work was supported by the Deutsche Forschungsgemeinschaft (Sonderforschungsbereich 439).
The computations have been carried out on the NEC-SX5 of the 
HLRS in Stuttgart (Germany). I thank P.~Strub for help with the X-ray data, and C.~Carilli for providing
the radio data.
\end{acknowledgements}

\bibliographystyle{apj}
\bibliography{/home/mkrause/KDesktop/text/LATEX/texinput/references}

\appendix
\section{Late phase of spherical blastwaves}
\label{bw}
For which radii is it possible to use the blastwave approximation, i.e. the solutions
of (\ref{globeqmot})? This equation was derived under the assumption of vanishing external
pressure, which is equal to a strongly supersonic shock. 
In the simulations we get weak bow shock Mach numbers very soon.
Then, the external pressure ($p_\mathrm{ext}$)
and possibly also the gravitation (remember that the atmosphere
is held in equilibrium of pressure and gravity) has to be taken into account.
This will be discussed in the following for the case of isothermal power law atmospheres,
applicable to the presented work.

\begin{figure}[t]
\centering
\rotatebox{-90}{\includegraphics[height=0.48\textwidth]{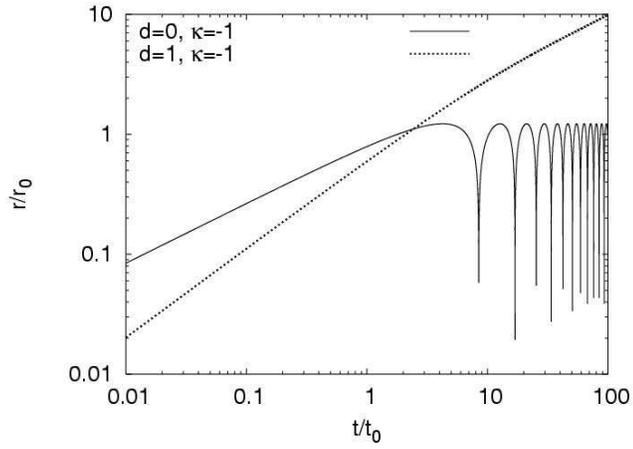}}\\
%
%
\caption{\small Numerical solutions of the force balance equation for late spherical blast waves. The blast wave with
constant energy injection changes the power law slope at the critical radius. The one with finite energy falls back.}
\label{bll}       
\end{figure}

The force ballance equation for a shell, driven by energy input $E(t)={\cal L}t^d$ 
into an environment
with gas mass profile ${\cal M}(r)=\int_0^r 4\pi r^2 \rho(r) \mathrm{d}r$, 
where the density distribution is given by $\rho(r)=\rho_0 (r/r_0)^\kappa$, 
can be written as follows:
\begin{equation}
\frac{\partial}{\partial t} ({\cal M} v)= S(p_\mathrm{int}-p_\mathrm{ext})
        -\frac{G {\cal M}_\mathrm{DM} {\cal M}}{r^2}\enspace.
\end{equation}
 Here, $S=4 \pi r^2$, $G$ is the gravitational constant, $ {\cal M}_\mathrm{DM}$
denotes the gravitating Mass profile of the dark matter halo,
and the internal pressure is given by $p_\mathrm{int}=2(E(t)-{\cal M}v^2/2)/3V$, where $V=4\pi r^3/3$.
Using hydrostatic equilibrium, this equation can be nondimensionalised and rearranged in the following way:
\begin{equation}\label{nondim}
\frac{\partial^2}{\partial^2 (t/t_0)}\left( \frac{r}{r_0} \right)^{\kappa+5}
        =(t/t_0)^d -\left( \frac{r}{r_0} \right)^{\kappa+3} \enspace .
\end{equation}
The scales $r_0$ and $t_0$ are given by:
\begin{eqnarray}
r_0&=& \sqrt{9(\kappa+5)/5} c_\mathrm{s} t_0 \nonumber\\
t_0^{(d-3)/5}&=& \left(\frac{9}{5}\right)^{1/2} \left(\frac{2\pi}{\kappa+3}\right)^{1/5} (\kappa+5)^{3/10}
\left(\frac{\rho_0}{\cal L}\right)^{1/5} c_\mathrm{s}
\enspace . \nonumber
\end{eqnarray}
For typical values of galaxy clusters, say $\rho_0=10^{-26}$~g/cm$^3$, constant sound speed 
of $c_\mathrm{s}= 1000$~km/s, and ${\cal L}=10^{47}$~erg/s for $d=1$ the typical scale is Mpc, 
reducing by a factor of a few when the jet turns off and the energy stays constant at some $10^{62}$~erg ($d=0$).
The force balance equation was numerically integrated, an example is shown in Fig.~\ref{bll}.
The general result is that blast waves with finite energy fall back after reaching the critical radius, 
the formal solution being oscillatory.
The ones with constant energy injection break and change the slope of their power law at that radius.
Up to that radius, the propagation follows the same power law that can be derived using the strong bow shock hypothesis.
The conclusion is that for the simulated jets, and the typical real sources, the critical radius has not yet been reached,
and the blastwave equation of motion can be expected to be a good approximation as long as non-spherical 
effects remain small.

\end{document}